\documentclass[preprint,12pt]{elsarticle}

\usepackage{amssymb}
\usepackage{amsmath}
\usepackage{graphicx}
\usepackage{booktabs}
\usepackage{array}
\usepackage{placeins}
\usepackage{xurl}
\usepackage{lineno}
\usepackage[bottom]{footmisc}
\usepackage[hypertexnames=false]{hyperref}
\newcolumntype{L}[1]{>{\raggedright\arraybackslash}p{#1}}
\newcommand{\WBCPanel}[3]{%
    \begin{tabular}{@{}c@{}}
        \includegraphics[width=#1]{#3}\\[1.5pt]
        \parbox[t]{#1}{\centering\scriptsize #2}
    \end{tabular}}
\newcommand{\WBCPanelGap}{\\[4pt]}

\graphicspath{
  {./}
  {./figures/}
}

\journal{Sustainable Cities and Society}

\begin{document}

\begin{frontmatter}

\title{Urban Wind Effects on UAV Operations in Building-Dense Low-Altitude Airspace}

\author[2,6]{Yue Cao}

\author[5]{Huanxia Wei\corref{cor1}}
\ead{huanxia.wei@manchester.ac.uk}
\author[4]{Chao Xia}
\author[1]{Qing Jia}
\author[6]{Yingying Xing}
\author[1,7]{Zhigang Yang}

\affiliation[2]{organization={Department of Civil and Environmental Engineering, University of California Berkeley},
         city={Berkeley},
         postcode={94704},
         country={United States}}

\affiliation[5]{organization={Department of Mechanical and Aerospace Engineering, The University of Manchester},
         city={Manchester},
         postcode={M13 9PL},
         country={UK}}

\affiliation[1]{organization={College of Automotive and Energy Engineering, Tongji University},
         city={Shanghai},
         postcode={201804},
         country={China}}
         
\affiliation[4]{
  organization={Department of Vehicle Engineering and Autonomous Systems, Chalmers University of Technology},
  city={Gothenburg},
  postcode={SE-412 58},
  country={Sweden}}

\affiliation[6]{
  organization={The Key Laboratory of Road and Traffic Engineering of Ministry of Education, Tongji University},
  city={Shanghai},
  postcode={201804},
  country={China}}

\affiliation[7]{
  organization={Beijing Aeronautical Science and Technology Research Institute, The Commercial Aircraft Corporation of China},
  city={Beijing},
  postcode={102211},
  country={China}}

\cortext[cor1]{Corresponding author.}

\begin{abstract}
Urban low-altitude uncrewed aerial vehicle (UAV) operations are shaped by building-modified wind and constrained terrain-building clearance. Spatially heterogeneous wind can perturb short-horizon motion, but its clearance consequence depends on local airspace: similar perturbations may remain benign in open areas yet become clearance-sensitive near building walls, roof edges, or elevated terrain. This study develops a common-basis method for early clearance assessment before route buffers or corridor alignments are fixed. For each position-height state, nominal and wind-affected vehicle-sized trajectory envelopes are generated with a shared heading and sampling design, separating fixed-support wind loading, initial clearance exposure, and wind-induced clearance change. Applied to Shanghai and Beijing, the method uses lattice-Boltzmann method (LBM) wind fields co-registered with 20 m terrain-building scene grids, height layers of 60-340 m, and a representative light UAV. Results show distinct spatial and vertical patterns in baseline wind loading and clearance exposure. Wind-affected envelopes produce little domain-wide median shift in clearance exposure but generate spatially concentrated upper-tail responses in lower-altitude layers on the normalized clearance-exposure scale. Kilometer-Tile-level analysis links these responses mainly to baseline clearance sensitivity and horizontal wind-gradient magnitude, with fixed-support wind loading and mean wind speed playing secondary roles. Low-altitude planning should distinguish wind-loaded locations from clearance-sensitive locations. As an upstream diagnostic, the method identifies units where nominal trajectory envelopes approach terrain-building margins and local wind variation may further alter clearance exposure, with increases concentrated in the upper tail.
\end{abstract}

\begin{highlights}
\item Paired local envelopes separate wind burden from clearance response.
\item Shanghai and Beijing show small medians but lower-layer response tails.
\item Upper-tail response links to clearance sensitivity and wind variation.
\item Wind-loading maps miss clearance-sensitive route-screening locations.
\end{highlights}


\end{frontmatter}

\section{Introduction\label{sec:introduction}}

As a critical part of urban air mobility (UAM) systems, urban low-altitude uncrewed aerial vehicles (UAVs) fly in the part of the atmospheric boundary layer most strongly modified by buildings and terrain. Wind speeds up around building walls and roof edges, turns in street canyons, and forms wakes, shear layers, and gusty flow behind irregular city blocks. These flow changes can move an aircraft away from its intended local path. The clearance consequence depends on the space around the aircraft. The same shift matters little over open ground due to the correction of flight control unit, but it can reduce the distance to a building wall, roof edge, or raised terrain in a tight part of the city. Early low-altitude airspace assessment therefore needs to evaluate where building-modified wind and nearby obstacles jointly increase terrain-building clearance exposure.

Urban wind studies explain why this localized view is needed by introducing the coupling effect from a flow dynamics perspective, especially with the help of computational fluid dynamics (CFD). Belcher \cite{belcher_mixing_2005} described urban-canopy transport as the combined effect of mean-flow advection, turbulent mixing, and street-building geometry, showing why canopy-scale wind exposure varies strongly with urban form. Barlow et al. \cite{barlow2008relating} related urban surface-layer structure to upwind terrain in the Salford experiment, showing that incoming wind history and surrounding roughness influence the wind field inside a district. Tominaga et al. \cite{tominaga_aij_2008} set out practical CFD guidelines for pedestrian wind around buildings. Blocken et al. \cite{blocken_cfd_2012} combined meteorological statistics, numerical results, and validation data in a campus wind-comfort and wind-safety framework. Toparlar et al. \cite{toparlar_cfd_2015} coupled wind flow, radiation, convection, and conduction in a Rotterdam district and validated urban microclimate CFD against satellite-derived surface temperatures, showing why flow resolution matters for district-scale screening. Palusci et al. \cite{palusci_impact_2022} used Reynolds-averaged Navier-Stokes (RANS) simulations in a compact Rome district and found strong links between building-wall area density, building density, and reduced ventilation. Vanky et al. \cite{vanky_evaluation_2024} tested an immersed-boundary CFD framework against wind-tunnel and body-fitted simulations for irregular urban terrain, showing the need to resolve terrain when elevation changes are comparable with building-scale flow structures. These studies explain how urban form creates local flow structures and how those structures should be modelled.

Existing UAV and UAM studies employed high-resolution numerical and observational wind assets to investigate the impact of environmental wind conditions on UAV flight response and dynamic airspace management. Dai et al. \cite{dai_urban_2025} reviewed aerodynamic interactions, thermal effects, and simulation challenges in the built environment, identifying building-induced turbulence, wake vortices, wind shear, rooftop effects, and vertiport-scale airflow as safety-relevant UAM concerns. Jia et al. \cite{jia_ten_2026} organised the urban-wind challenge for UAM around flight stability, rooftop operation, airflow prediction, wind-tunnel and CFD assessment, and wind-aware route planning. Steiner \cite{steiner_urban_2019} identified urban-scale observing, microscale prediction, and impact-based weather guidance as weather-community needs for UAM operations. Reiche et al. \cite{reiche_initial_2021} used climate observations and public-perception surveys to assess weather barriers across candidate UAM operating heights in ten metropolitan areas, showing that weather constraints vary by city and time of day. Giersch et al. \cite{giersch_atmospheric_2022} used the Parallelized Large-Eddy Simulation Model (PALM), a large-eddy simulation (LES) model, to resolve meter-scale turbulent wind over an urban district and argued that turbulence-resolving urban wind databases can support safe drone mission planning. Chrit \cite{chrit_reconstructing_2023} developed a reduced-order data-assimilation system that combines RANS predictions with sparse observations and LES-based reference fields, reducing urban wind-field error and changing hazardous-area estimates for UAV navigation. Lee et al. \cite{lee_efficient_2026} proposed dictionary-learning and gated recurrent unit (GRU)-based super-resolution of urban wind fields for UAM flight-safety assessment, showing one path from costly microscale simulations to faster operational wind products. Vuppala et al. \cite{vuppala_modeling_2024} coupled representative UAM aircraft dynamics with LES and reduced-order realistic urban winds, finding that time-mean wind fields under-predict lateral response and control-surface activity for the studied wing-borne case. Asignacion et al. \cite{asignacion_accurate_2023} combined drone-mounted wind sensing, nonlinear disturbance observation, and robust quadrotor control experiments, linking local wind observation with vehicle-level disturbance rejection. Jeong et al. \cite{jeong_hazardous_2021} combined Weather Research and Forecasting-LES (WRF-LES), multicopter flight simulation, and a deep neural network to predict hazardous flight regions based on wind-induced path-deviation distance. Seon et al. \cite{seon_urban_2026} extended this direction by training long short-term memory (LSTM) networks on WRF-LES and UAM dynamics simulations to predict a flight-hazard index in corridors, with crosswind and height-dependent wind speed shaping the predicted response. These UAV and UAM studies establish that turbulent city wind can produce vehicle responses missed by pointwise wind speed or uniform inflow, which can bias flight-safety assessment.

For UAV clearance assessment, a wind response becomes meaningful only when it is evaluated against the surrounding geometry. The relevant output is the terrain-building clearance exposure of the local flight envelope after considering wind perturbation. A local deviation over an open plaza, above a river, or over a wide road can leave enough clearance. A similar deviation near a roof edge or terrain break can move the aircraft toward a nearby building or ground surface. Geometry-aware UAV risk studies address parts of this coupling with different outputs. Roseman and Argrow \cite{roseman_weather_2020} proposed a weather hazard risk framework for small uncrewed aircraft systems (sUAS) operations that combines forecast weather, population density, structure density, and aircraft information in a Bayesian safety-risk model. So et al. \cite{so_integrated_2023} used vertiport obstacle analysis as a screening step before multicriteria mobility-hub location selection, showing how geometric constraints can enter early urban mobility planning. Zhao et al. \cite{zhao_flight_2024} represented advanced low-altitude transportation safety through a flight-risk-field model that combines UAV cruising interactions with static building and obstacle effects. Gao et al. \cite{gao_developing_2025} converted public ground-risk constraints into 3D risk-informed no-fly zones for urban UAV operations, showing how detailed urban exposure and geometry can be turned into no-fly volumes. Patil and Garc\'ia-S\'anchez \cite{patil_quantifying_2025} used RANS simulations with different building levels of detail to forecast UAM wind risk from velocity and turbulence fields, finding that higher geometric detail can produce more conservative high-risk maps. Geng and Gou \cite{geng_urban_2026} provide the closest urban-form comparator: they combined CFD-ready urban geometry, a wind-speed and turbulence-based UAV risk index, spatial autocorrelation, and GeoShapley analysis to link urban form with fixed-location risk clustering. That framework answers how urban form relates to local wind-risk index values. The remaining question for clearance assessment is how wind-affected motion changes terrain-building clearance exposure relative to a nominal local path.

Path-planning, corridor-design, and airspace-network studies address a later planning step. Doole et al. \cite{doole_constrained_2021} compared limited one-way and two-way urban drone-delivery airspace concepts above a Manhattan-like street grid, showing how airspace structure changes conflict and intrusion patterns in high-density traffic simulations. Chan et al. \cite{chan_wind_2023} combined Voronoi decomposition, Dijkstra search, energy modelling, and particle swarm optimisation to plan wind-aware, energy-efficient UAV paths in dense urban airspace. Aldao et al. \cite{aldao_dynamic_2025} integrated probabilistic UAV failure paths and detailed 3D urban geometry into multi-objective parcel-delivery trajectory optimisation with third-party risk reduction. He et al. \cite{he_route_2022} formulated tube-based urban air-delivery route-network planning by decomposing a NP-hard network problem into priority-structured single-path problems for dense corridor design. He et al. \cite{he_air_2025} later analysed the computational complexity of city-scale air-corridor planning through multi-commodity network flow and graph-search formulations. Stuive and Gzara \cite{stuive_airspace_2024} designed road-based 3D UAV corridor networks under budget, demand, battery, capacity, and congestion constraints. Chen et al. \cite{chen_optimizing_2026} used Alpha-Shape obstacle extraction and space-syntax analysis to identify free-route airspace and access control points, offering a spatial alternative to fixed road-following corridors. Reviews of UAM airspace concepts and low-altitude airspace management further emphasise that route structures, capacity, separation assurance, communication, navigation, and surveillance performance, certification, and urban planning constraints must be coordinated before large-scale operations \cite{bauranov_designing_2021,pongsakornsathien_advances_2025}. These studies optimise or evaluate routes, corridors, and networks after candidate routes or corridors have been defined. An upstream clearance assessment is therefore needed to flag positions and height layers that require more clearance margin, avoidance, or more detailed analysis before those route buffers and corridor alignments are fixed.

Across these research streams, prior work has established four links: urban form changes local wind; local wind affects aircraft motion; urban geometry shapes flight consequences; and risk information supports route and airspace design. These links are usually analysed with different spatial units, sampling bases, and outputs. It remains unclear whether high wind loading along nominal motion identifies the same positions as high terrain-building clearance exposure after wind-affected motion. The two conditions can differ: a wind-exposed position can retain ample surrounding space, and a more moderately loaded position can be clearance-sensitive when local wind variation moves the aircraft toward a nearby obstacle.

This distinction leads to three research questions. First, do positions with high wind loading along nominal motion also show high terrain-building clearance exposure? Second, how does short-horizon wind-affected motion change clearance exposure across height layers and between cities? Third, which inflow wind and geometric conditions are associated with the upper tail of the resulting clearance response?

Our study addresses these questions by comparing two versions of the same local flight situation before route buffers or corridor alignments are selected. For each position and height, short nominal paths are first traced over the same set of possible horizontal headings. Wind-affected paths are then generated from the same starting point and heading set. Because a UAV has finite size, each centreline path is expanded into a vehicle-sized flight envelope. Terrain-building clearance is evaluated on these envelopes to determine whether wind-affected motion brings the UAV closer to nearby obstacles. Wind loading is sampled on the same envelopes. The paired comparison separates three quantities: the wind burden along nominal motion, the initial terrain-building clearance condition, and the clearance change caused by wind-affected motion.
The same design is applied to Shanghai and Beijing using common grid, height-layer, trajectory-sampling, vehicle-class, and scoring settings. The resulting clearance-exposure and wind-loading fields are analysed through domain-wide vertical summaries, tile-height baseline state space, and explanatory models of upper-tail clearance response. The outputs are intended for early clearance assessment and corridor pre-design. Accident probability, formal operational risk, and regulatory separation compliance remain outside the scope of the method.

The study makes three main contributions. First, it formulates local UAV clearance assessment as a coupled wind-geometry problem and separates wind exposure from geometric clearance sensitivity. Second, it introduces a paired finite-trajectory envelope for computing baseline clearance exposure, trajectory-induced clearance-exposure change, and sampled wind loading on a common basis. Third, it provides a same-design test in two megacities and identifies baseline wind-geometry conditions associated with upper-tail clearance response across multiple height layers.

The remainder of this paper is organised as follows. \S\ref{sec:2_methods} formulates the coupled wind-geometry screening problem and presents the trajectory propagation model, clearance check, and paired exposure measures. \S\ref{sec:3_exp} introduces the inputs and cross-city analysis design. \S\ref{sec:3_ResDis} reports the baseline wind and clearance structures, analyses trajectory-induced responses across height layers, and identifies local conditions associated with upper-tail clearance change. Finally, \S\ref{sec:4_concl} summarises our main findings.

\section{Methodology\label{sec:2_methods}}

The method begins by tracing short nominal paths from each local cell and height across sampled headings. Each centreline path is swept by the representative vehicle footprint; the resulting finite envelope is the swept support used below. The same cell-height state is then evaluated on nominal and wind-perturbed supports. Comparing the two supports gives clearance and wind-loading score fields, separating wind burden on the nominal envelope from clearance changes caused by wind-affected motion.

\subsection{Wind-geometry inputs and local states}

The diagnostic uses a shared wind-geometry scene. Wind fields were generated using \textsc{LatticeUrbanWind} \cite{wei2026latticeurbanwind,LUW01,LUW02}, our high-performance GPU-accelerated lattice Boltzmann method (LBM)-LES platform built on the \textsc{FluidX3D} kernel \cite{fluidx3d}. The wind runs cover each city at the configured lattice resolution and were computed on an $8\times$ NVIDIA A800 80G GPU server. For more details on numerical approaches, verifications and validaionts, see our previous work \cite{LUW01}.

Building footprints \cite{3dglobfp} and terrain from a digital elevation model (DEM) \cite{NASADEM_2021} are voxelised into a geometric model represented by triangular surface elements. The coordinate system is converted to Universal Transverse Mercator (UTM), with the grid convergence angle included at the target location. The resulting geometry is nondimensionalised and scaled under the LBM framework by mapping the reference wind speed $5~\mathrm{m/s}$ to LBM Mach number $Ma_{\mathrm{LBM}} = 0.1$. Using a three-direction ray-tracing voxeliser, the geometry is imprinted onto an isotropic Cartesian grid to flag fluid and solid cells. The ground and buildings are modelled as no-slip walls. The top and lateral grids include sponge layers and buffer zones. Wind vectors, occupancy, and clearance are sampled from the native 20 m wind and solid-mask products.

\begin{figure}[b]
    \centering
    \includegraphics[width=\linewidth]{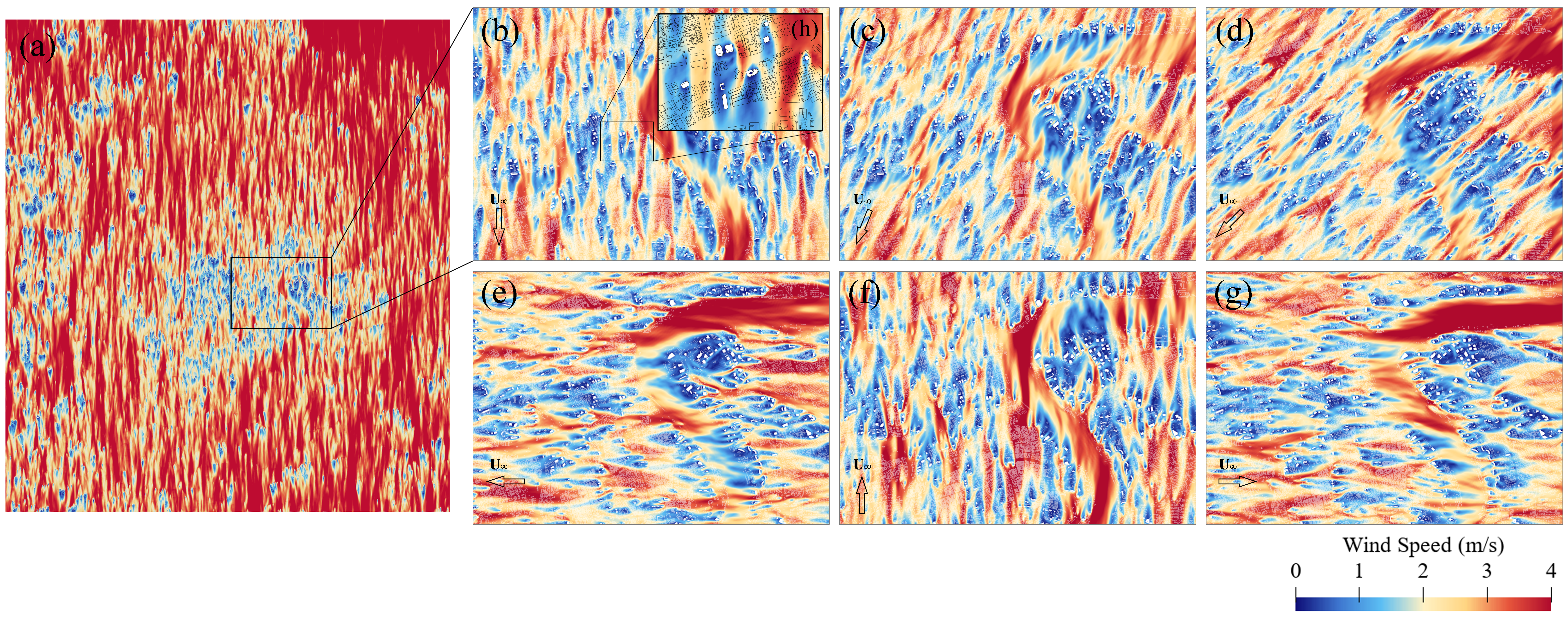}
    \caption{Time-averaged wind speed distribution at 150m height with zoomed-in views. (a) Employed Shanghai domain about 30 km$\times$33 km, with magnified insets under different wind directions: (b) N, (c) NNE, (d) NE, (e) E, (f) S, (g) W. (h) A further zoom-in of panel (b) for local building footprint and wind field details. Only 6 of 16 inflow directions are presented here.}
    \label{fig:winds}
\end{figure}

Wind profiles and direction probabilities use ERA5 data \cite{Olauson2018ERA5}. Mean speed at each vertical grid level is fitted with the atmospheric boundary-layer logarithmic law. Sixteen inflow directions are simulated at $22.5^\circ$ intervals and combined with wind-rose weights in Eq.~\eqref{eq:annual_mean_wind}, extracted from ERA5. This gives the annual representative forcing field. The LBM solver is based on the three-dimensional, 19-velocity (D3Q19) mesoscopic discrete velocity model and uses a single-relaxation-time collision operator. Subgrid-scale stresses are modelled using the Smagorinsky-Lilly model within an LES framework. The inflow turbulence intensity is fixed at $5\%$. A synthetic turbulent inflow is imposed using a GPU-based on-the-fly Fourier boundary to realise a von K\'arm\'an inflow. Thermal buoyancy effects and associated atmospheric-stability differences are omitted. The Coriolis force is formulated with respect to the centre of the computational domain and incorporated through Guo's source-term method \cite{Guo2002}. Momentum-flux correction is applied to all wind boundaries before turbulence inflow synthesis is imposed. The numerical results of wind speed under different wind directions are shown in Figure~\ref{fig:winds}. \ref{app:vv} summarises numerical verifications and validations for the wind simulations. For more details on numerical approaches, verifications and validaionts, see our previous work \cite{LUW01}.

Effective wind speed is then computed with a standard-deviation-based formulation \cite{BanksDenoon2006PracticalIssues,BlockenCarmeliet2004PedestrianWind,AkahoshiUematsu2019Criterion}. According to existing standards, guidelines and prior studies in wind engineering \cite{HKSAR2006AVA,Ng2009AVA,KwokTseTsangWong2012AVA,DuMakKwokTseLeeAiLiuNiu2017LowWindCriteria},
the annual representative mean wind field is obtained under linear-combination assumption:
\begin{equation}
\bar{u}_{a,\ell}(\mathbf{x})
=
\sum_{j=1}^{N_\theta}
p_j\,\mathrm{VR}_{j,\ell}(\mathbf{x})\,U_{\mathrm{ref},j},
\label{eq:annual_mean_wind}
\end{equation}
where $\mathbf{x}$ denotes the position vector in the 3D airspace, $\ell$ denotes the coordinate direction, $N_\theta$ is the number of wind directions, $p_j$ is the occurrence probability of wind direction $j$ derived from ERA5 \cite{Olauson2018ERA5}, $U_{\mathrm{ref},j}$ is the corresponding reference wind speed, and $\mathrm{VR}_{j,\ell}(\mathbf{x})$ is the CFD-derived velocity ratio for component $\ell$ at $\mathbf{x}$. The effective wind vector used by the event model preserves the annual-mean direction and scales its magnitude with a speed-standard-deviation field:
\begin{equation}
\begin{aligned}
U_{\mathrm{eff}}(\mathbf{x}) &=
\left\|\bar{\mathbf{u}}_a(\mathbf{x})\right\|+\lambda\sigma_U(\mathbf{x}),\\
\mathbf{u}_{\mathrm{eff}}(\mathbf{x}) &=
U_{\mathrm{eff}}(\mathbf{x})
\frac{\bar{\mathbf{u}}_a(\mathbf{x})}{\max\left(\left\|\bar{\mathbf{u}}_a(\mathbf{x})\right\|,\varepsilon\right)},
\end{aligned}
\end{equation}
where $\sigma_U(\mathbf{x})$ is the local speed-standard-deviation field, $\varepsilon$ prevents division by zero, and $\lambda=3.5$ follows Melbourne's gust-speed criterion \cite{Melbourne1978Criteria,Melbourne1978Australia}. This annual representative vector is used only for screening. Annual exceedance probabilities require direction- or regime-specific propagation before aggregation. For each city, coordinates are georeferenced, sampled by row-column and analysis-height index, and rotated back to the geographic grid. Wind and obstacle geometry use the same CFD vertical coordinate. Basal layers below the reported analysis range are retained only as runtime wind-scene support and are excluded from reported summaries. Let $k(\mathbf{x})$ denote the turbulent kinetic energy (TKE) field retained from the wind asset. We define the trajectory-disturbance proxy
\begin{equation}
    q(\mathbf{x})
    =
    \sqrt{k(\mathbf{x})}\Delta t ,
    \label{eq:tke_displacement_proxy}
\end{equation}
where $\sqrt{k}$ has velocity units and $q$ has length units for one propagation step. This proxy maps the sampled TKE field into the stochastic displacement scale used for vehicle-trajectory spread. The wind-loading score below keeps $k$ as the environmental TKE component.

Each airspace cell $\mathbf{x}_i$ at height $h$ is evaluated as a local state under heading $\psi_m$ and short horizon $T$. Airspace cells are non-solid, non-restricted, in-domain cells with finite wind and clearance values. The local state is therefore $(i,h)$, where $i$ indexes the native scene cell, $h$ indexes the native CFD/scene height layer.

\subsection{Short-horizon swept-support propagation}

For each airspace cell-height state $(i,h)$, short-horizon trajectories are propagated from the cell centre $\mathbf{x}_{i,h}$. The method samples paired trajectory realisations indexed by $j$, each with a planar cruise heading $\psi_j$ and a shared displacement-score sequence for the nominal and wind-perturbed rollouts. The intended horizontal cruise direction for heading $j$ is
\begin{equation}
    \mathbf{e}_{\parallel}(\psi_j)
    =
    (\cos\psi_j,\sin\psi_j,0)^{\top}.
\end{equation}
Here $\mathbf{e}_{\parallel}(\psi_j)$ is a unit vector in the local horizontal coordinate frame, and $(\cdot)^{\top}$ denotes transpose. The rollout horizon is $T$, the time step is $\Delta t$, and $N_T=\lceil T/\Delta t\rceil$ is the number of propagation steps. The intended air-relative cruise speed is $v_0$. The time step is capped by 1 m along-track spacing; for the light-UAV response class used here, $v_0=8.0$ m s$^{-1}$ and $\Delta t=0.125$ s.

The paired baseline and wind-perturbed rollouts use the same trajectory realisation $j$. Both recurrences are evaluated for $n=0,\ldots,N_T-1$. The baseline rollout is
\begin{equation}
\label{eq:baseline_rollout}
\begin{aligned}
    \mathbf{x}^{0}_{0,i,h,j}
    &=
    \mathbf{x}_{i,h},\\
    \mathbf{x}^{0}_{n+1,i,h,j}
    &=
    \mathbf{x}^{0}_{n,i,h,j}
    +
    \Delta t\,v_0\mathbf{e}_{\parallel}(\psi_j)
    +
    \boldsymbol{\eta}^{0}_{n,i,h,j}.
\end{aligned}
\end{equation}
The wind-perturbed rollout adds start-relative drift from local effective-wind variation and turbulence-driven spread. The response term is used as a planning-scale displacement basis because CFD-informed path-planning evidence shows that headwind, tailwind, crosswind, and turbulence can materially alter urban UAV paths \cite{cao_estimating_2026}, and vehicle-control studies show measurable gust- or strong-wind tracking disturbances in multirotor platforms \cite{bangura_thrust_2017,oconnell_neural-fly_2022}. Define
\begin{equation}
    \Delta\mathbf{u}_{n,i,h,j}
    =
    \mathbf{u}_{\mathrm{eff}}
    \left(\mathbf{x}^{w}_{n,i,h,j}\right)
    -
    \mathbf{u}_{\mathrm{eff}}
    \left(\mathbf{x}_{i,h}\right).
\end{equation}
The wind-perturbed rollout is
\begin{equation}
\label{eq:wind_rollout}
\begin{aligned}
    \mathbf{x}^{w}_{0,i,h,j}
    &=
    \mathbf{x}_{i,h},\\
    \mathbf{x}^{w}_{n+1,i,h,j}
    &=
    \mathbf{x}^{w}_{n,i,h,j}
    +
    \Delta t
    \left[
    v_0\mathbf{e}_{\parallel}(\psi_j)
    +
    \mathbf{G}_{u}
    \Delta\mathbf{u}_{n,i,h,j}
    \right]
    +
    \boldsymbol{\eta}^{w}_{n,i,h,j}.
\end{aligned}
\end{equation}
Here $\mathbf{u}_{\mathrm{eff}}=(u,v,w)^{\top}$ is the three-component effective wind vector. The start-cell value defines the local airspeed reference, so deterministic displacement comes from wind-vector changes sampled along the path. The gain matrix
\begin{equation}
    \mathbf{G}_{u}
    =
    \operatorname{diag}(g_x,g_y,g_z)
\end{equation}
provides a reduced-order axis-wise response for the screening method. Within this diagnostic formulation, vehicle-level disturbance tolerance is represented at screening scale through a response-amplitude envelope: smaller drift gains and dispersion terms represent stronger disturbance rejection, and larger values represent a more disturbance-sensitive light-UAV class.

Stochastic displacement uses standard-normal score sequences scaled by support-specific covariance:
\begin{equation}
    \boldsymbol{\eta}^{\kappa}_{n,i,h,j}
    =
    \mathbf{L}^{\kappa}_{n,i,h,j}\boldsymbol{\xi}_{n,j},
    \qquad
    \mathbf{L}^{\kappa}_{n,i,h,j}
    (\mathbf{L}^{\kappa}_{n,i,h,j})^{\top}
    =
    \boldsymbol{\Sigma}^{\kappa}_{n,i,h,j},
    \qquad
    \kappa\in\{0,w\}.
\end{equation}
The covariance is
\begin{equation}
\begin{aligned}
    \boldsymbol{\Sigma}^{0}_{n,i,h,j}
    &=
    \boldsymbol{\Sigma}_{\mathrm{veh}},\\
    \boldsymbol{\Sigma}^{w}_{n,i,h,j}
    &=
    \boldsymbol{\Sigma}_{\mathrm{veh}}
    +
    q^2\left(\mathbf{x}^{w}_{n,i,h,j}\right)
    \operatorname{diag}
    \left(\alpha_x^2,\alpha_y^2,\alpha_z^2\right),
\end{aligned}
\end{equation}
where $\boldsymbol{\Sigma}_{\mathrm{veh}}=\operatorname{diag}(\sigma_x^2,\sigma_y^2,\sigma_z^2)$. The baseline covariance represents vehicle-class spread, positioning error, and reduced-order kinematic uncertainty. The wind-perturbed covariance adds spread from the trajectory-disturbance proxy $q$, which maps the sampled TKE field $k$ into step-level displacement variance through Eq.~\eqref{eq:tke_displacement_proxy}. The light-UAV response parameters define a nominal screening class, with response-amplitude sensitivity examined in ~\ref{app:uav_response_scale_sensitivity}. The parameter values are reported in Appendix Table~\ref{tab:vehicle_scoring_params}.

Wind vectors, TKE, and wind-loading fields are sampled at continuous trajectory positions by trilinear interpolation from the native wind lattice. Occupancy, restriction, contact, and clearance checks are evaluated on the native terrain-building scene lattice and its derived clearance field. When a rollout first reaches an occupied or restricted cell, or reaches non-positive effective swept clearance, the first-contact state is retained for the remaining horizon. Rollouts leaving the co-registered wind-scene domain are treated as outside-domain samples. Each cell-height field uses the valid trajectory set for its own support.

\begin{figure}[t]
    \centering
    \begingroup
    \setlength{\tabcolsep}{3pt}
    \renewcommand{\arraystretch}{0.86}
    \begin{tabular}{@{}cc@{}}
        \WBCPanel{0.468\textwidth}{\textbf{(a)} Nominal support without wind displacement\\~}{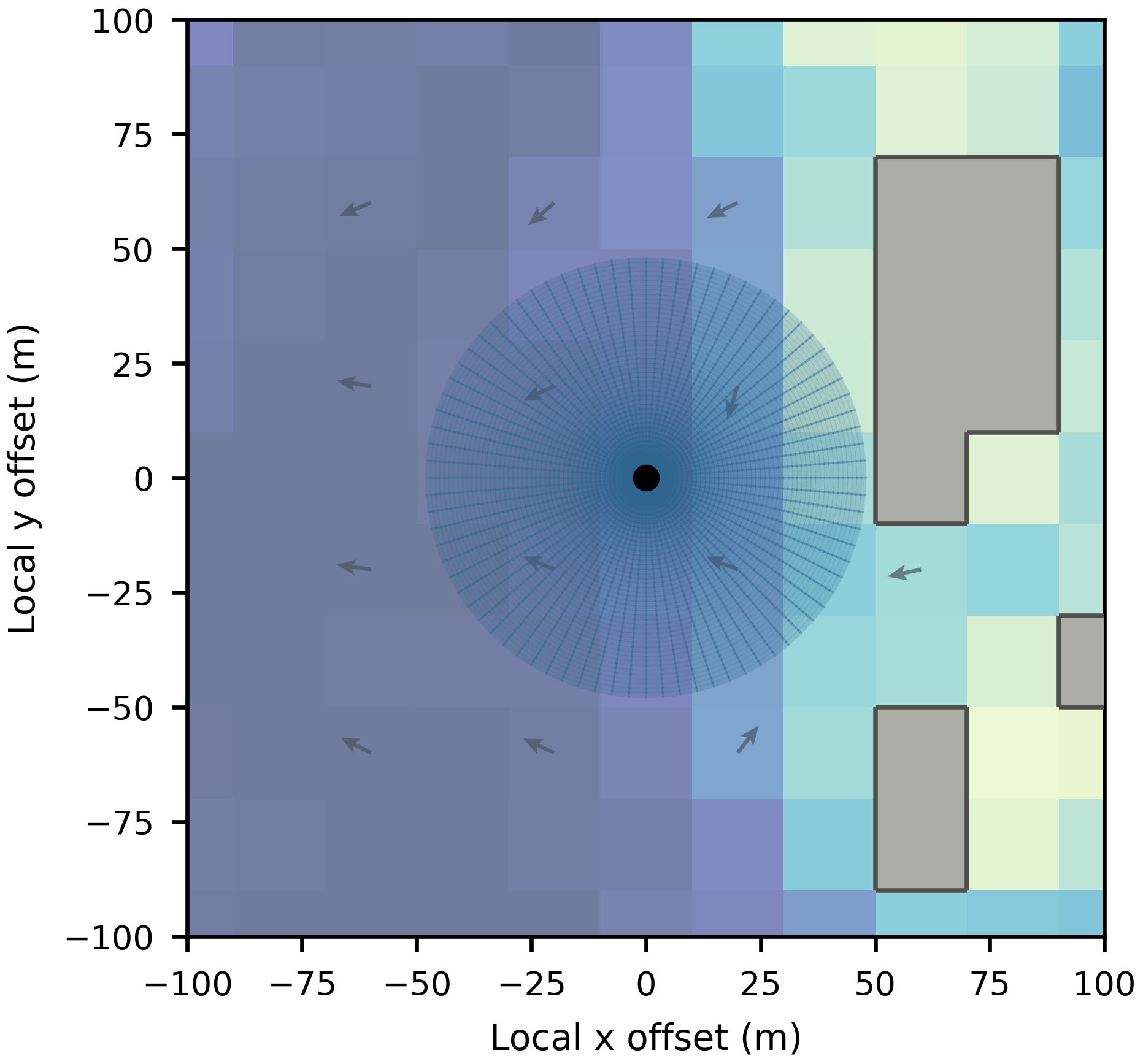} &
        \WBCPanel{0.49\textwidth}{\textbf{(b)} Wind-perturbed support with first-contact states}{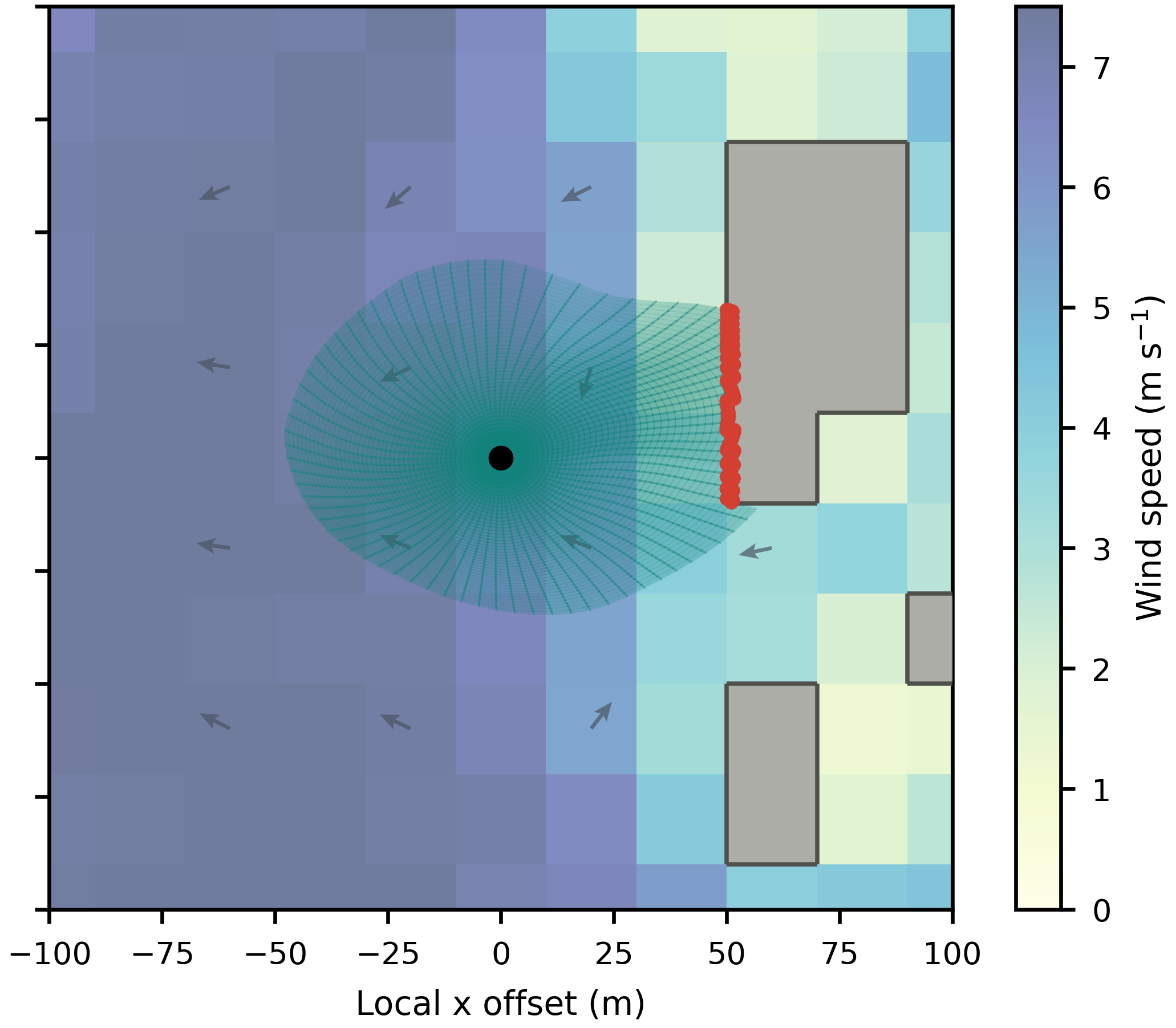}
    \end{tabular}
    \endgroup
    \caption{Example paired trajectory supports for one Shanghai 80 m cell in a locally sheared wind patch. \textbf{(a)} Nominal support without wind displacement. \textbf{(b)} Wind-perturbed support propagated by local wind-vector increments relative to the start cell. Both supports start from the same cell centre and use 2000 fixed horizontal headings over a 6 s horizon. The example cell is about 60 m from the nearest occupied or restricted building cell in the displayed window; the no-wind support remains collision-free, whereas a subset of wind-perturbed headings reaches the building edge and stops at first contact. A regular subset of headings is overplotted more strongly to make individual trajectories visible. The pale background gives the absolute local wind-speed magnitude, arrows mark horizontal wind direction, grey voxel cells mark nearby occupied or restricted building cells, and red points mark first building contact where present. }
    \label{fig:method_trajectory_example}
\end{figure}

Each ordered trajectory is converted into a finite swept support by sweeping the vehicle-class footprint along the path:
\begin{equation}
    \Gamma^{\kappa}_{i,h,j}
    =
    \bigcup_{n=0}^{N_T}
    \mathcal{B}_{r_{\mathrm{veh}}}
    \left(\mathbf{x}^{\kappa}_{n,i,h,j}\right),
    \qquad
    \kappa\in\{0,w\}.
\end{equation}
Here $\mathcal{B}_{r_{\mathrm{veh}}}(\cdot)$ is applied in the horizontal plane at the sampled trajectory height. The primary domain-wide configuration uses $N=16384$ paired trajectory samples per evaluated cell-height state. Figure~\ref{fig:method_trajectory_example} illustrates the same support construction for one Shanghai 80 m cell using $N_{\mathrm{traj}}=2000$ fixed, equally spaced headings.

\subsection{Exposure scores on swept supports}

Two bounded scores are evaluated on each swept support: a clearance score $c(\Gamma)$ and a wind-loading score $b(\Gamma)$. Both are reported on a 0-1 scale.

The clearance score measures how close the swept support comes to terrain-building solids and clearance margins. Contact and clearance-margin intrusion are retained as event channels:
\begin{equation}
\begin{aligned}
    I_{\mathrm{contact}}
    &=
    \mathbb{I}
    \left[
    d_{\min}\left(\Gamma^{\kappa}_{i,h,j}\right)\leq 0
    \right],\\
    I_{\mathrm{margin}}
    &=
    \mathbb{I}
    \left[
    d_{\min}\left(\Gamma^{\kappa}_{i,h,j}\right)\leq d_{\mathrm{clear}}
    \right].
\end{aligned}
\end{equation}
These binary channels are retained as event indicators. The primary clearance exposure scale uses a continuous near-clearance score.

Let $D_{\mathrm{clr}}(\mathbf{y})$ be the 3D clearance field to terrain-building solids. The minimum clearance over the swept support is
\begin{equation}
    d_{\min}\left(\Gamma^{\kappa}_{i,h,j}\right)
    =
    \min_{\mathbf{y}\in\Gamma^{\kappa}_{i,h,j}}
    D_{\mathrm{clr}}(\mathbf{y}).
\end{equation}
The support is already expanded by the vehicle footprint, and the resulting minimum clearance is compared with the class-specific clearance margin $d_{\mathrm{clear}}$. The planning-envelope radius $r_{\mathrm{veh}}$ and clearance margin $d_{\mathrm{clear}}$ are nominal planning-envelope parameters listed in Appendix Table~\ref{tab:experimental_config}. Define the clearance surplus beyond the margin as
\begin{equation}
    s_{\mathrm{clr}}
    =
    \max
    \left[
    d_{\min}\left(\Gamma^{\kappa}_{i,h,j}\right)
    -
    d_{\mathrm{clear}},
    0
    \right].
\end{equation}
The bounded clearance score is then
\begin{equation}
    c\left(\Gamma^{\kappa}_{i,h,j}\right)
    =
    \exp
    \left[
    -
    \left(
    \frac{s_{\mathrm{clr}}}{\ell}
    \right)^2
    \right],
    \qquad
    \ell
    =
    \frac{2\Delta x}{\sqrt{\log 20}} .
\end{equation}
Here $\Delta x$ is the native horizontal grid spacing specified in \S\ref{sec:3_exp}. Thus $c=1$ for contact or clearance-margin intrusion, and $c\approx0.05$ when the swept support is two native grid cells beyond the clearance margin. The decay length provides a finite near-obstacle screening zone tied to the native data lattice. The score is dimensionless and nonlinear. A difference such as $\Delta C^{\mathrm{traj}}=0.012$ is a difference between two 0-1 exposure scores; it has no fixed conversion to collision probability, percentage clearance loss, or metres of clearance, and its distance meaning depends on the baseline value of $C^0$. In the discrete evaluation, the support minimum is approximated with trajectory samples, vehicle-footprint expansion, and a grid-crossing guard. Alternative finite-support hinge and exponential-decay definitions are retained for sensitivity checks.

The wind-loading score measures the wind-loading environment sampled along the same finite support. For any field $f$, let $\mathcal{S}(\Gamma)$ denote the finite set of valid trajectory and raster support samples used to approximate $\Gamma$. The support aggregation operator is
\begin{equation}
    \mathcal{A}_{\Gamma}[f]
    =
    \frac{1}{|\mathcal{S}(\Gamma)|}
    \sum_{\mathbf{y}\in\mathcal{S}(\Gamma)} f(\mathbf{y}) .
\end{equation}
Samples outside the co-registered wind-scene support are omitted from the denominator. If no valid raster sample remains, the trajectory is excluded from the support-level wind-loading mean for that support. Upper-quantile support aggregation is retained as a sensitivity definition.

Four bounded wind-loading components are evaluated: effective wind speed, TKE magnitude, horizontal speed-gradient magnitude, and vertical shear:
\begin{equation}
\begin{aligned}
    r_U(\mathbf{y})
    &=
    \min\left(\frac{U_{\mathrm{eff}}(\mathbf{y})}{U_{\mathrm{score}}},1\right),&
    r_k(\mathbf{y})
    &=
    \min\left(\frac{k(\mathbf{y})}{k_{\mathrm{ref}}},1\right),\\
    r_g(\mathbf{y})
    &=
    \min\left(\frac{\left\|\nabla_{xy} U_{\mathrm{eff}}(\mathbf{y})\right\|_2}{g_{\mathrm{ref}}},1\right),&
    r_s(\mathbf{y})
    &=
    \min\left(\frac{\left|\partial_z U_{\mathrm{eff}}(\mathbf{y})\right|}{s_{\mathrm{ref}}},1\right).
\end{aligned}
\end{equation}
Here $U_{\mathrm{eff}}=\left\|\mathbf{u}_{\mathrm{eff}}\right\|_2$ is the effective wind-speed magnitude, $k$ is TKE, $\nabla_{xy}U_{\mathrm{eff}}$ is the horizontal speed-gradient vector, and $\partial_zU_{\mathrm{eff}}$ is the vertical speed derivative. Separating the horizontal gradient from the vertical derivative avoids counting the same vertical shear contribution in both components.

The support-level wind-loading score is
\begin{equation}
    b(\Gamma)
    =
    \mathcal{A}_{\Gamma}
    \left[
    \frac{
    w_U r_U+w_k r_k+w_g r_g+w_s r_s
    }{
    w_U+w_k+w_g+w_s
    }
    \right].
\end{equation}
For the light-UAV response class used here, the score reference values are $U_{\mathrm{score}}=8.0$ m s$^{-1}$, $k_{\mathrm{ref}}=1.0$ m$^2$ s$^{-2}$, $g_{\mathrm{ref}}=1.0$ s$^{-1}$, and $s_{\mathrm{ref}}=1.0$ s$^{-1}$. The baseline analysis uses equal component weights, while sensitivity checks vary the weights and the support aggregation operator.

\subsection{Aggregation and paired contrasts}

For each paired trajectory sample, the clearance and wind-loading score functions are evaluated on the baseline and wind-perturbed supports:
\begin{equation}
\begin{aligned}
    c^0_{i,h,j}
    &=
    c\left(\Gamma^0_{i,h,j}\right),
    &
    b^0_{i,h,j}
    &=
    b\left(\Gamma^0_{i,h,j}\right),\\
    c^w_{i,h,j}
    &=
    c\left(\Gamma^w_{i,h,j}\right),
    &
    b^w_{i,h,j}
    &=
    b\left(\Gamma^w_{i,h,j}\right).
\end{aligned}
\end{equation}

The superscripts $0$ and $w$ denote the support on which the score is evaluated. The field $B^0$ measures fixed-support wind loading because $b^0$ samples the wind-loading field on $\Gamma^0$. The fields $C^w$ and $B^w$ evaluate clearance and wind loading after support displacement and turbulence-driven spread.

Trajectory-level scores are aggregated to cell-height fields using support-specific sample means. Let $\mathcal{J}^{0}_{i,h}$ denote baseline-valid trajectories for cell $i$ and height $h$, and let $\mathcal{J}^{w}_{i,h}$ denote wind-perturbed-valid trajectories. The baseline fields use $\mathcal{J}^{0}_{i,h}$, and the wind-perturbed fields use $\mathcal{J}^{w}_{i,h}$.

For any trajectory-level quantity $z^\kappa_{i,h,j}$ on support $\kappa\in\{0,w\}$, define
\begin{equation}
    \langle z^\kappa\rangle^{\kappa}_{i,h}
    =
    \frac{1}{|\mathcal{J}^{\kappa}_{i,h}|}
    \sum_{j\in\mathcal{J}^{\kappa}_{i,h}}
    z^\kappa_{i,h,j}.
\end{equation}
The four aggregated fields are
\begin{equation}
\begin{aligned}
    C^0_{i,h}
    &=
    \langle
    c^0
    \rangle^{0}_{i,h},&
    B^0_{i,h}
    &=
    \langle
    b^0
    \rangle^{0}_{i,h},\\
    C^w_{i,h}
    &=
    \langle
    c^w
    \rangle^{w}_{i,h},&
    B^w_{i,h}
    &=
    \langle
    b^w
    \rangle^{w}_{i,h}.
\end{aligned}
\end{equation}
Upper-tail summaries over samples are used as auxiliary checks. The primary cell-height fields use the sample means above.

The primary metrics are $B^0$, $\Delta C^{\mathrm{traj}}$, and $\Delta B^{\mathrm{traj}}$:
\begin{equation}
\begin{aligned}
    \Delta C^{\mathrm{traj}}_{i,h}
    &=
    C^w_{i,h}-C^0_{i,h},\\
    \Delta B^{\mathrm{traj}}_{i,h}
    &=
    B^w_{i,h}-B^0_{i,h}.
\end{aligned}
\end{equation}
Here $B^0$ measures wind loading present on the baseline support. A positive $\Delta C^{\mathrm{traj}}$ indicates a higher sample-mean clearance score on the wind-perturbed support, and a negative value indicates a lower score. Similarly, $\Delta B^{\mathrm{traj}}$ compares the sample-mean wind-loading score sampled by the perturbed and baseline supports. These contrasts are formed after support-specific cell-level aggregation.

Auxiliary combined-exposure fields use the bounded-union operator
\begin{equation}
    \mathcal{U}(a,b)=1-(1-a)(1-b).
\end{equation}
The retained context fields are
\begin{equation}
    E^0=C^0,
    \qquad
    E^{\mathrm{fixed}}=\langle\mathcal{U}(c^0,b^0)\rangle^0,
    \qquad
    E^{\mathrm{traj}}=\langle\mathcal{U}(c^w,b^w)\rangle^w .
\end{equation}
The primary mechanism-related metrics are $B^0$, $\Delta C^{\mathrm{traj}}$, and $\Delta B^{\mathrm{traj}}$.

\section{Analysis design\label{sec:3_exp}}


The primary comparison applies the same design to Shanghai and Beijing. These two cases are used for same-protocol comparison: the grid, reported height layers, trajectory sampling, vehicle response class, and scoring parameters are held common. Shanghai represents a low-relief coastal megacity, whereas Beijing represents a large northern plain city with different urban form and wind setting.

The analysis uses co-registered wind-scene data with native horizontal grid spacing $\Delta x=20$ m and native scene-height spacing $\Delta h=20$ m. The primary vertical coordinate is the native scene height $h$ because wind vectors, occupancy, and clearance are co-registered on this height lattice. Reported start and evaluation layers run from 60 to 340 m. The runtime wind-scene stack spans 20 to 420 m: the 20 and 40 m layers provide lower interpolation and propagation support, and layers above 340 m provide upper-neighbour support. Reported summaries use the 60-340 m start/evaluation layers. Local above-ground-level (AGL) altitude is obtained by subtracting ground elevation from the native scene height and is reported only as contextual information for interpretation. Terrain and buildings are solid cells in the same lattice. The solid-mask volume defines contact and clearance-margin geometry:
\begin{equation}
    O(x,y,h)=\mathbb{I}\left[(x,y,h)\in\Omega_{\mathrm{solid}}\right],
\end{equation}
where $\Omega_{\mathrm{solid}}$ is the terrain-building solid set after basal-layer removal. Digital elevation model (DEM) and building rasters are used for visualisation and context; collision geometry comes from the ray-tracing-derived CFD solid-fluid mask.


Wind fields are stored on the native uniform 20 m scene-height layers. Each layer contains the effective wind-vector components, TKE, and derived wind-loading score fields. The effective wind-speed magnitude is $U_{\mathrm{eff}}=\|\mathbf{u}_{\mathrm{eff}}\|_2$. The horizontal speed-gradient and vertical-shear fields are computed as $g_U=\|\nabla_{xy}U_{\mathrm{eff}}\|_2$ and $s_U=|\partial_zU_{\mathrm{eff}}|$, respectively. During propagation, these wind variables are sampled from the eight surrounding grid nodes by trilinear interpolation, and $U_{\mathrm{eff}}$ is recomputed from the interpolated vector components. Speed standard deviation is retained as an upstream gust-scaling field where used to construct $\mathbf{u}_{\mathrm{eff}}$; the wind-loading score components are $U_{\mathrm{eff}}$, $k$, $g_U$, and $s_U$. 
\begin{equation}
    \mathcal{W}(x,y,h)
    =
    \left\{
    \mathbf{u}_{\mathrm{eff}}(x,y,h),
    k(x,y,h),
    g_U(x,y,h),
    s_U(x,y,h)
    \right\}.
\end{equation}
Here $\mathbf{u}_{\mathrm{eff}}$ is the gust-scaled effective wind vector, $k$ is the TKE field used to derive $q$ for propagation covariance and to form the wind-loading score, $g_U$ is the horizontal speed-gradient magnitude, and $s_U$ is the vertical speed derivative.

The city fields are occupancy $O(x,y,h)$, restriction mask $R(x,y,h)$, and clearance $d_O(x,y,h)$. These fields define the airspace-cell set $\mathcal{F}_h$; occupied, restricted, out-of-domain, and wind-missing cells are excluded. A 160 m horizontal edge buffer further excludes outer-domain cells from the starting-state set, while those cells remain available as wind-geometry support for trajectories that pass through them. The same nominal planning-grid envelope parameters listed in Appendix Table~\ref{tab:experimental_config} are used for the clearance-margin screen defined in \S\ref{sec:2_methods}. Primary-comparison city domains and coordinate reference system (CRS) details are in Appendix Table~\ref{tab:city_assets}.

\begin{figure}[b]
    \centering
    \begingroup
    \setlength{\tabcolsep}{2pt}
    \renewcommand{\arraystretch}{0.80}
    \begin{tabular}{@{}ccc@{}}
        \WBCPanel{0.32\textwidth}{\textbf{(a)} Shanghai airspace mask}{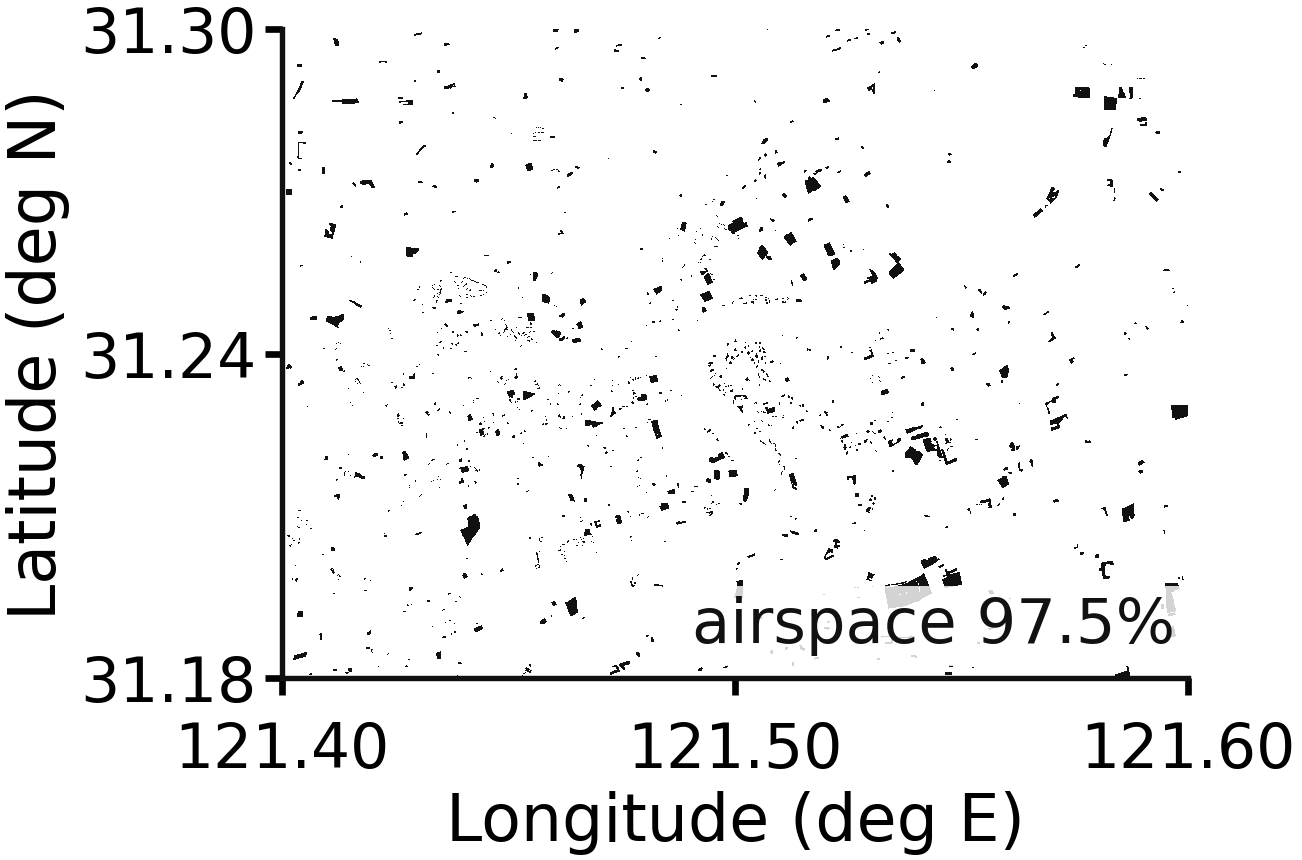} &
        \WBCPanel{0.32\textwidth}{\textbf{(b)} Shanghai $|\mathbf{u}_{\mathrm{eff}}|$\\~}{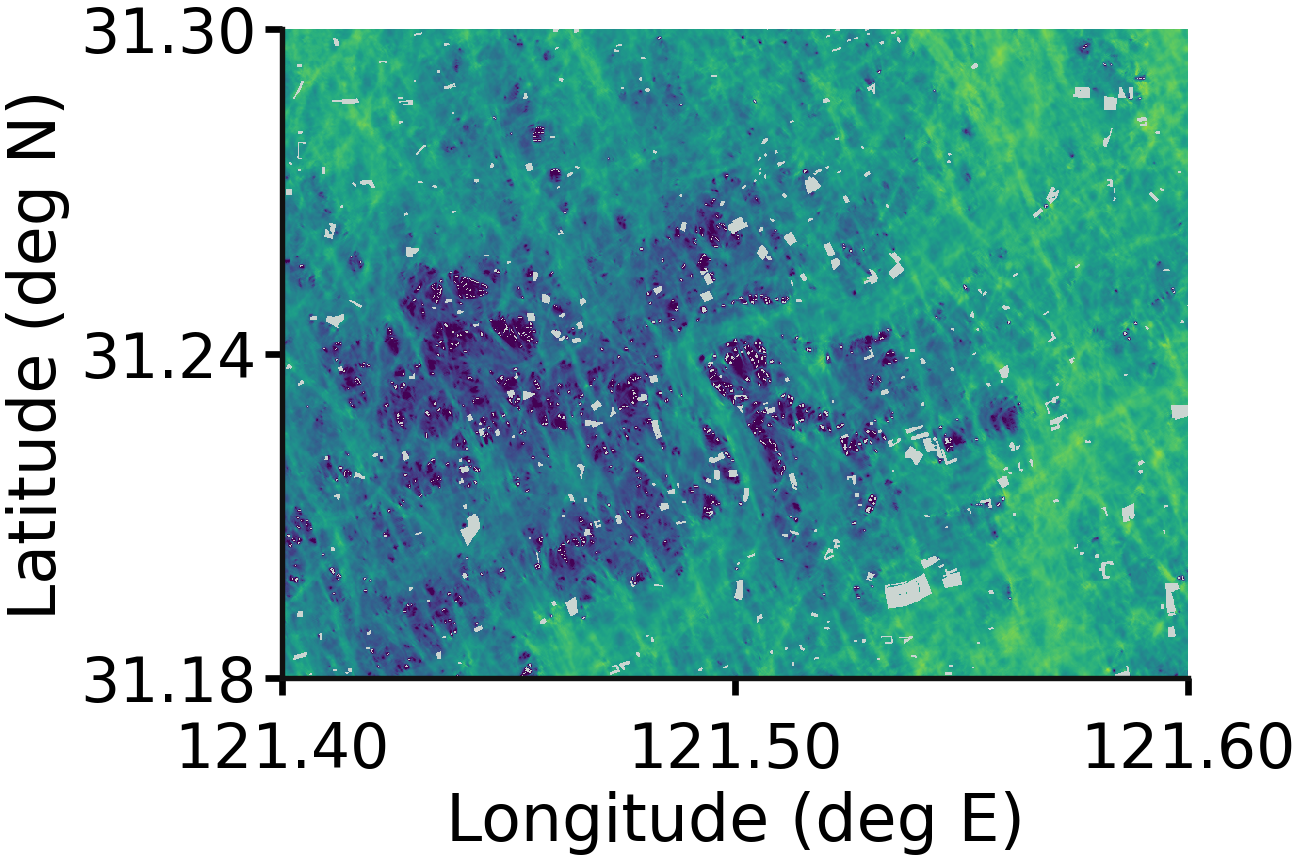} &
        \WBCPanel{0.32\textwidth}{\textbf{(c)} Shanghai TKE $k$\\~}{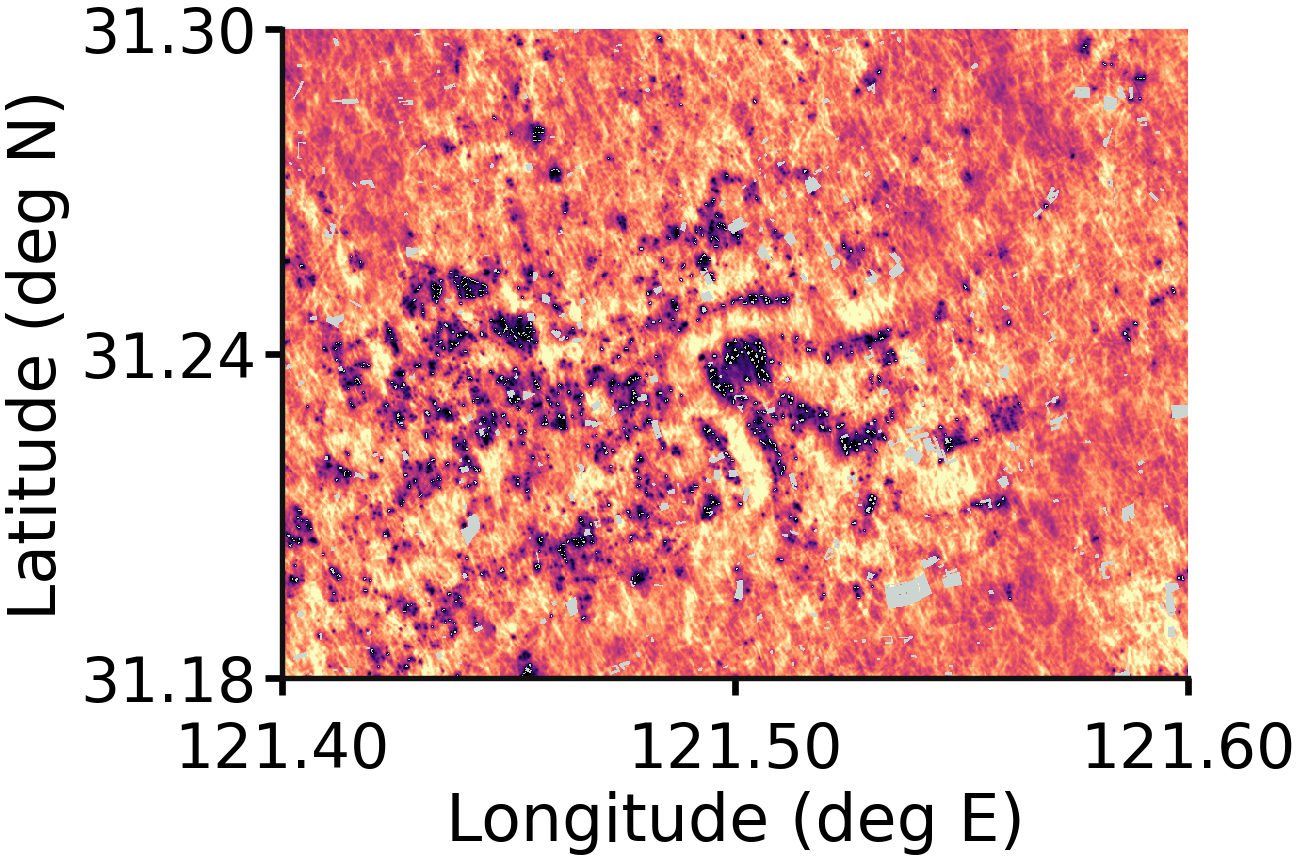} \WBCPanelGap
        \WBCPanel{0.32\textwidth}{\textbf{(d)} Beijing airspace mask}{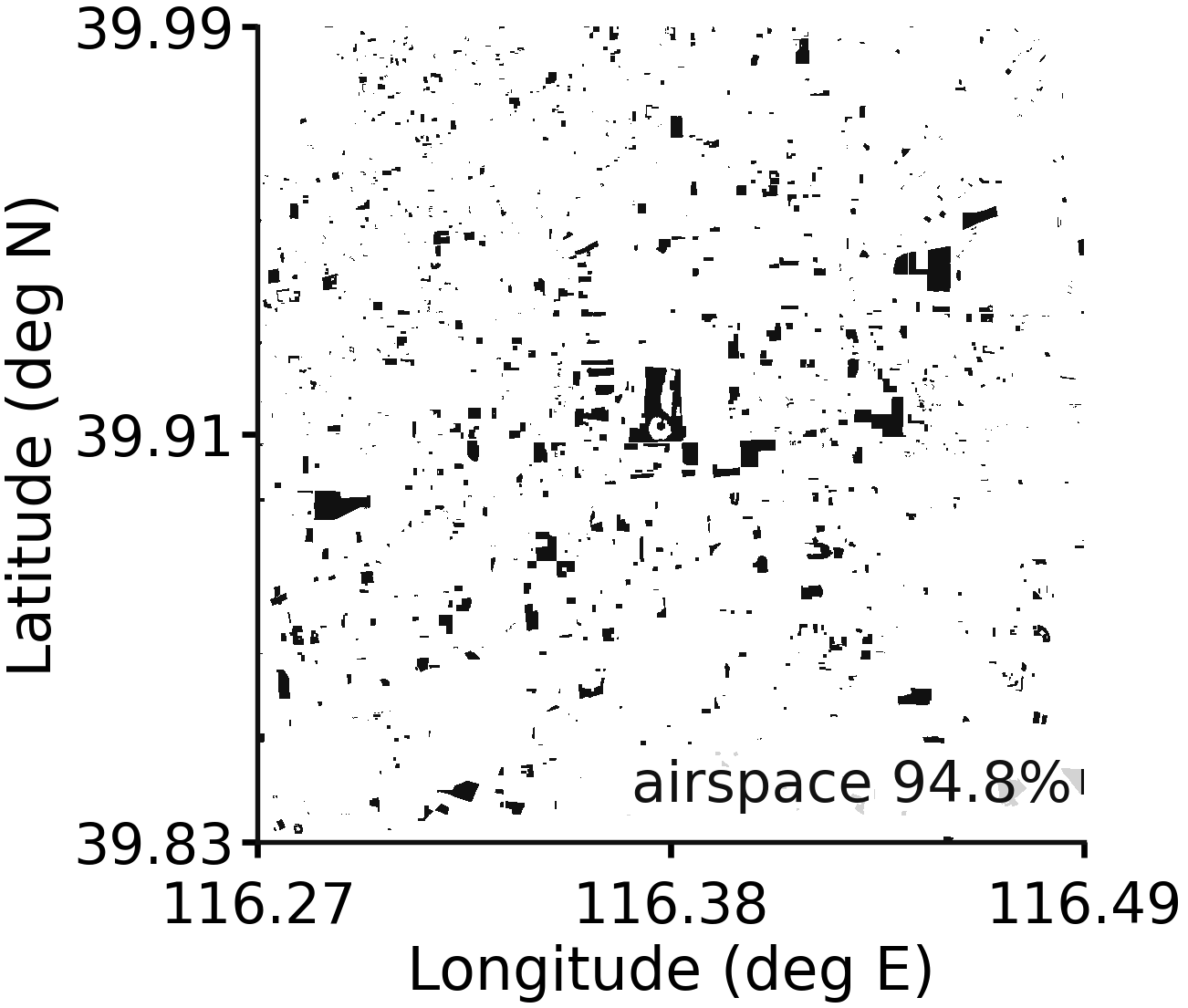} &
        \WBCPanel{0.32\textwidth}{\textbf{(e)} Beijing $|\mathbf{u}_{\mathrm{eff}}|$\\~}{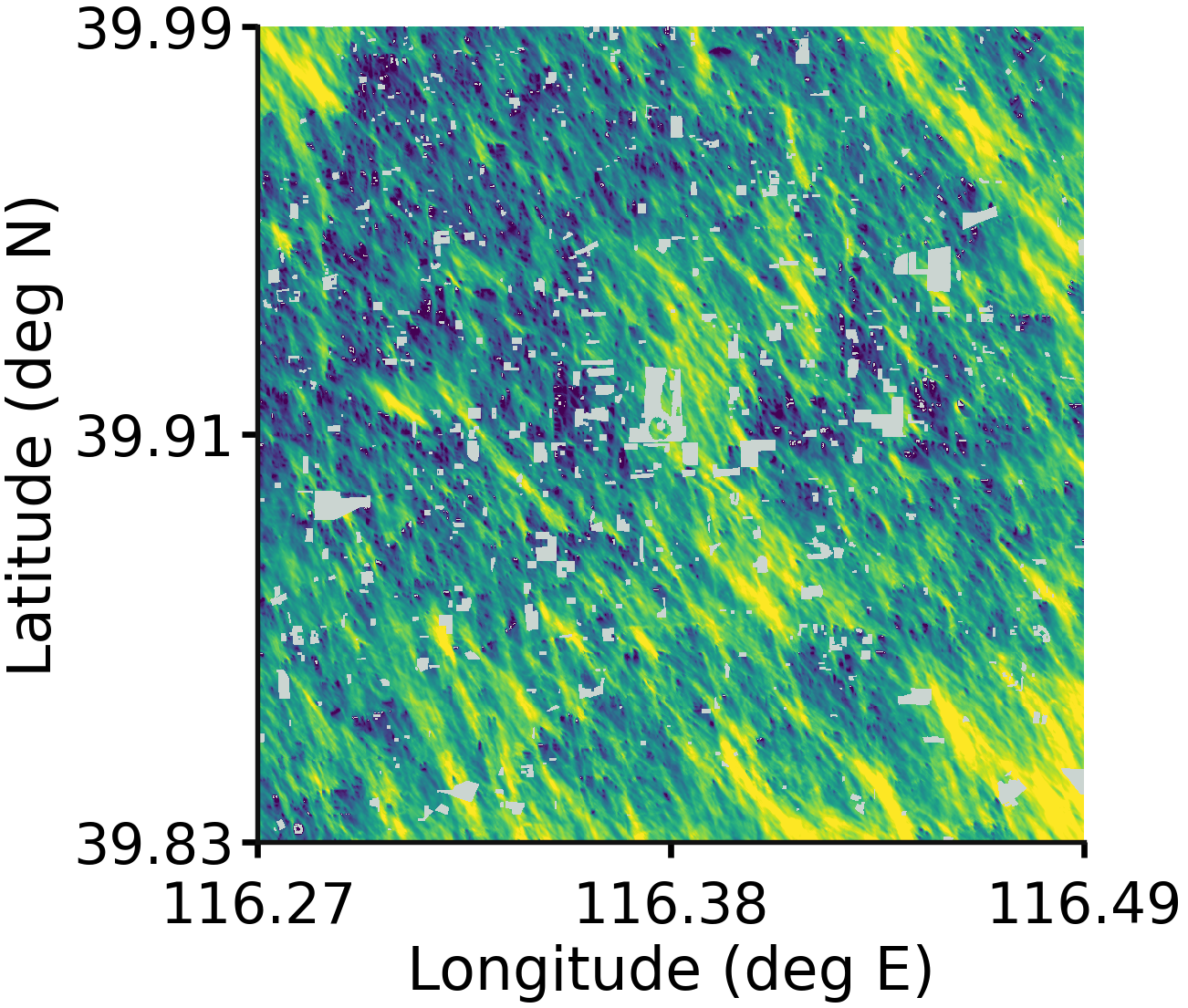} &
        \WBCPanel{0.32\textwidth}{\textbf{(f)} Beijing TKE $k$\\~}{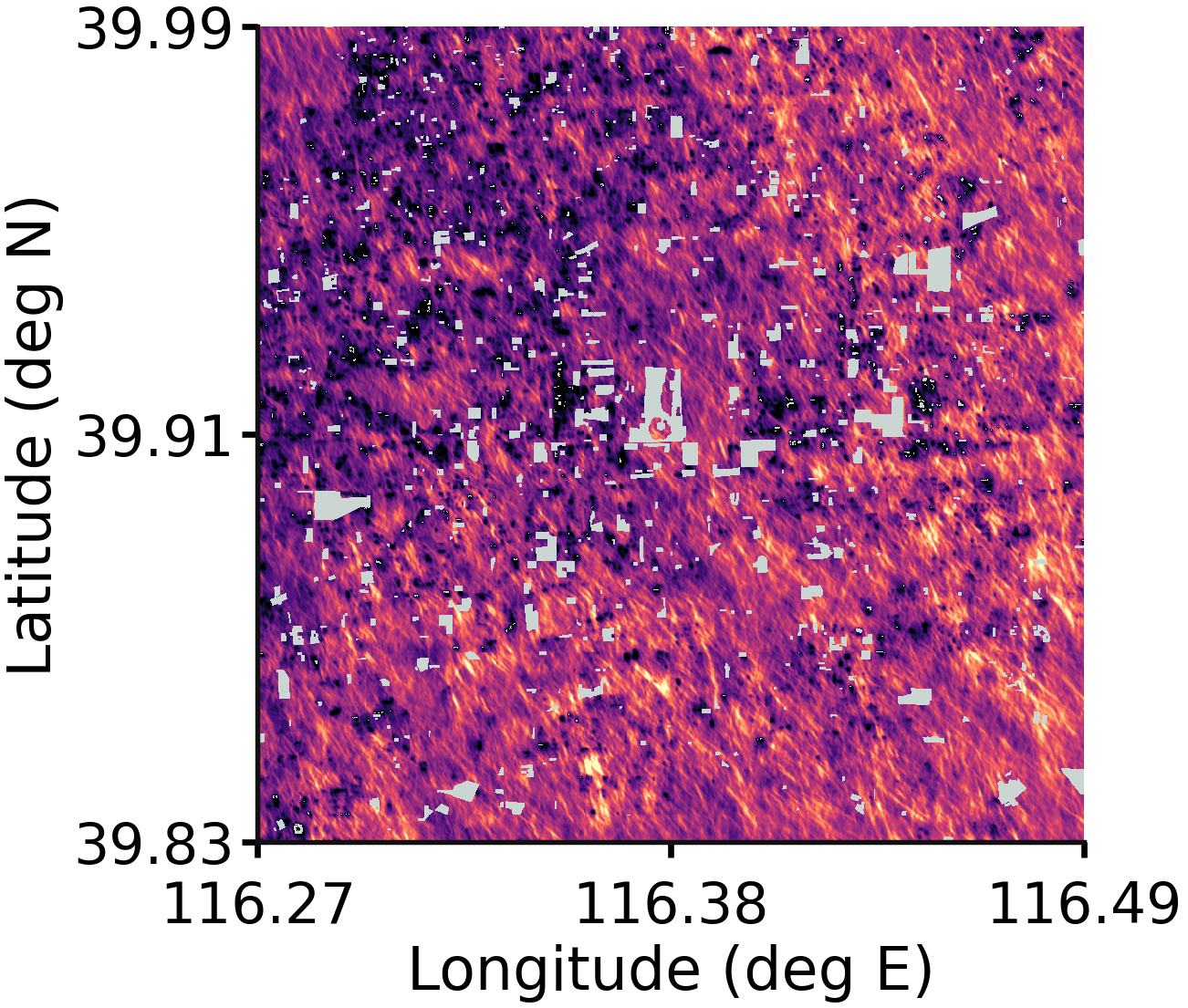} \WBCPanelGap
        \WBCPanel{0.29\textwidth}{}{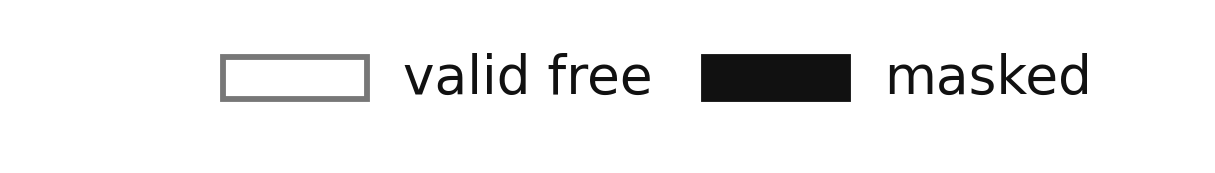} &
        \WBCPanel{0.29\textwidth}{}{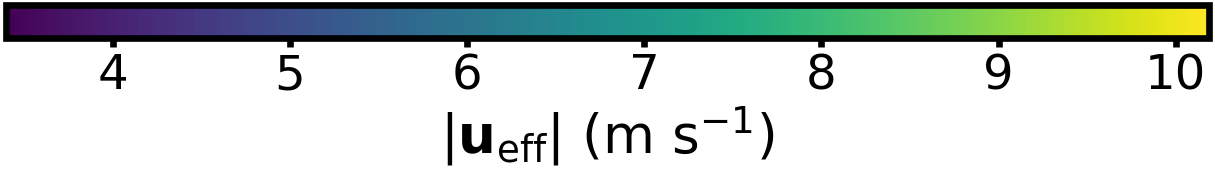} &
        \WBCPanel{0.29\textwidth}{}{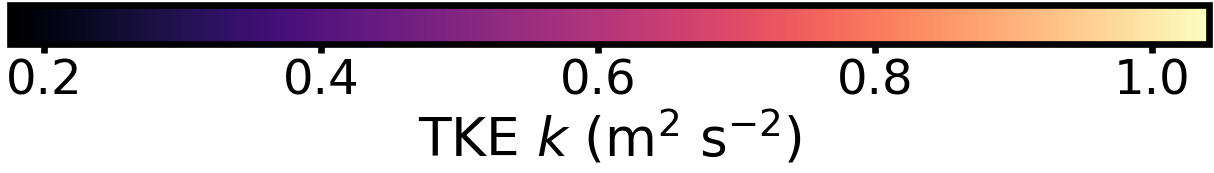}
    \end{tabular}
    \endgroup
    \caption{Input assets for the Shanghai-Beijing primary comparison at the 100 m native scene height. \textbf{(a,d)} Airspace-cell masks on the native 20 m city grids, with valid-free fractions annotated. \textbf{(b,e)} Magnitude of the gust-scaled effective wind vector. \textbf{(c,f)} TKE field $k$.}
    \label{fig:input_context}
\end{figure}

Figure~\ref{fig:input_context} provides the airspace-cell and wind-field context used to define the Shanghai-Beijing primary comparison. The representative 100 m native layer serves as a design-context layer before the Results section reports sample-mean fields and aggregate response contrasts.


Motivated by an isomorphic basis, the same propagation, vehicle, and geometry-screening settings are used for Shanghai and Beijing. The primary domain-wide configuration uses $N=16384$ paired trajectory samples, reported native scene heights from 60 to 340 m, runtime wind-scene support from 20 to 420 m, trilinear wind-field sampling on the native wind lattice, a 160 m starting-state edge buffer, $v_0=8.0$ m/s, $\Delta t=0.125$ s, the envelope parameters in Appendix Table~\ref{tab:experimental_config}, cell-level sample-mean aggregation over support-specific valid trajectories, 1 km $\times$ 1 km local tiles, and a 50\% tile airspace-cell threshold. Compact settings and vehicle-parameter tables are provided in Appendix Tables~\ref{tab:experimental_config} and~\ref{tab:vehicle_scoring_params}.


For any cell-level field $Z_{i,h}$, domain-wide height summaries are computed over the airspace-cell set $\mathcal{F}_h$ at each reported native scene height:
\begin{equation}
    \bar{Z}^{\mathrm{city}}_{h}
    =
    \frac{1}{|\mathcal{F}_h|}
    \sum_{i\in\mathcal{F}_h}
    Z_{i,h}.
\end{equation}
Spatial quantile summaries use the same airspace-cell set:
\begin{equation}
    Q_p^{\mathrm{city},h}[Z]
    =
    Q_p
    \left(
    \left\{
    Z_{i,h}:i\in\mathcal{F}_h
    \right\}
    \right).
\end{equation}
The same definitions are applied to $C^0$, $B^0$, $\Delta C^{\mathrm{traj}}$, $\Delta B^{\mathrm{traj}}$, and the combined exposure context fields. Cell-level maps use the native-cell values directly, with no interpolation or spatial smoothing. Although the computational support extends to 420 m, domain-wide profiles, cell-level maps, and standardized local-space summaries are filtered to 340 m and below.
In the Results figures, domain-wide profile curves use the spatial median $Q_{0.5}^{\mathrm{city},h}$ across finite airspace cells at each height. The cell-level fields themselves remain sample-mean fields over support-specific valid trajectories; the median profile is used only to summarise the spatial distribution across cells. Upper quantile curves or shaded quantile bands are retained to show whether local high-exposure subsets differ from the domain-wide median.

Local-space summaries use fixed 1 km $\times$ 1 km spatial tiles evaluated separately by native scene height. On the native 20 m grid, each full tile contains 50 $\times$ 50 cells before trimming at the city-domain boundary. A tile is included when the airspace-cell fraction is at least 50\% of the nominal tile cell set. For tile $t$, the tile-level value is
\begin{equation}
    \bar{Z}_{t,h}
    =
    \frac{1}{|\mathcal{F}_{t,h}|}
    \sum_{i\in\mathcal{F}_{t,h}}
    Z_{i,h},
\end{equation}
where $\mathcal{F}_{t,h}$ is the airspace-cell set inside tile $t$ at height $h$. These standardized local-space summaries are compared using $C^0$, $B^0$, $\Delta C^{\mathrm{traj}}$, and $\Delta B^{\mathrm{traj}}$. They are separate from domain-wide height profiles.

\section{Results and discussion\label{sec:3_ResDis}}

\subsection{Baseline swept-support context}

\begin{figure}[!htbp]
    \centering
    \begingroup
    \setlength{\tabcolsep}{3pt}
    \renewcommand{\arraystretch}{0.86}
    \begin{tabular}{@{}cc@{}}
        \WBCPanel{0.47\textwidth}{\textbf{(a)} $B^0$ map at 80 m}{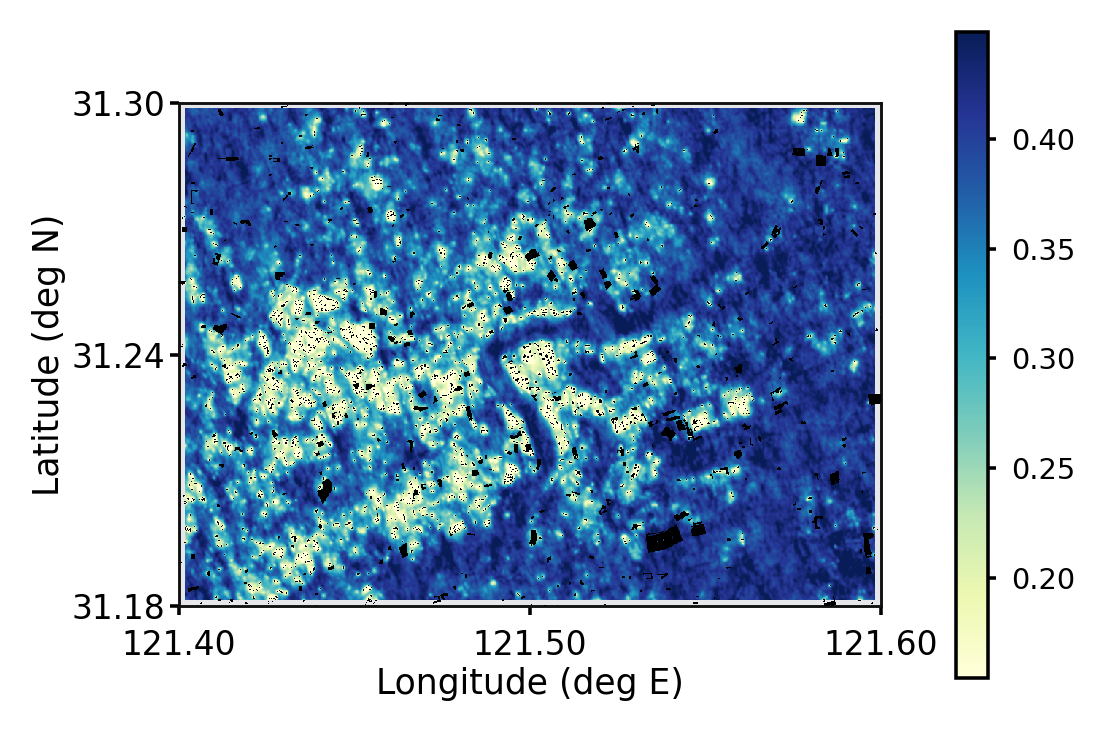} &
        \WBCPanel{0.47\textwidth}{\textbf{(b)} $C^0$ map at 80 m}{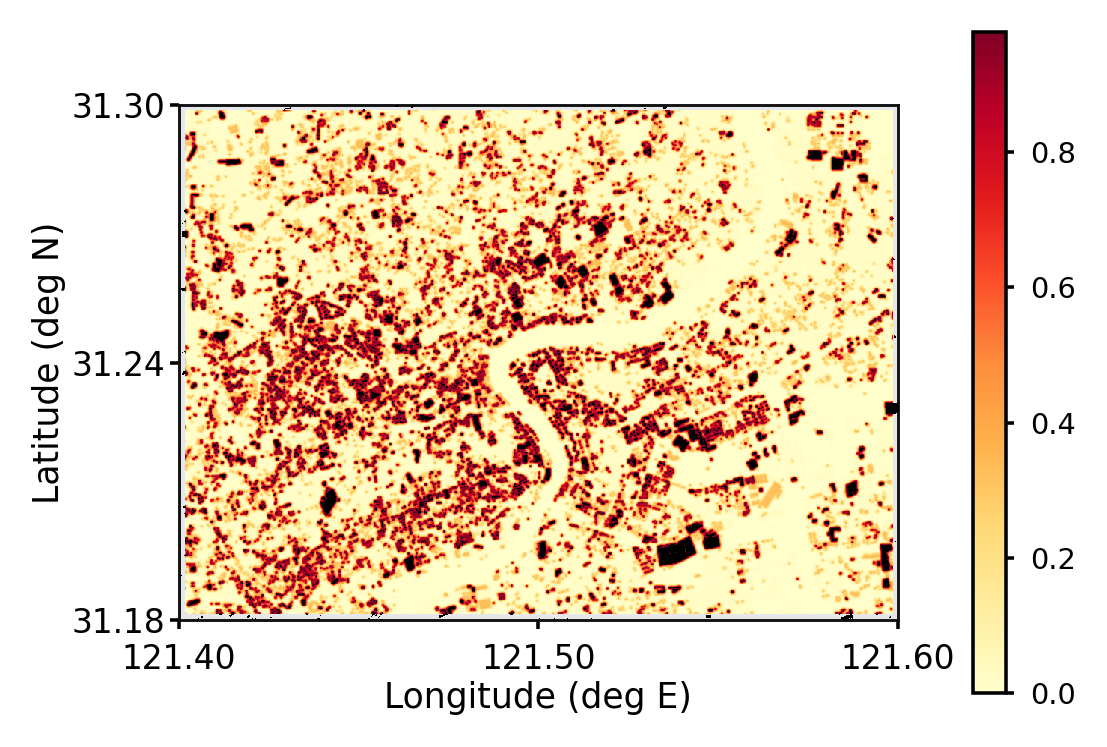} \WBCPanelGap
        \WBCPanel{0.47\textwidth}{\textbf{(c)} $B^0$ spatial-median profile}{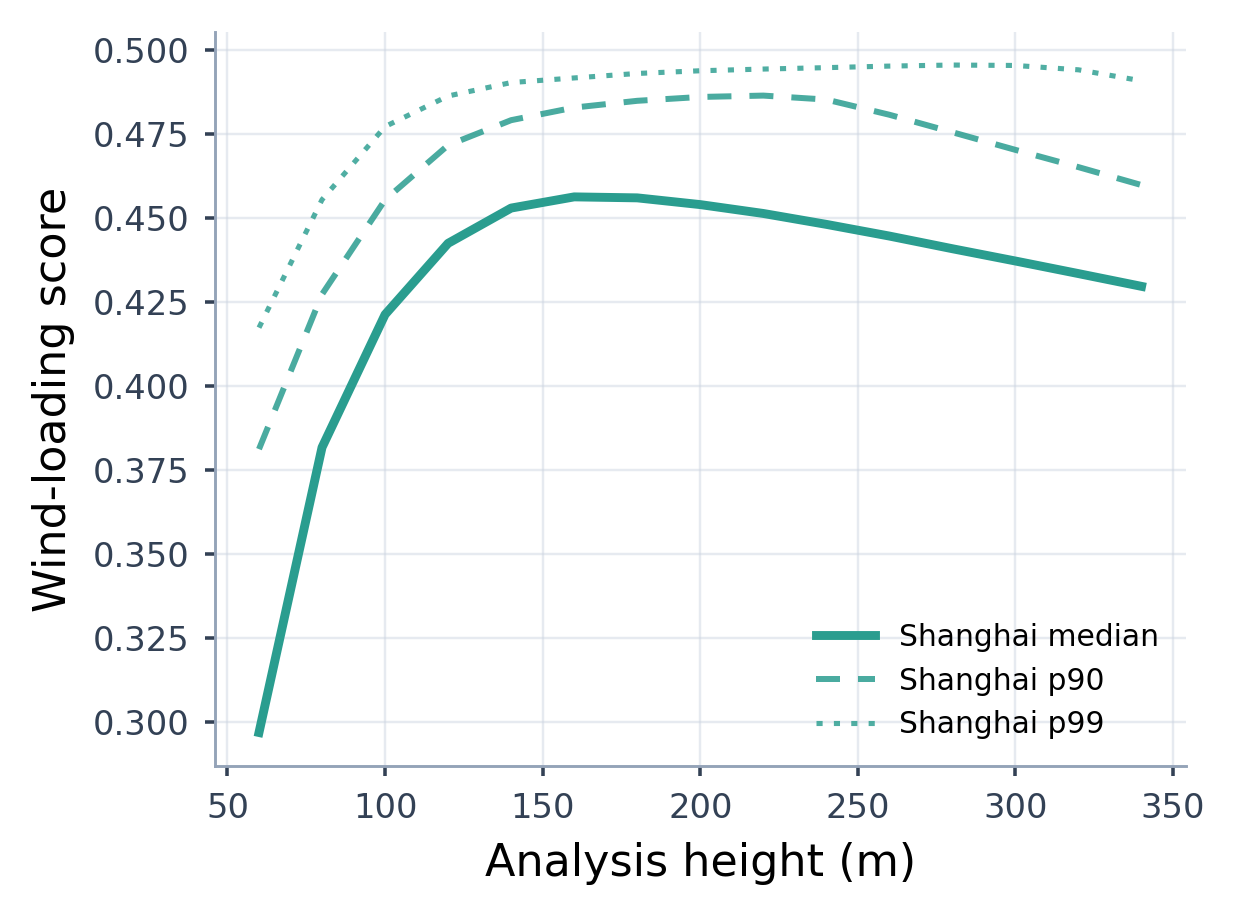} &
        \WBCPanel{0.47\textwidth}{\textbf{(d)} $C^0$ spatial-median profile}{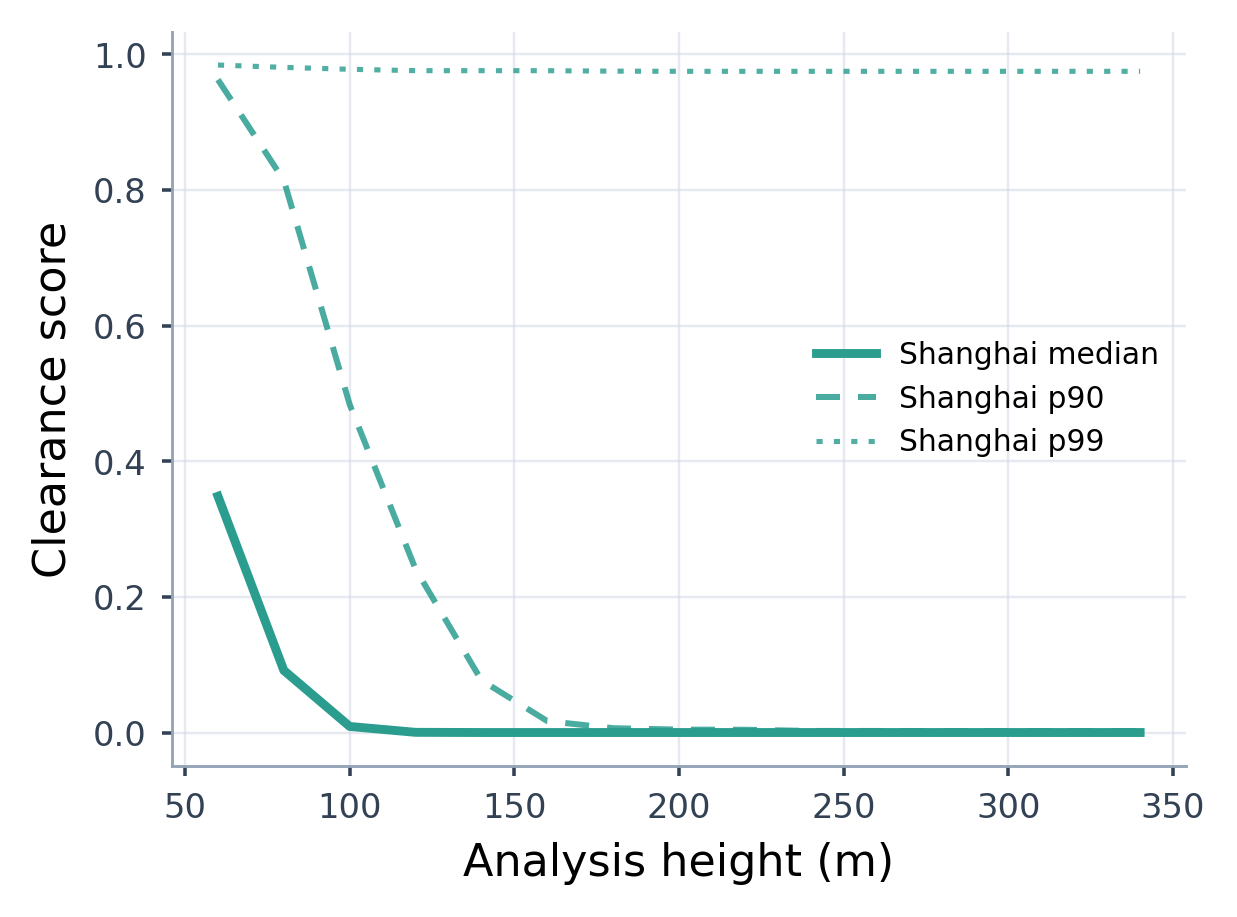}
    \end{tabular}
    \endgroup
    \caption{Shanghai nominal-support baseline context from the $N=16384$ paired-support analysis. \textbf{(a-b)} 80 m cell-level fields for fixed-support wind loading $B^0$ and nominal-support clearance exposure $C^0$; black cells are occupied or restricted cells. \textbf{(c-d)} Spatial-median height profiles across finite airspace cells, with upper spatial quantiles retained as distribution context.}
    \label{fig:results_shanghai_baseline}
\end{figure}

Figure~\ref{fig:results_shanghai_baseline} tests whether a fixed-support wind-loading layer can represent nominal clearance sensitivity. The fixed-support wind-loading field $B^0$ is the wind burden sampled on the baseline swept support. The nominal clearance-exposure field $C^0$ is the terrain-building proximity exposure of the same baseline support. They share the same support, height, and evaluation basis, but describe different screening quantities: $B^0$ is an environmental burden on nominal motion, while $C^0$ is near-obstacle exposure of that nominal motion.

At 80 m in Shanghai, the spatial structures of $B^0$ and $C^0$ are visibly different. High wind-loading areas form broad patches that follow the wind field, while high clearance-exposure areas concentrate near dense or tall terrain-building features. The two maps select different locally tight cells. Some locations with elevated $B^0$ remain away from terrain-building margins, and some clearance-sensitive locations sit away from the strongest fixed-support wind-loading patches. Wind-loaded space and low-clearance space are different screening layers; $B^0$ alone is an incomplete low-altitude clearance screen.

The height profiles show the same separation in the vertical dimension. In Shanghai, the spatial median of $B^0$ rises from 0.297 at 60 m to 0.456 at 160 m and remains high through the upper reported layers. By contrast, the median $C^0$ decreases from 0.348 at 60 m to 0.092 at 80 m, then falls rapidly after 100 m and approaches zero above the main building-height band. The upper quantiles of $C^0$ remain non-zero at higher layers, showing that localized tall or constrained building clusters continue to produce nominal clearance exposure after the domain-wide median support has moved away from terrain-building proximity.

The Beijing case in \ref{app:beijing_diagnostic} repeats the same baseline separation under the same design, with a stronger lower-altitude layer $C^0$ background: the Beijing median $C^0$ is 0.963 at 60 m and 0.468 at 80 m before decreasing toward zero above 140-160 m. The Beijing $B^0$ profile also rises into the mid-height range, reaching a median of 0.431 at 140 m. Nominal wind loading and nominal clearance exposure are different screening layers, and the clearance-exposure response must be evaluated on wind-displaced supports.

\subsection{Trajectory-induced changes and height structure}

\begin{figure}[!htbp]
    \centering
    \begingroup
    \setlength{\tabcolsep}{3pt}
    \renewcommand{\arraystretch}{0.86}
    \begin{tabular}{@{}cc@{}}
        \WBCPanel{0.47\textwidth}{\textbf{(a)} $\Delta B^{\mathrm{traj}}$ map at 80 m}{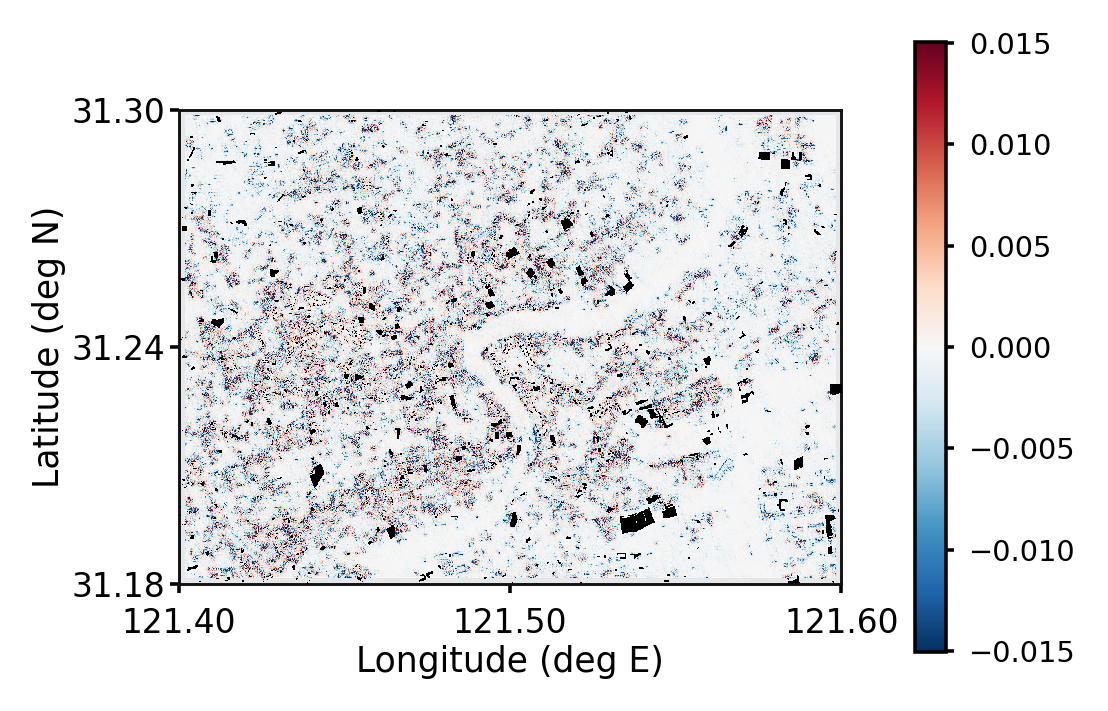} &
        \WBCPanel{0.47\textwidth}{\textbf{(b)} $\Delta B^{\mathrm{traj}}$ map at 160 m}{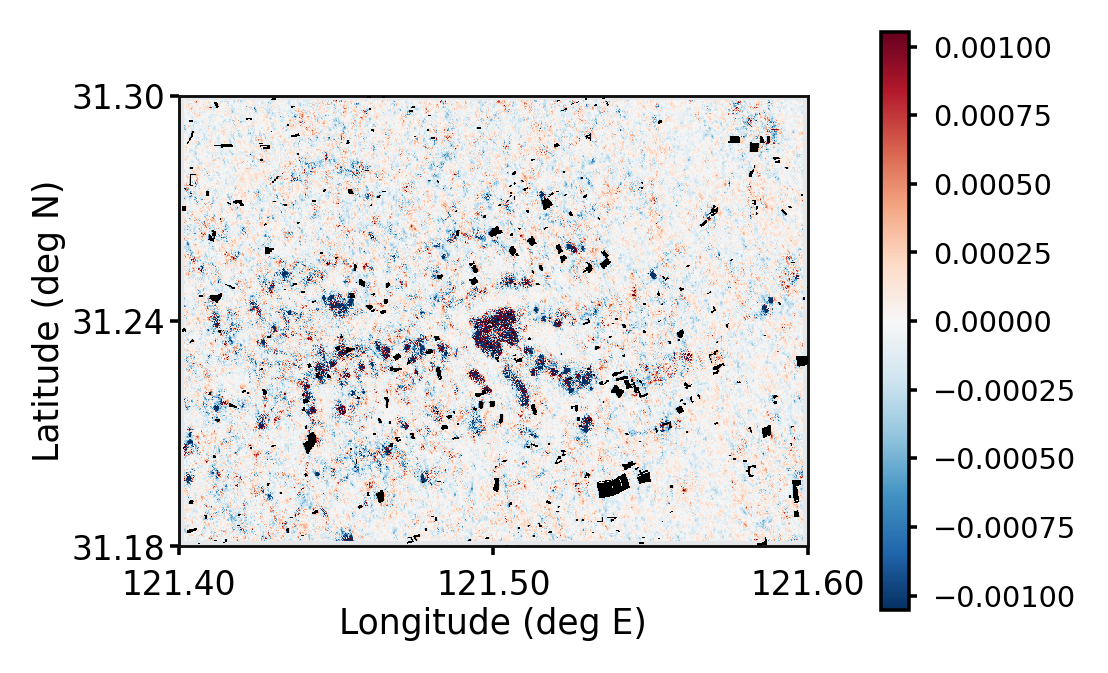} \WBCPanelGap
        \WBCPanel{0.47\textwidth}{\textbf{(c)} $\Delta C^{\mathrm{traj}}$ map at 80 m}{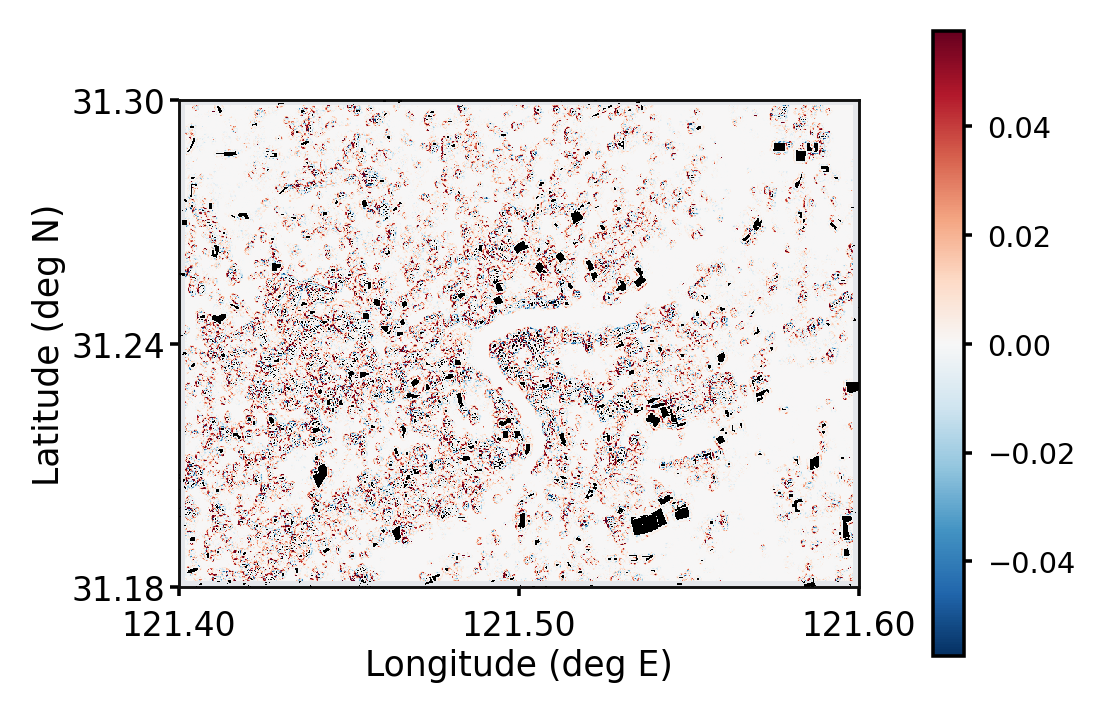} &
        \WBCPanel{0.47\textwidth}{\textbf{(d)} $\Delta C^{\mathrm{traj}}$ map at 160 m}{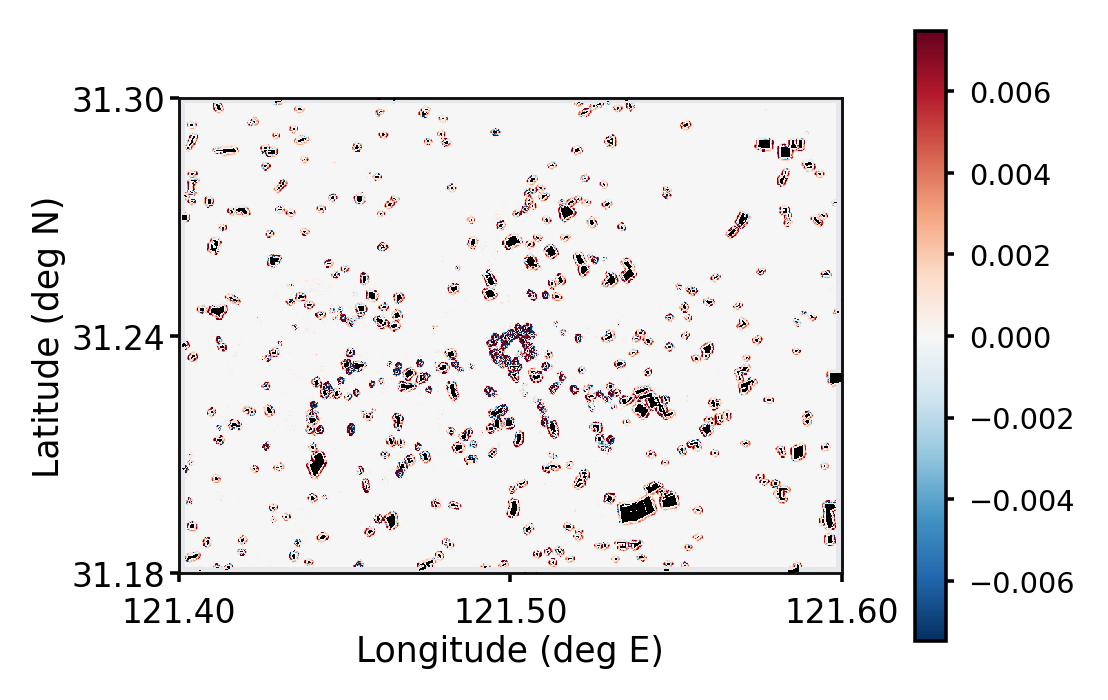}
    \end{tabular}
    \endgroup
    \caption{Shanghai trajectory-induced wind-loading and clearance-exposure changes from the $N=16384$ paired-support analysis. \textbf{(a-b)} Cell-level $\Delta B^{\mathrm{traj}}$ fields at 80 m and 160 m. \textbf{(c-d)} Cell-level $\Delta C^{\mathrm{traj}}$ fields at 80 m and 160 m. Black marks occupied or restricted cells.}
    \label{fig:results_shanghai_delta_maps}
\end{figure}

Figure~\ref{fig:results_shanghai_delta_maps} compares the wind-perturbed swept support with the baseline support. The resulting increments $\Delta B^{\mathrm{traj}}$ and $\Delta C^{\mathrm{traj}}$ are much smaller than $B^0$ and $C^0$, and their signs are mixed across space. Wind-induced displacement produces localized positive, negative, and near-zero changes. In Shanghai at 80 m, the spatial median of $\Delta B^{\mathrm{traj}}$ is $-3.1\times10^{-4}$, and the spatial median of $\Delta C^{\mathrm{traj}}$ is $5.8\times10^{-5}$. On the dimensionless clearance-exposure scale, the 90th percentile of $\Delta C^{\mathrm{traj}}$ reaches 0.012, showing that the small median coexists with a local upper tail.

\FloatBarrier

\begin{figure}[!htbp]
    \centering
    \begingroup
    \setlength{\tabcolsep}{3pt}
    \renewcommand{\arraystretch}{0.86}
    \begin{tabular}{@{}cc@{}}
        \WBCPanel{0.47\textwidth}{\textbf{(a)} $\Delta C^{\mathrm{traj}}$ spatial-median profile}{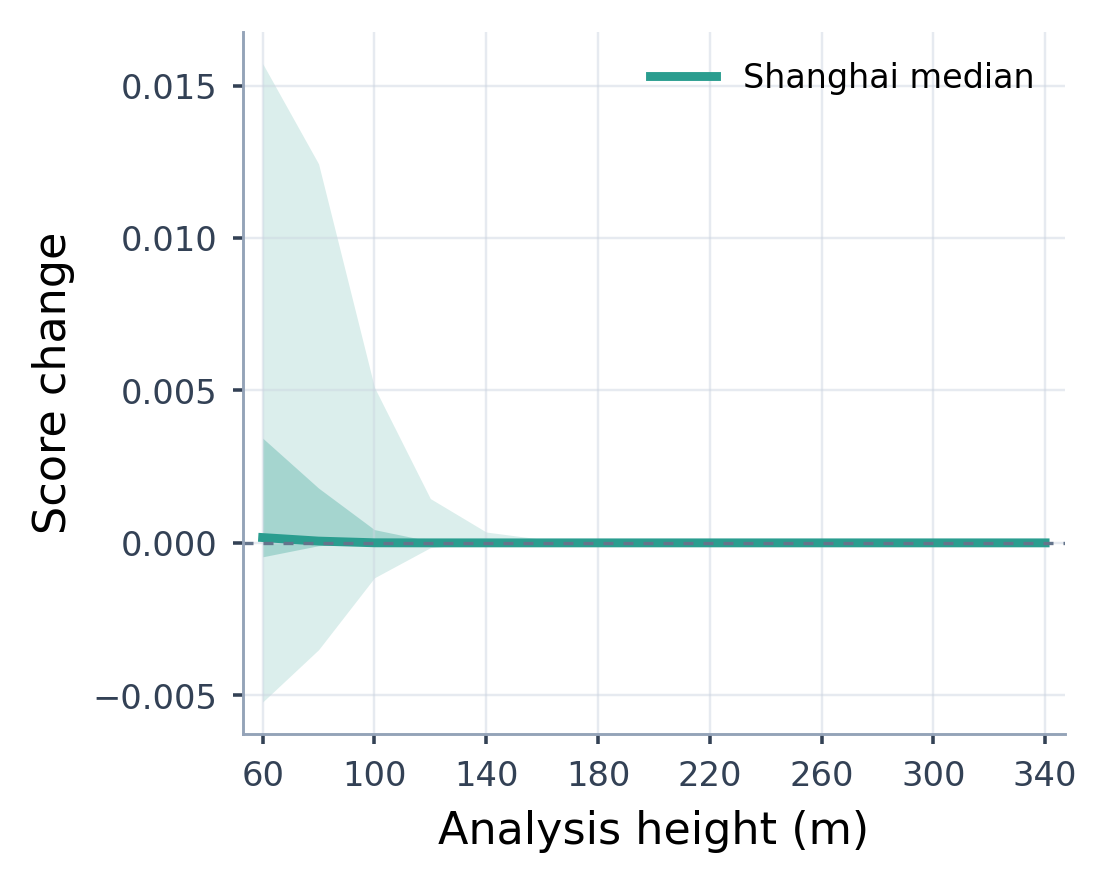} &
        \WBCPanel{0.47\textwidth}{\textbf{(b)} $\Delta B^{\mathrm{traj}}$ spatial-median profile}{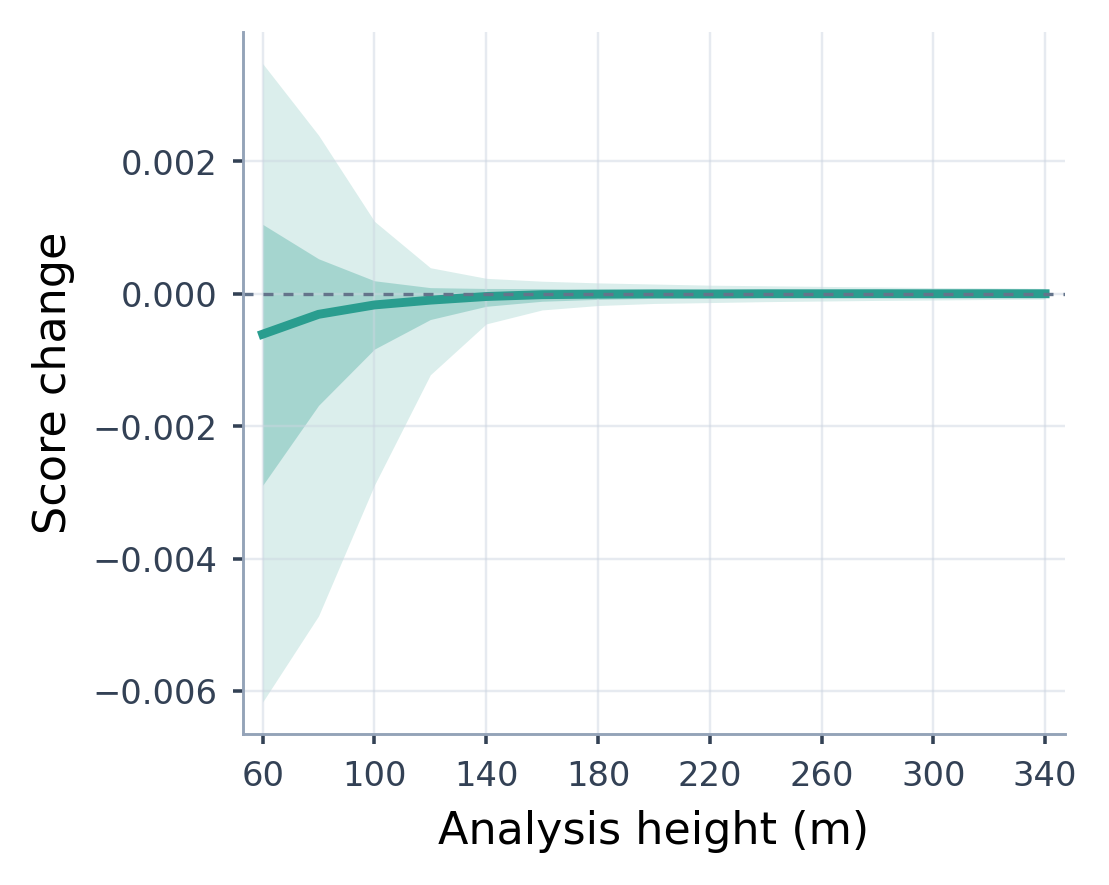} \WBCPanelGap
        \multicolumn{2}{c}{\WBCPanel{0.55\textwidth}{\textbf{(c)} Sign fractions for $\Delta C^{\mathrm{traj}}$}{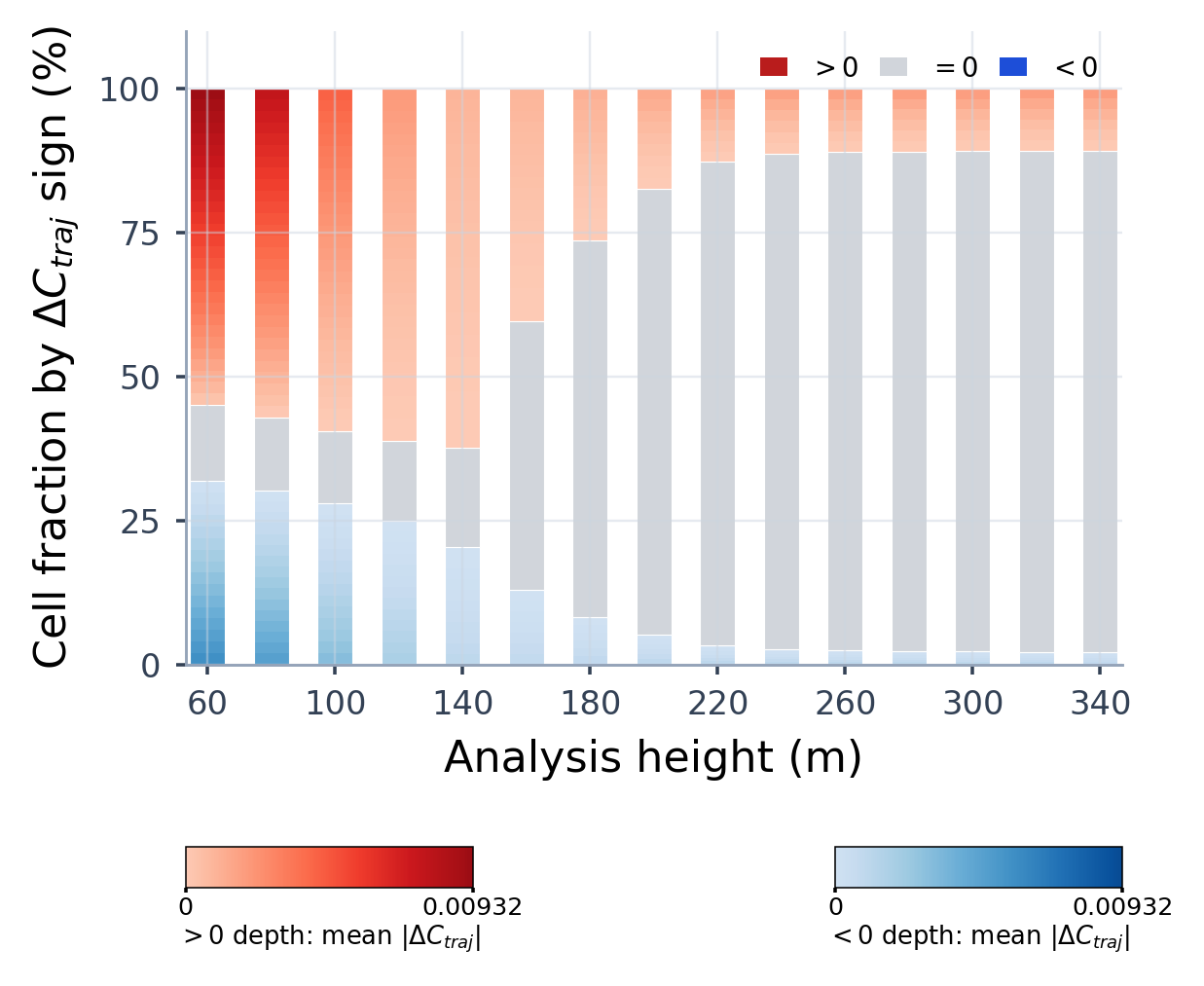}}
    \end{tabular}
    \endgroup
    \caption{Shanghai vertical summaries for trajectory-induced contrasts. \textbf{(a-b)} Spatial-median profiles for $\Delta C^{\mathrm{traj}}$ and $\Delta B^{\mathrm{traj}}$ across reported heights, with interquartile, 10th-90th, and 1st-99th percentile bands. \textbf{(c)} Fraction of finite airspace cells with positive, zero, and negative clearance-exposure change by height.}
    \label{fig:results_shanghai_delta_profiles}
\end{figure}

The vertical response profile in Figure~\ref{fig:results_shanghai_delta_profiles} has a near-zero median $\Delta C^{\mathrm{traj}}$ by 100-120 m, while the upper tail remains in the low layers: the 90th percentile is 0.012 at 80 m and 0.0051 at 100 m. The sign-fraction profile gives a related but distinct view. Positive $\Delta C^{\mathrm{traj}}$ cells account for 54.9\% of finite airspace cells at 60 m, 57.2\% at 80 m, and 59.5\% at 100 m, whereas zero-change cells rise to 46.6\% at 160 m and 77.4\% at 200 m. The lower-altitude layer response has broad sign coverage but concentrated magnitude: many cells show small positive changes, while a smaller subset forms the upper-tail clearance response. The planning signal is the location of low-clearance points within the lower-altitude layer response field.

\FloatBarrier

\begin{table*}[!htbp]
\centering
\scriptsize
\caption{Same-design Shanghai--Beijing summary for trajectory-induced clearance-exposure change. Fractions are computed over finite airspace cells at the stated height.}
\label{tab:cross_city_delta_c_summary}

\begin{tabular*}{\textwidth}{@{\extracolsep{\fill}}lcccc@{}}
\toprule
City &
\begin{tabular}[c]{@{}c@{}}80 m median $\Delta C^{\mathrm{traj}}$\end{tabular} &
\begin{tabular}[c]{@{}c@{}}80 m $Q_{90}(\Delta C^{\mathrm{traj}})$\end{tabular} &
\begin{tabular}[c]{@{}c@{}}80 m positive cells\end{tabular} &
\begin{tabular}[c]{@{}c@{}}200 m zero-change cells \end{tabular} \\
\midrule
Shanghai & $5.8\times10^{-5}$ & 0.012 & 57.2\% & 77.4\% \\
Beijing  & $8.3\times10^{-5}$ & 0.024 & 52.8\% & 68.9\% \\
\bottomrule
\end{tabular*}

\end{table*}

The Beijing case has the same response pattern with a different response strength (Table~\ref{tab:cross_city_delta_c_summary}). The Beijing appendix reports a 90th percentile $\Delta C^{\mathrm{traj}}$ of 0.024 at 80 m, higher than the Shanghai value under the same design (\ref{app:beijing_diagnostic}). At 160 m, the Beijing median is effectively zero, but the 95th and 99th percentiles remain 0.0060 and 0.026. Under the same design, both cities show small domain-wide medians and lower-altitude layer upper-tail clearance responses, with different response strength and height persistence. Wind-induced displacement creates local lower-altitude layer clearance-response tails.

\subsection{Tile-level baseline screening state space}

\begin{figure}[!htbp]
    \centering
    \begingroup
    \setlength{\tabcolsep}{3pt}
    \renewcommand{\arraystretch}{0.86}
    \begin{tabular}{@{}cc@{}}
        \WBCPanel{0.47\textwidth}{\textbf{(a)} Low layer, 60-100 m}{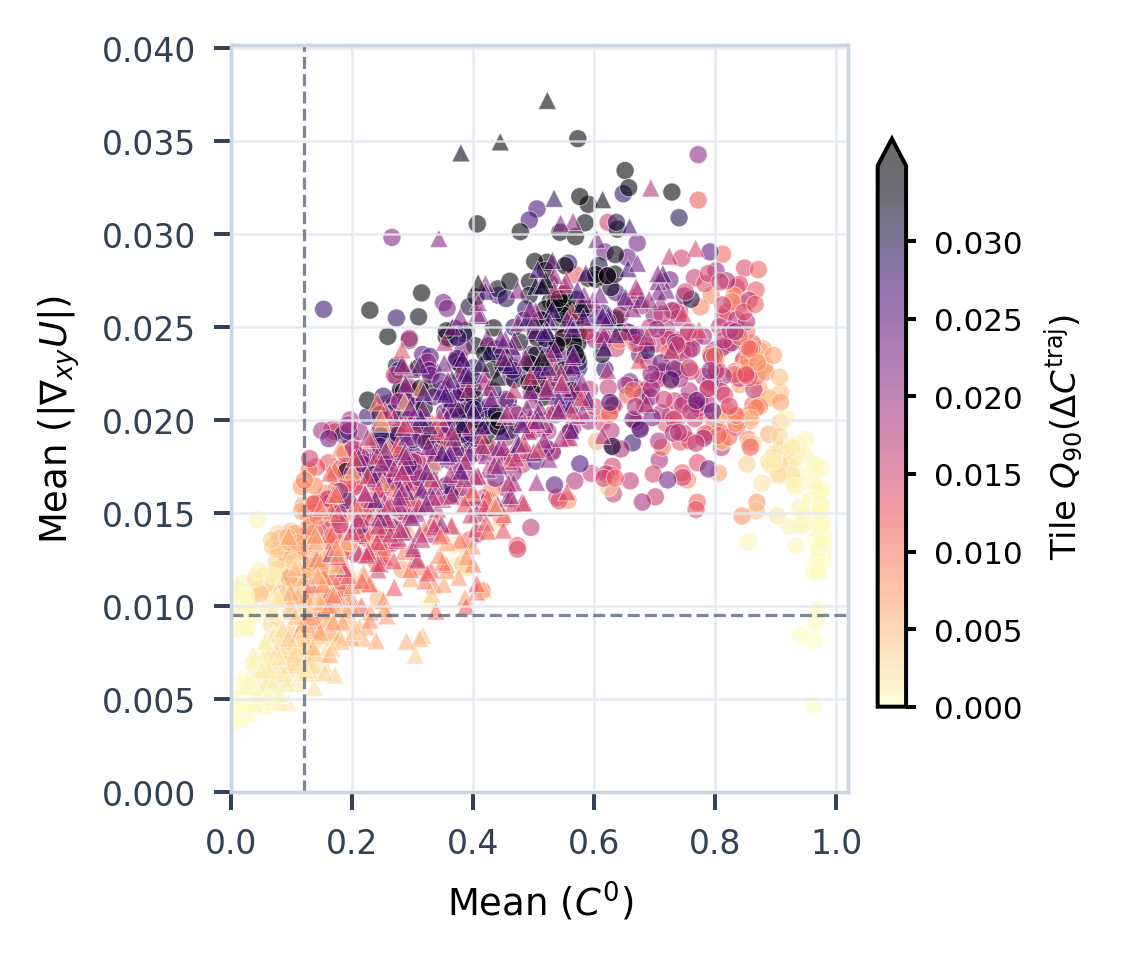} &
        \WBCPanel{0.47\textwidth}{\textbf{(b)} Middle layer, 120-180 m}{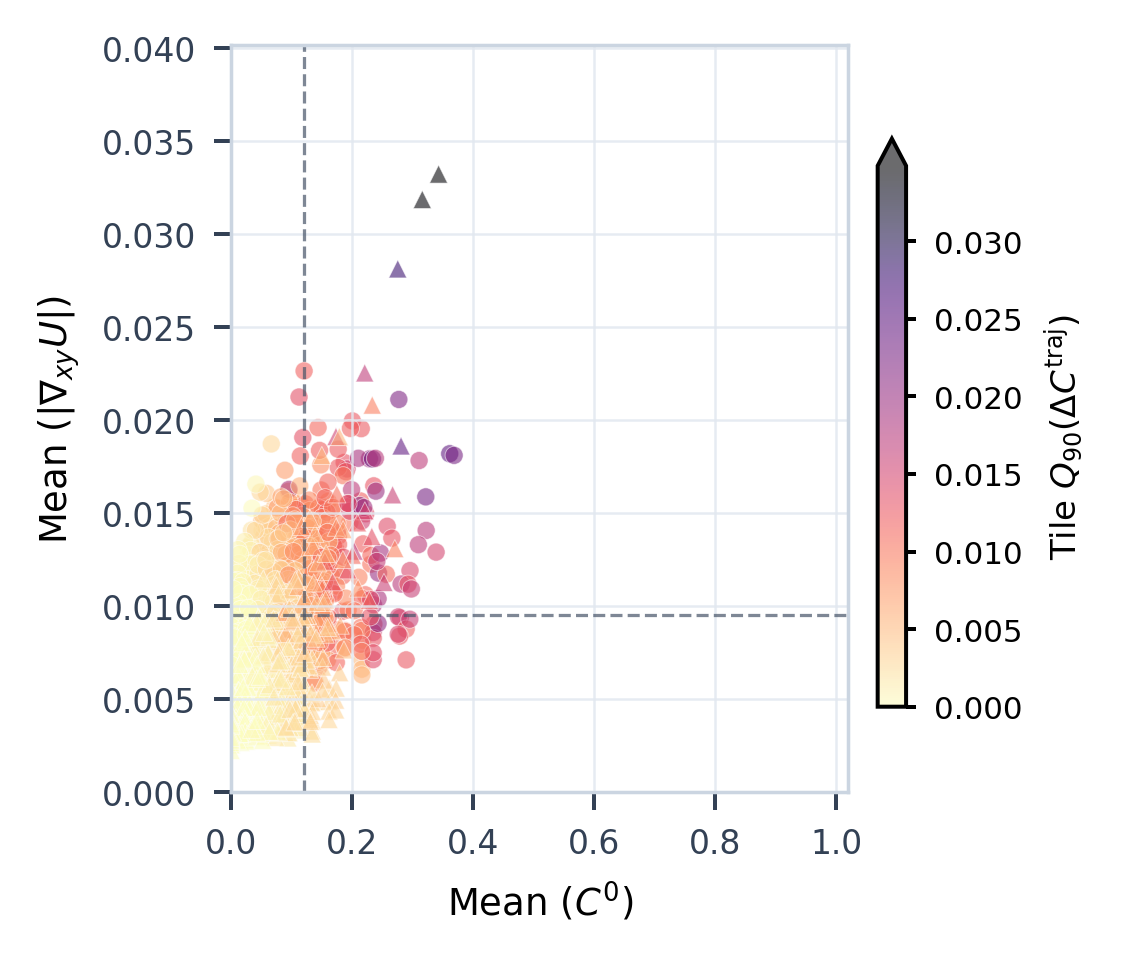} \WBCPanelGap
        \multicolumn{2}{c}{\WBCPanel{0.47\textwidth}{\textbf{(c)} Upper layer, 200-340 m}{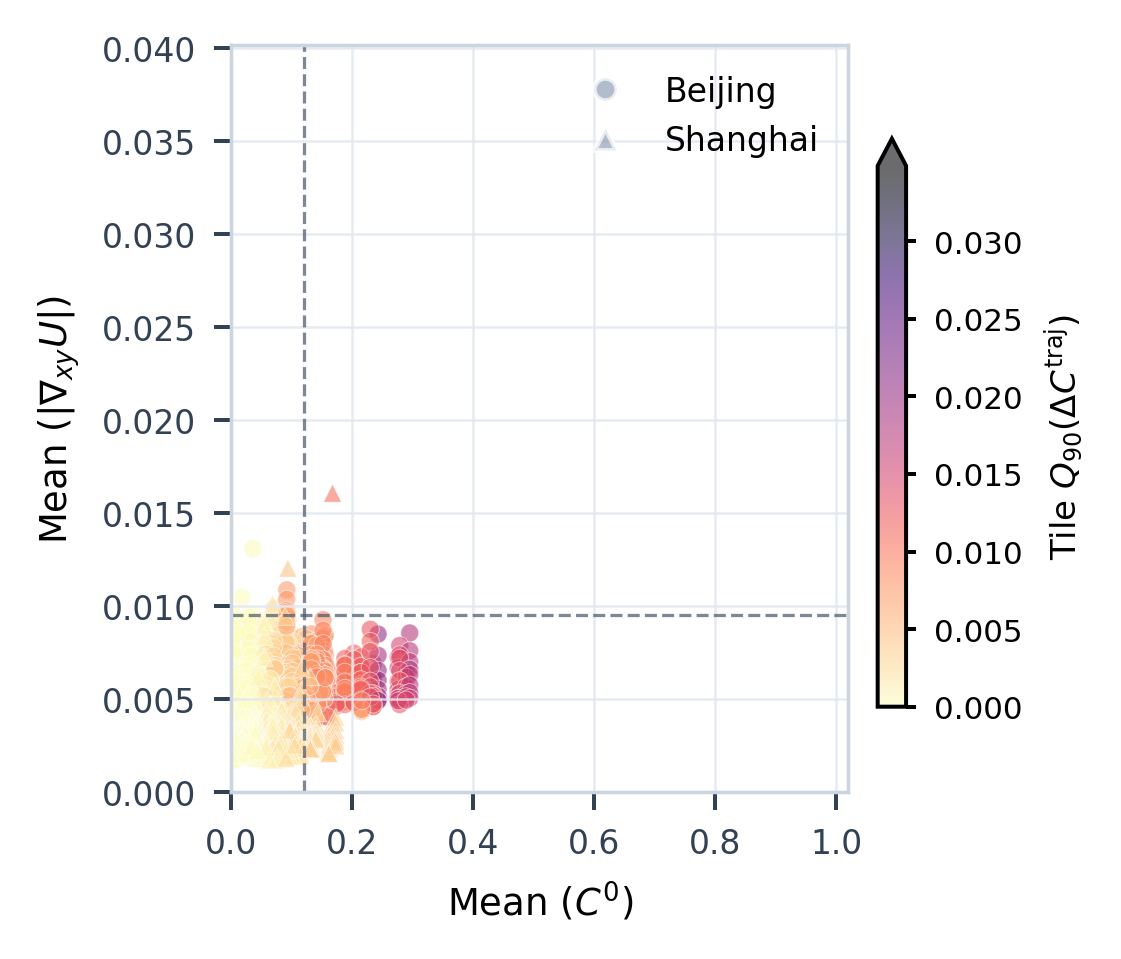}}
    \end{tabular}
    \endgroup
    \caption{Tile-level baseline screening state space by height band. \textbf{(a)} Low layer, 60-100 m. \textbf{(b)} Middle layer, 120-180 m. \textbf{(c)} Upper layer, 200-340 m. Each point is one 1 km tile-height unit passing the 50\% airspace-cell threshold. The x-axis is mean nominal clearance exposure $C^0$, the y-axis is mean horizontal wind-speed-gradient magnitude $|\nabla_{xy}U|$, and colour is the tile-level $Q_{90}(\Delta C^{\mathrm{traj}})$. Dashed lines mark pooled upper-quartile reference values for the two baseline coordinates.}
    \label{fig:results_tile_screening_state_space}
\end{figure}

The next planning question is which 1 km tile-height units should be inspected before route or corridor geometry is fixed. Figure~\ref{fig:results_tile_screening_state_space} moves from city-height summaries to a baseline screening state space. For each tile $t$ and native height $h$, the plotted coordinates are
\begin{equation}
    x_{t,h}=\operatorname{mean}_{i\in\mathcal{F}_{t,h}} C^0_{i,h},
    \qquad
    y_{t,h}=\operatorname{mean}_{i\in\mathcal{F}_{t,h}} |\nabla_{xy}\mathbf{u}_{\mathrm{eff}}|_{i,h},
\end{equation}
and colour shows the resulting upper-tail clearance response $Q_{90}(\Delta C^{\mathrm{traj}})_{t,h}$. Each point is one tile-height unit passing the 50\% airspace-cell threshold. The marker shape distinguishes cities, and panels separate low, middle, and upper height bands. The plotted state space links pre-response conditions to the observed upper-tail clearance response.

The high-colour points concentrate mainly where nominal clearance exposure is non-negligible and horizontal wind-field variation is strong. Lower-altitude layer tile-height units span the widest range of both baseline coordinates and contain most of the high $Q_{90}(\Delta C^{\mathrm{traj}})$ points. Middle layers retain some response where the two baseline conditions still overlap, and upper layers are mostly low-response except for isolated tile-height units. Beijing and Shanghai occupy overlapping screening state space, but the extent and organization of their high-response points differ. The practical screening entry point is the overlap between baseline clearance sensitivity and horizontal wind variation.

\subsection{Baseline drivers of upper-tail clearance response}

\begin{figure}[!htbp]
    \centering
    \begingroup
    \setlength{\tabcolsep}{3pt}
    \renewcommand{\arraystretch}{0.86}
    \begin{tabular}{@{}cc@{}}
        \WBCPanel{0.43\textwidth}{\textbf{(a)} Observed vs. predicted $Q_{90}(\Delta C^{\mathrm{traj}})$}{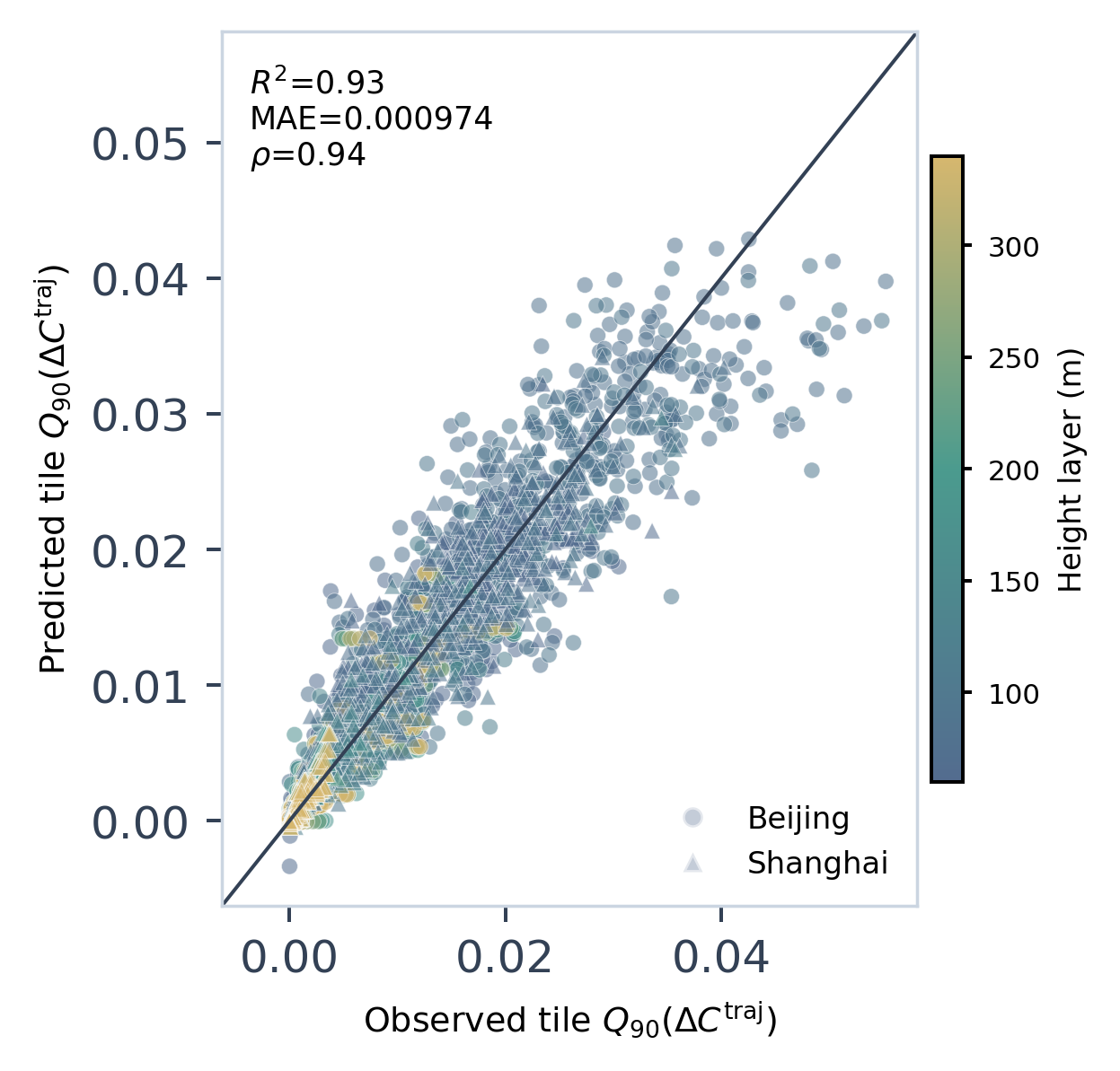} &
        \WBCPanel{0.54\textwidth}{\textbf{(b)} Global SHAP distribution}{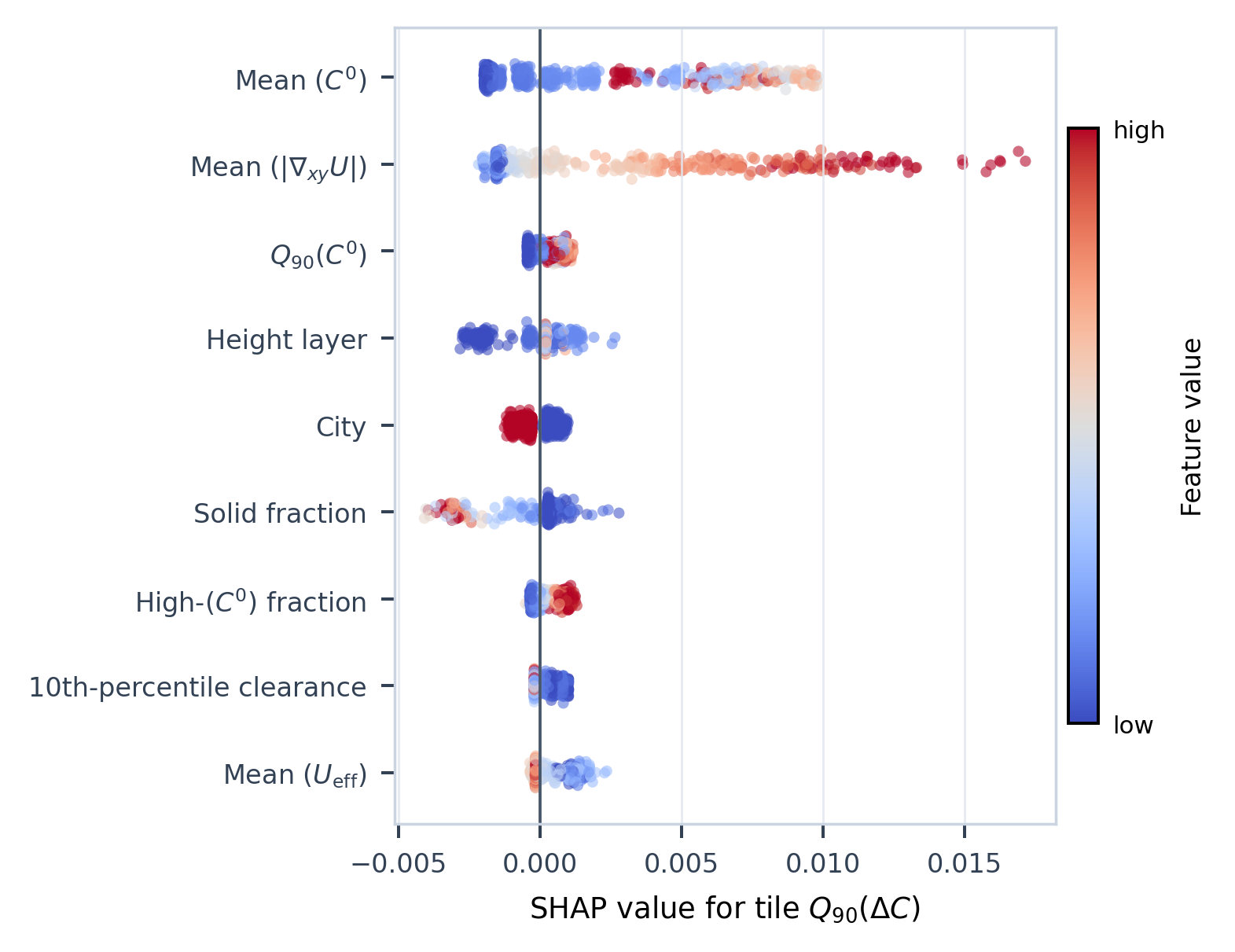} \WBCPanelGap
        \WBCPanel{0.43\textwidth}{\textbf{(c)} City-stratified mean absolute SHAP}{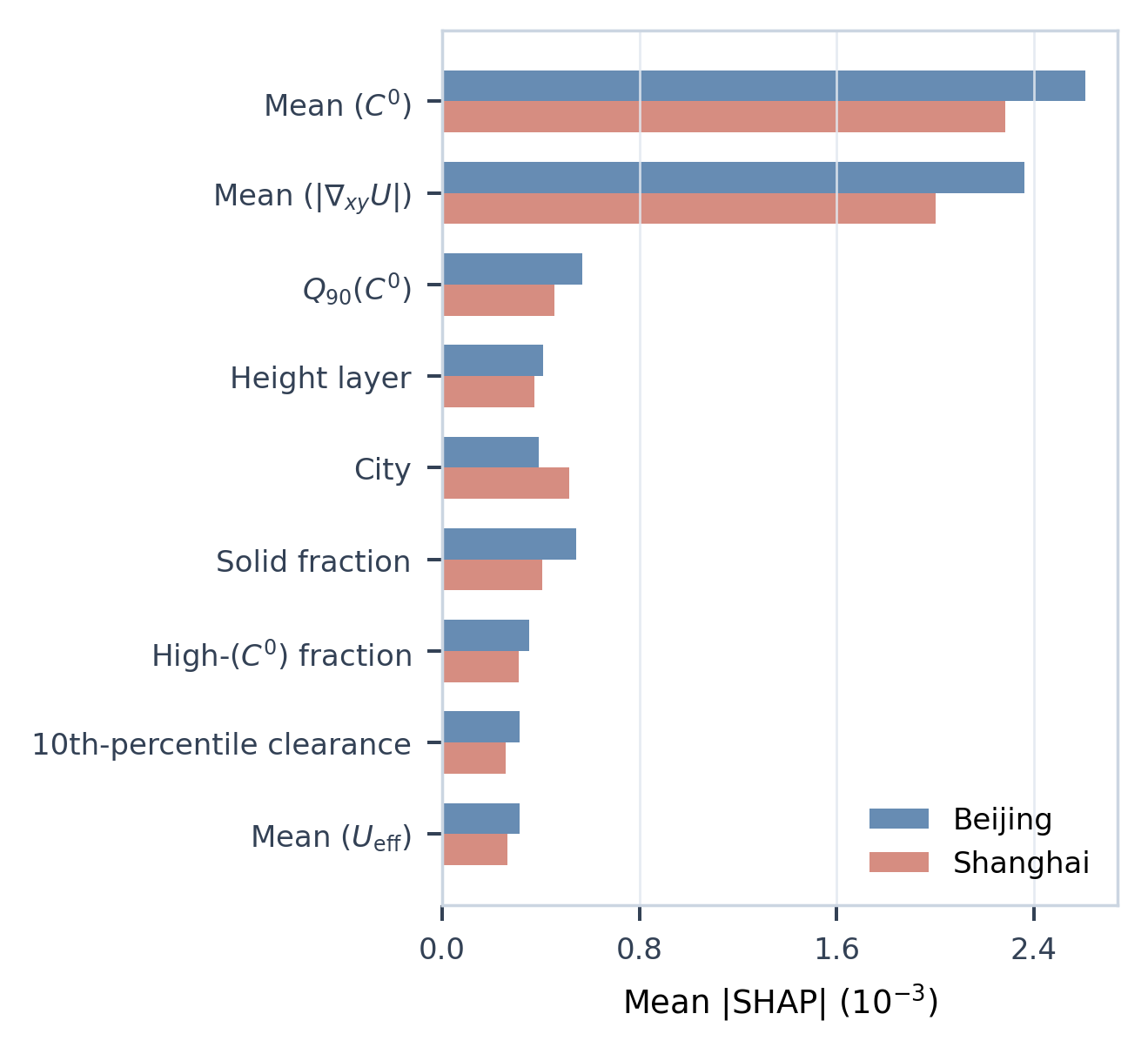} &
        \WBCPanel{0.54\textwidth}{\textbf{(d)} SHAP dependence for dominant features}{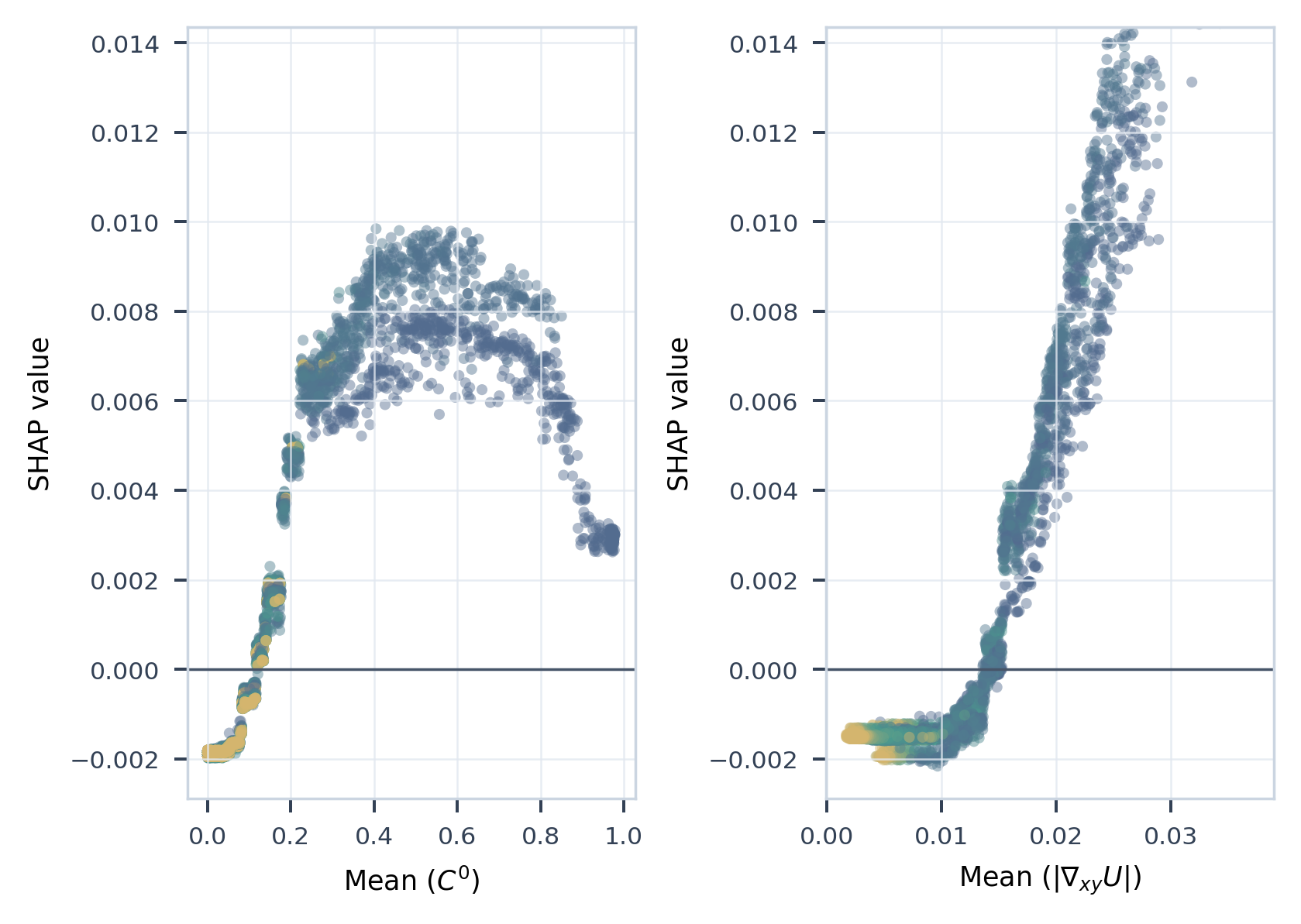}
    \end{tabular}
    \endgroup
    \caption{Baseline wind-geometry explanation of tile-level upper-tail clearance response. \textbf{(a)} Observed and eXtreme Gradient Boosting (XGBoost)-predicted tile-level $Q_{90}(\Delta C^{\mathrm{traj}})$ for pooled Shanghai and Beijing tile-height units; marker shape distinguishes cities and marker color indicates height layer. \textbf{(b)} Global SHAP summary for the pooled regression model, showing the baseline features most strongly associated with predicted upper-tail clearance response. \textbf{(c)} City-stratified mean absolute SHAP values for the leading predictors. \textbf{(d)} SHAP dependence for the two dominant features, Mean ($C^0$) and Mean ($|\nabla_{xy}U|$). The model associates upper-tail clearance response mainly with baseline clearance sensitivity and horizontal wind-field variation; fixed-support wind-loading magnitude, mean wind speed, and city identity are secondary.}
    \label{fig:results_tile_clearance_shap}
\end{figure}

Figure~\ref{fig:results_tile_clearance_shap} extends the bivariate screening state space with a broader SHapley Additive exPlanations (SHAP) explanatory model. The practical rule emerging from the tile analysis is to prioritise tile-height units where nominal clearance exposure and horizontal wind variation coincide. The model uses $Q_{90}(\Delta C^{\mathrm{traj}})_{t,h}$ as the target and baseline or pre-response variables as predictors: $C^0$ and $B^0$ summaries, wind speed, TKE, horizontal speed-gradient and vertical shear summaries, height, city, and tile urban-form or clearance descriptors. The pooled model uses Shanghai and Beijing tile-height units together, encodes city as a feature, and assigns all height layers from the same tile to the same cross-validation (CV) fold.

The grouped-CV performance supports retaining the continuous response target. The pooled regression gives $R^2=0.927$, mean absolute error (MAE) $=0.000974$, and Spearman $\rho=0.935$ (Figure~\ref{fig:results_tile_clearance_shap}a). A binary high-response check using the top 20\% of tile-height units gives an area under the receiver-operating-characteristic curve (AUC) of 0.994 and a precision-recall AUC of 0.973, giving the same separation as a robustness check.

The global SHAP distribution identifies mean baseline clearance score $C^0$ and mean horizontal wind-speed-gradient magnitude $|\nabla_{xy}U|$ as the two leading correlates of upper-tail clearance response (Figure~\ref{fig:results_tile_clearance_shap}b). Mean $C^0$ indicates whether a tile is already clearance-sensitive under the nominal support. Mean $|\nabla_{xy}U|$ indicates the strength of local horizontal wind-field variation. Other contributors include $Q_{90}(C^0)$, height, city, solid fraction, and the fraction of cells above the baseline-clearance score threshold. Fixed-support wind-loading magnitude $B^0$, mean effective wind speed, and the city label remain secondary to the two dominant baseline descriptors.

The city-stratified SHAP panel shows that Beijing and Shanghai share the same dominant driver structure, while the magnitude of some feature contributions differs by city (Figure~\ref{fig:results_tile_clearance_shap}c). The city indicator represents a residual city-specific component in the pooled model. The dependence plots further separate the two leading driver signatures (Figure~\ref{fig:results_tile_clearance_shap}d). Mean $C^0$ has a nonlinear contribution: tile-height units with very low nominal clearance exposure rarely form high $\Delta C^{\mathrm{traj}}$ tails, while intermediate and higher $C^0$ values increase the predicted response before the relation bends. Mean $|\nabla_{xy}U|$ has a more consistently positive dependence, linking stronger horizontal wind-field variation with higher predicted upper-tail clearance response.

The auxiliary model for $Q_{90}(\Delta B^{\mathrm{traj}})$ gives a strong grouped-CV fit ($R^2=0.971$, MAE $=0.000116$, Spearman $\rho=0.950$). Its leading features shift toward mean effective wind speed, horizontal wind-gradient summaries, mean $C^0$, and baseline $B^0$. The clearance-response channel is led by baseline clearance sensitivity and horizontal wind variation, while the wind-loading-response channel is led by wind-speed and gradient descriptors.

\subsection{Discussion and limitations}

Urban wind affects UAV clearance through the combined state of wind, vehicle motion, and surrounding terrain-building geometry. The largest upper-tail clearance responses occur where the nominal vehicle envelope is already close to terrain-building margins and where local wind varies sufficiently over space to move that envelope toward those margins. By separating nominal-envelope wind burden, clearance-exposure change caused by wind-affected motion, and the corresponding change in wind burden, the analysis shows that wind loading, clearance exposure, and trajectory response have distinct spatial and vertical structures.

Relative to pointwise wind-risk maps, the dependent variable here is clearance-exposure change. Patil and Garc\'ia-S\'anchez \cite{patil_quantifying_2025} compared UAM wind-risk maps from velocity and turbulence fields under different building levels of detail, while Geng and Gou \cite{geng_urban_2026} linked urban form to fixed-location UAV wind-risk clustering. These studies evaluate wind risk at locations or map cells. The present study evaluates a paired change in clearance exposure on nominal and wind-perturbed vehicle-sized trajectory envelopes. This difference in dependent variable explains why fixed-support wind-loading magnitude is secondary (Figure~\ref{fig:results_tile_clearance_shap}). A location can be wind loaded while still having enough surrounding space, and a more moderate wind-loading location can be clearance-sensitive when the nearby geometry is tight.

These results clarify wind-cost definitions used before route search. Wind costs based on cumulative speed exposure or fixed-support loading are burden proxies. Clearance-sensitive cost requires the wind-displaced vehicle envelope and surrounding terrain-building geometry. A windy open segment and a moderate-wind constrained segment can therefore carry different clearance implications.

Trajectory-deviation hazard defines the response around aircraft motion. Jeong et al. \cite{jeong_hazardous_2021} combined WRF-LES, flight simulation, and a neural network to predict hazardous flight regions from wind-induced path-deviation distance. The present response measure is centred on terrain-building clearance exposure: it asks whether the wind-affected trajectory envelope changes near-obstacle exposure relative to the nominal envelope. A similar deviation can therefore produce different clearance responses in open space, near a building edge, or near raised terrain.

These clearance patterns align with studies showing that spatially varying wind matters for UAV and UAM operations. Giersch et al. \cite{giersch_atmospheric_2022} argued that meter-scale turbulent urban wind databases can support safer drone mission planning. Vuppala et al. \cite{vuppala_modeling_2024} showed that time-mean wind fields under-predict lateral response and control activity for a representative AAM aircraft case. The present results add the geometry side of that argument. Spatially varying wind produces the strongest clearance response where the baseline local flight envelope is already clearance-sensitive.

The diagnostic operates upstream of route optimization. It flags cell-height and tile-height units that deserve more clearance margin, avoidance, or detailed local analysis before route buffers and corridor alignments are fixed. Route optimization remains a later step that must also handle origin-destination demand, path continuity, energy use, traffic separation, vertiport access, temporal scheduling, and route-level constraints. The present outputs are best used as pre-design evidence for where route search or corridor design should apply greater clearance caution.

The input bounds define the interpretation. The annual representative wind field describes a planning wind state, so annual exceedance probability and storm-regime risk require separate wind-regime analysis. The wind-geometry data resolution reported in \S\ref{sec:3_exp} represents district-scale building effects, while narrow street gaps, roof-edge details, and small terrain breaks may be smoothed. The planning-envelope radius and clearance margin are nominal class parameters within this planning-grid assessment and are reported in Appendix Table~\ref{tab:experimental_config}. The reduced-order vehicle response gives a common paired displacement basis; aircraft-specific control laws, gust response, and flight-envelope limits require dedicated vehicle models. These limits make the results suitable for ranking clearance-sensitive areas and explaining response patterns, with calibrated safety probability and route-level operation left to later analysis.

\FloatBarrier

\section{Conclusions\label{sec:4_concl}}

Building-modified urban wind changes near-obstacle clearance for UAV-sized flight envelopes through a local wind-geometry response. At each location and height, the analysis compared nominal motion with wind-affected local motion. The comparison separates three quantities: the wind burden encountered along the nominal envelope, the envelope's initial proximity to terrain and buildings, and the additional near-obstacle exposure introduced by wind-affected displacement. This clearance response also differs from the change in wind burden sampled after displacement.

Applied to Shanghai and Beijing under common grid, height, sampling, vehicle-class, and scoring settings, the analysis shows small median clearance changes over the full domain. The main response is a localized upper tail in lower height layers and around tight terrain-building structures. Beijing repeats the same height-decay pattern but shows a stronger lower-altitude upper-tail clearance response than Shanghai under the same design.

The km-scale tile analysis indicates that high upper-tail clearance response is associated mainly with two pre-response conditions: the nominal envelope is already close to terrain-building margins, and the local horizontal wind field varies enough to shift that envelope toward those margins. Fixed-support wind-loading magnitude, mean wind speed, and city identity are secondary predictors. The model for wind-loading change has a different driver profile, reinforcing that clearance response and wind-loading response should be analysed as separate channels.

The combined evidence defines the route-pre-design contribution. Costs based on cumulative speed exposure, fixed-support wind-loading layers, static building-clearance layers, and direct overlays of these layers are incomplete bases for clearance screening. A wind-loading map identifies where the nominal flight envelope samples a stronger aerodynamic environment. A clearance map identifies where that envelope is already close to terrain-building margins. Wind-affected near-obstacle exposure requires the paired trajectory response. Wind-aware route optimization therefore needs the clearance consequence of the wind-displaced envelope alongside cumulative wind exposure. Route and corridor studies should treat wind loading, nominal clearance, and wind-affected clearance response as distinct but connected screening layers.

In low-altitude planning for UAM systems, the results identify locations and heights where nominal flight envelopes are already near terrain-building margins and horizontal wind gradients can move them closer. These locations are candidates for greater clearance caution, local refinement, or avoidance before route buffers and corridor alignments are fixed. 

Accident probability, regulatory separation compliance, and route-level operational performance require additional modelling. Future work should add wind-regime-specific propagation, vehicle-response validation, higher-resolution clearance geometry, and route-level aggregation.

\section*{Declarations}
{
\small
\begin{itemize}

    \item \textbf{Funding:}
    \begin{sloppypar}
    This work was supported by the Shanghai Key Laboratory of Aerodynamics and Thermal Environment Simulation for Ground Vehicles (Grant No.~23DZ2229029).
    \end{sloppypar}

    \item \textbf{Acknowledgements:} \\
    Huanxia Wei would like to thank Dalin Liu from National University of Singapore and Shuai Han from The National Meteorological Information Centre of the China Meteorological Administration for helpful discussions on WRF-CFD coupling.

    \item \textbf{Conflicts of interest:} \\
    The authors have no conflicts of interest to declare that are relevant to this article.

    \item \textbf{Data availability:} \\
    Data will be made available on request.
\end{itemize}
}

\appendix
\renewcommand{\thetable}{A.\arabic{table}}
\renewcommand{\thefigure}{A.\arabic{figure}}
\renewcommand{\theHtable}{appendix.\arabic{table}}
\renewcommand{\theHfigure}{appendix.\arabic{figure}}
\setcounter{table}{0}
\setcounter{figure}{0}

\appendix
\clearpage

\section{Wind-asset verification and validations}\label{app:vv}

A lattice convergence study uses vertical profiles as convergence criteria. Three lattice sizes are tested: 40~m, 20~m, and 10~m. The 40~m lattice shows a modest bias, and the 20~m and 10~m cases are closely aligned. The reported method uses the native 20 m grid.

Two turbulence-statistics checks are used for the LES mesh: normalised two-point correlation and spectral energy distribution. Following Davidson \cite{Davin_validation}, the normalised two-point correlation is defined as
\begin{equation}{
C_{U_i}^{\mathrm{norm}}\left(\vec{x}_0,\vec{x}\right)
=
\frac{\overline{{U_i}'\left(\vec{x}_0\right)\,{U_i}'\left(\vec{x}\right)}}
{{U_i}'_{\mathrm{RMS}}\left(\vec{x}_0\right)\,{U_i}'_{\mathrm{RMS}}\left(\vec{x}\right)},~~ \\
{U_i}'_{\mathrm{RMS}}\left(\vec{x_i}\right)
=
\overline{{U_i}'\left(\vec{x_i}\right)^2}^{1/2},
}\end{equation}
where $\vec{x}_0$ and $\vec{x}$ are the reference and sampling locations, respectively, and $U_i$ is the velocity component in the $i^\text{th}$ direction. The reference point is set at the geographical centre, with 12 successive probes placed along the $x$-axis at one-mesh-size intervals.

\begin{figure}[!htbp]
    \centering
    \includegraphics[width=\linewidth]{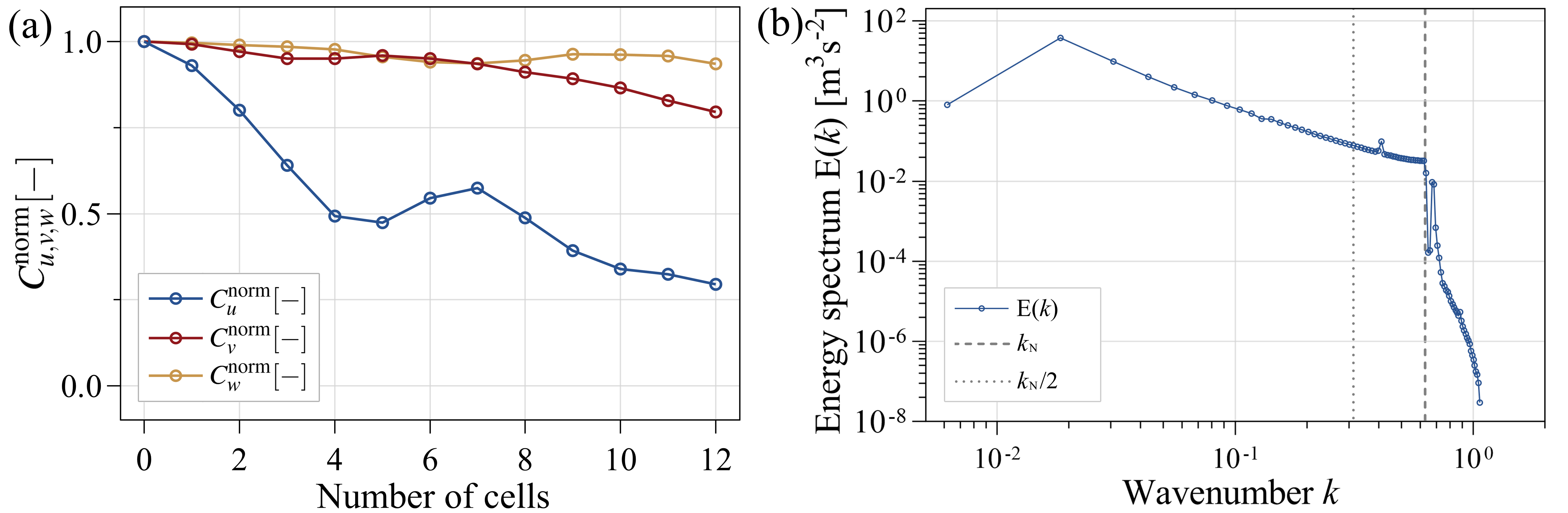}
    \caption{Turbulence statistics verification. \textbf{(a)} Convergence of two-point correlation $C_{u,v,w}^{\mathrm{norm}}$ versus stride number of cells. \textbf{(b)} Global energy wavenumber spectrum $E(k)$.}
    \label{fig:Mesh_kn}
\end{figure}

In Figure~\ref{fig:Mesh_kn}(a), the two-point correlation decays across the 12-cell probe sequence and retains resolved spatial structure beyond the commonly used eight-cell energetic-eddy criterion. The decay is consistent with use of the 20 m lattice for the resolved turbulence checks in this planning study.

The spectral kinetic energy density at wavenumber vector $\mathbf{k}$ is given by
\begin{equation}
e(\mathbf{k})=\frac{1}{2}\left(\left|\widehat{u}(\mathbf{k})\right|^{2}+\left|\widehat{v}(\mathbf{k})\right|^{2}+\left|\widehat{w}(\mathbf{k})\right|^{2}\right),
\end{equation}
where $\widehat{\mathbf{u}}(\mathbf{k})=\mathcal{F}\{\mathbf{u}'(\mathbf{x})\}$ and $\mathbf{u}'=\mathbf{u}-\langle\mathbf{u}\rangle$. The isotropic energy spectrum is obtained by averaging over wavevectors with identical magnitude $k=\|\mathbf{k}\|$, satisfying $\int_{0}^{\infty} E(k)\,\mathrm{d}k=\int_{\mathbb{R}^{3}} e(\mathbf{k})\,\mathrm{d}\mathbf{k}$:
\begin{equation}
E(k)=\int_{\|\mathbf{k}\|=k} e(\mathbf{k})\,\mathrm{d}S_{\mathbf{k}}
= k^{2}\int_{4\pi} e\!\left(k,\Omega\right)\,\mathrm{d}\Omega.
\end{equation}

Figure~\ref{fig:Mesh_kn}(b) compares $E(k)$ with the Kolmogorov reference slope $k^{-5/3}$, together with the Nyquist wavenumber $k_N$ and the limit $k_N/2$. The spectrum remains positive and smooth across the resolved range. For $k<k_N/2$, it follows an approximate power-law decay and a broad compensated range, consistent with the Kolmogorov inertial-range scaling. The high-wavenumber range has no pile-up, blocking, or aliasing-induced energy accumulation. The resolved spectrum has no grid-scale instability.

\begin{figure}[!htbp]
    \centering
    \includegraphics[width=\linewidth]{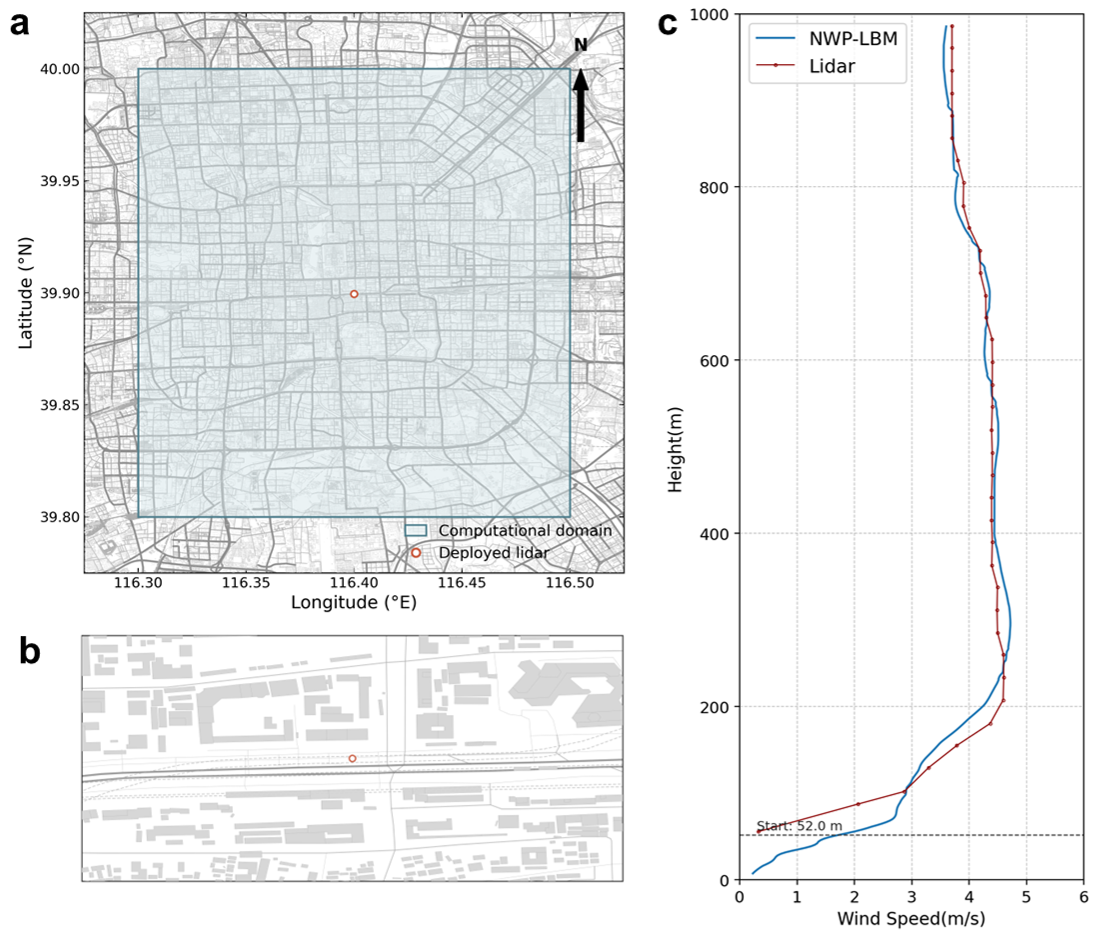}
    \caption{Example profile comparison against lidar observation under NWP-LBM coupling. \textbf{(a-b)} Computational region and lidar location. \textbf{(c)} Vertical wind profiles.}
    \label{fig:vl}
\end{figure}

Field observations rarely provide uniform inflow and an exact annual-mean wind profile, so an example comparison is made against observations from a lidar profiler located in Beijing under numerical weather prediction-LBM (NWP-LBM) coupling mode. The \textsc{LatticeUrbanWind} run uses China Meteorological Administration (CMA) model output as the mesoscale upper boundary. The boundary condition is generated with Kriging interpolation for 14:00 UTC on 2025-09-03.
Figure~\ref{fig:vl}(a)-(b) locates the computational domain, lidar site, and nearby buildings, and Figure~\ref{fig:vl}(c) compares the vertical profiles. The comparison provides a qualitative check on the vertical-profile structure used by the urban wind inputs. UAV-specific validation, multi-city quantitative error assessment, vehicle-response validation, and wind-regime reliability analysis would use dedicated datasets.

\section{Study inputs and analysis settings}\label{app:assets_notation}

Table~\ref{tab:experimental_config} summarises the shared analysis settings. Table~\ref{tab:vehicle_scoring_params} reports the light-UAV response and scoring parameters. Table~\ref{tab:city_assets} reports the primary-comparison city computational domains and CRS details.

\begin{table}[!htbp]
\centering
\scriptsize
\caption{Shared analysis settings for the primary comparison.}
\label{tab:experimental_config}
\begin{tabular}{@{}L{0.22\textwidth}L{0.40\textwidth}L{0.30\textwidth}@{}}
\toprule
Setting & Value & Role \\
\midrule
Data resolution and vertical support & Native horizontal grid spacing: $\Delta x=20$ m; reported start/evaluation heights: 60-340 m; runtime wind-scene support: 20-420 m; $\Delta h=20$ m & Separates reported analysis layers from support-only vertical layers. \\
Wind fields & $\mathbf{u}_{\mathrm{eff}}$, $k$, $g_U$, and $s_U$; trilinear sampling during rollout; $\sigma_U$ only for upstream gust scaling & Inputs to drift, spread, and wind-loading score. \\
Trajectory sampling & Primary domain-wide configuration: $N=16384$ paired trajectory samples; Method schematic: $N_{\mathrm{traj}}=2000$ fixed headings & Common basis for comparing no-wind and wind-perturbed supports. \\
Aggregation & Cell-level sample mean over support-specific valid trajectories & Primary field definition for $C^0$, $B^0$, $C^w$, and $B^w$. \\
Propagation & $v_0=8.0$ m s$^{-1}$; $\Delta t=0.125$ s; $T=6.0$ s; 1 m along-track spacing cap; first-contact freeze & Short-horizon continuous-rollout settings. \\
Vehicle envelope and tiles & $r_{\mathrm{veh}}=2.5$ m; $d_{\mathrm{clear}}=3.0$ m; 1 km $\times$ 1 km tiles; 50\% airspace-cell threshold & Planning-envelope radius and clearance-screen margin; standardized local-space summaries. \\
\bottomrule
\end{tabular}
\end{table}

\begin{table}[!htbp]
\centering
\scriptsize
\caption{Light-UAV response and scoring parameters.}
\label{tab:vehicle_scoring_params}
\begin{tabular}{@{}L{0.25\textwidth}L{0.36\textwidth}L{0.31\textwidth}@{}}
\toprule
Parameter group & Value & Role \\
\midrule
Vehicle response class & Light UAV & Nominal light-UAV screening class for diagnostic propagation. \\
Wind drift gains & $g_x=1.0$, $g_y=1.0$, $g_z=1.2$ & Map start-relative effective-wind variation into diagnostic drift. \\
Baseline dispersion & $\sigma_x=\sigma_y=0.05$ m; $\sigma_z=0.025$ m & Step-level vehicle, positioning, and reduced-model spread. \\
Turbulence-spread gains & $\alpha_x=\alpha_y=1.0$; $\alpha_z=1.2$ & Map $q=\sqrt{k}\Delta t$ into directional spread. \\
Wind-score references & $U_{\mathrm{score}}=8.0$ m s$^{-1}$; $k_{\mathrm{ref}}=1.0$ m$^2$ s$^{-2}$; $g_{\mathrm{ref}}=1.0$ s$^{-1}$; $s_{\mathrm{ref}}=1.0$ s$^{-1}$ & Normalise and clip the wind-loading score components. \\
Wind-score weights & $w_U=w_k=w_g=w_s=1$ & Equal component weighting in the baseline analysis. \\
\bottomrule
\end{tabular}
\end{table}

\begin{table}[!htbp]
\centering
\scriptsize
\caption{Primary-comparison city computational domains and CRS details.}
\label{tab:city_assets}
\begin{tabular}{@{}L{0.12\textwidth}L{0.16\textwidth}L{0.18\textwidth}L{0.18\textwidth}L{0.20\textwidth}@{}}
\toprule
City & Computation domain & Longitude range & Latitude range & Projected CRS \\
\midrule
Shanghai & 19.24 km $\times$ 13.60 km & 121.400-121.600$^\circ$E & 31.180-31.305$^\circ$N & UTM zone 51N (EPSG:32651) \\
Beijing & 18.51 km $\times$ 18.19 km & 116.275-116.490$^\circ$E & 39.825-39.990$^\circ$N & UTM zone 50N (EPSG:32650) \\
\bottomrule
\end{tabular}
\end{table}

\FloatBarrier

\subsection{Vehicle response-scale sensitivity}\label{app:uav_response_scale_sensitivity}

The vehicle response-scale sweep tests whether the Shanghai clearance-response pattern depends on the assumed light-UAV response strength. The scale factor $s$ multiplies the vehicle drift gains and step-level dispersion parameters in Appendix Table~\ref{tab:vehicle_scoring_params}; the environmental wind-speed field and wind-amplitude inputs remain fixed. The sweep can be read as a disturbance-rejection envelope: lower $s$ represents stronger rejection of wind-induced displacement in the diagnostic basis, and higher $s$ represents a more disturbance-sensitive response class. The same Shanghai grid, height layers, airspace-cell mask, edge buffer, trajectory sampling, wind-field sampling, and clearance-score definition are retained as in the primary run.

\begin{figure}[!htbp]
    \centering
    \includegraphics[width=\linewidth]{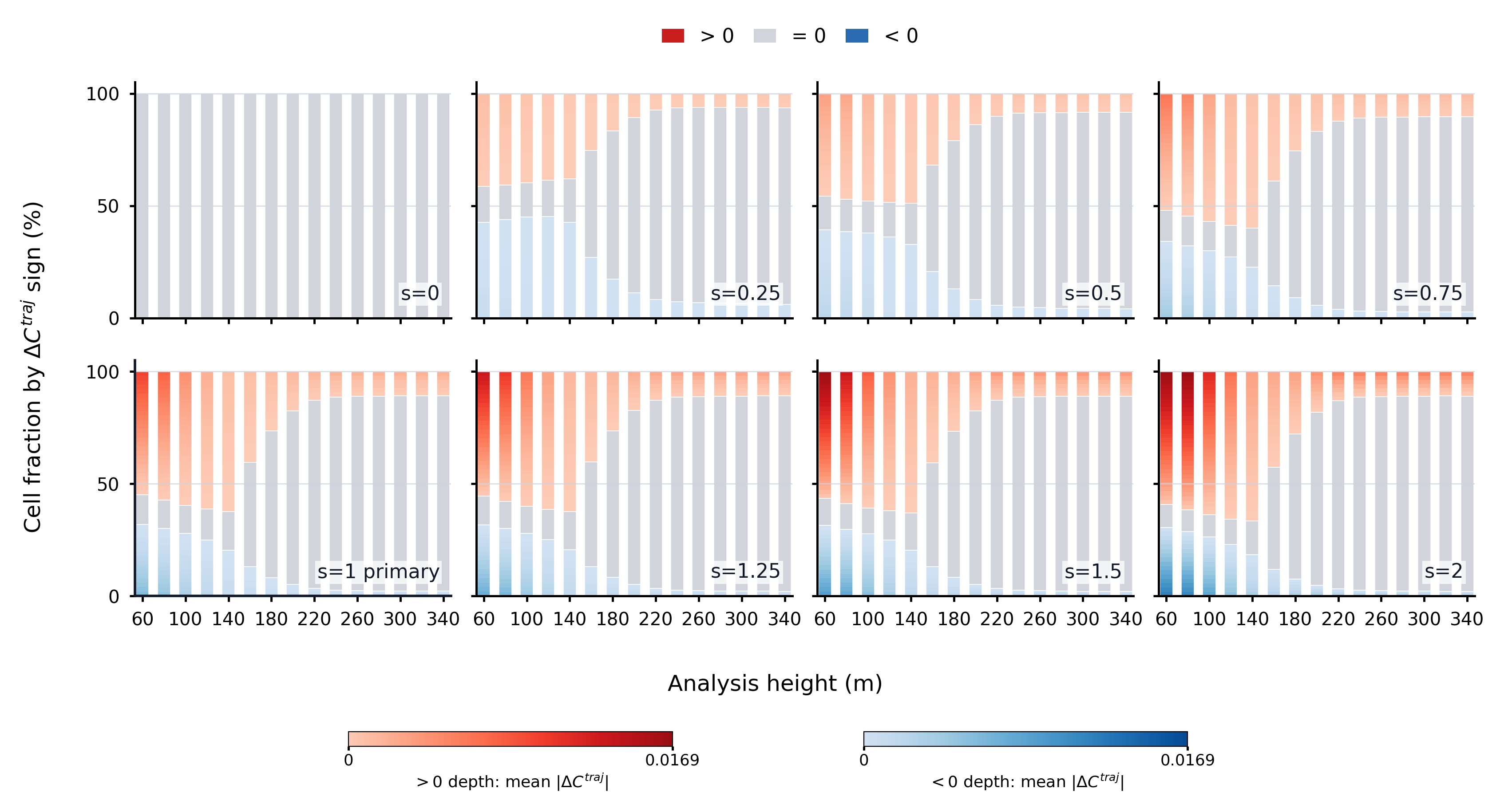}
    \caption{Shanghai vehicle response-scale sensitivity for signed clearance-exposure change. Each panel reports finite airspace-cell fractions by the sign of $\Delta C^{\mathrm{traj}}$ across analysis heights for one vehicle response scale $s$. Grey bars indicate zero change; red and blue bars indicate positive and negative cells, respectively. Colour depth within the red and blue portions gives the sign-specific mean $|\Delta C^{\mathrm{traj}}|$. The primary run is $s=1$.}
    \label{fig:app_uav_response_scale_delta_c_direction_fractions}
\end{figure}

The vehicle response scale changes the amplitude and signed composition of the response; the lower reported layers retain the strongest signed composition across the tested range (Figure~\ref{fig:app_uav_response_scale_delta_c_direction_fractions}). At $s=0$, the wind-perturbed and nominal supports coincide and $\Delta C^{\mathrm{traj}}=0$ across the reported heights. At 80 m, the positive-cell fraction increases from 40.9\% at $s=0.25$ to 57.2\% in the primary run and 61.6\% at $s=2$, while the zero-change fraction decreases from 15.3\% to 12.7\% and 9.7\%. The colour depth also increases with $s$, indicating stronger within-sign clearance-exposure changes. The upper reported layers remain dominated by zero change: at 340 m, zero-change cells account for 87.0\% of finite airspace cells in both the primary and $s=2$ runs. The response magnitude depends on the assumed vehicle response class, and the lower-layer concentration persists across the tested response amplitudes.

\FloatBarrier

\section{Beijing same-design check}\label{app:beijing_diagnostic}

\begin{figure}[!htbp]
    \centering
    \begingroup
    \setlength{\tabcolsep}{3pt}
    \renewcommand{\arraystretch}{0.86}
    \begin{tabular}{@{}cc@{}}
        \WBCPanel{0.47\textwidth}{\textbf{(a)} $B^0$ map at 80 m}{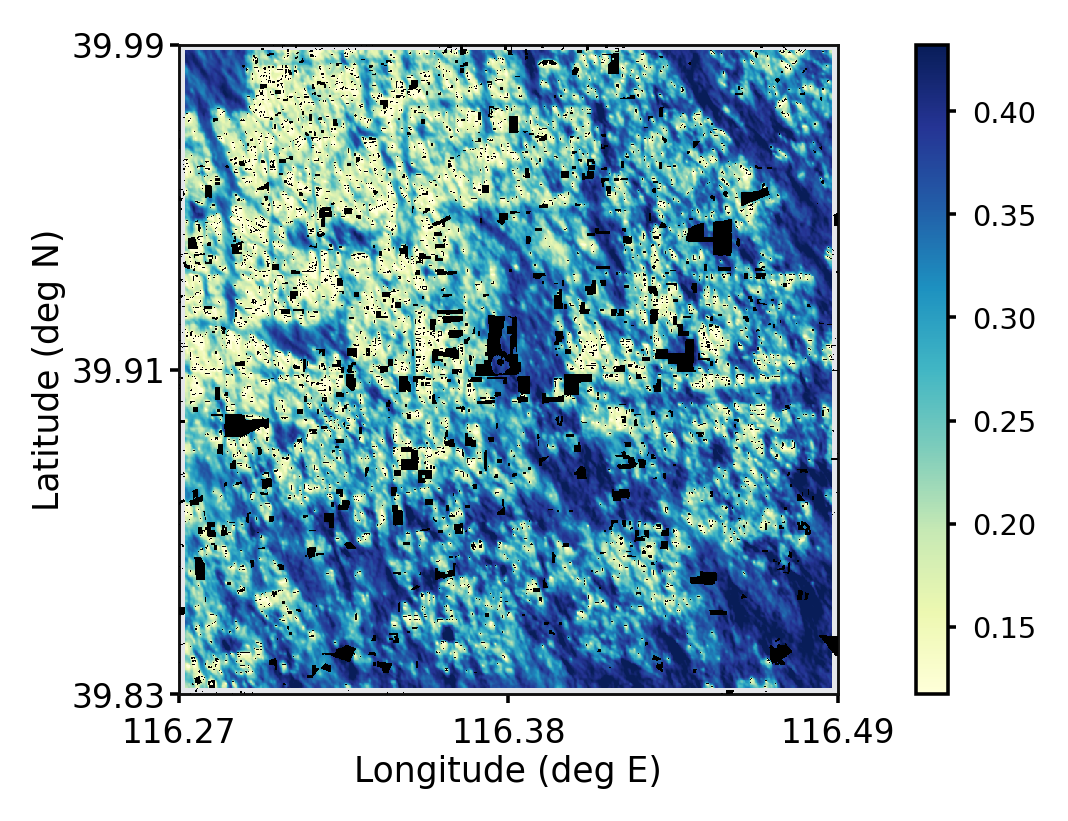} &
        \WBCPanel{0.47\textwidth}{\textbf{(b)} $C^0$ map at 80 m}{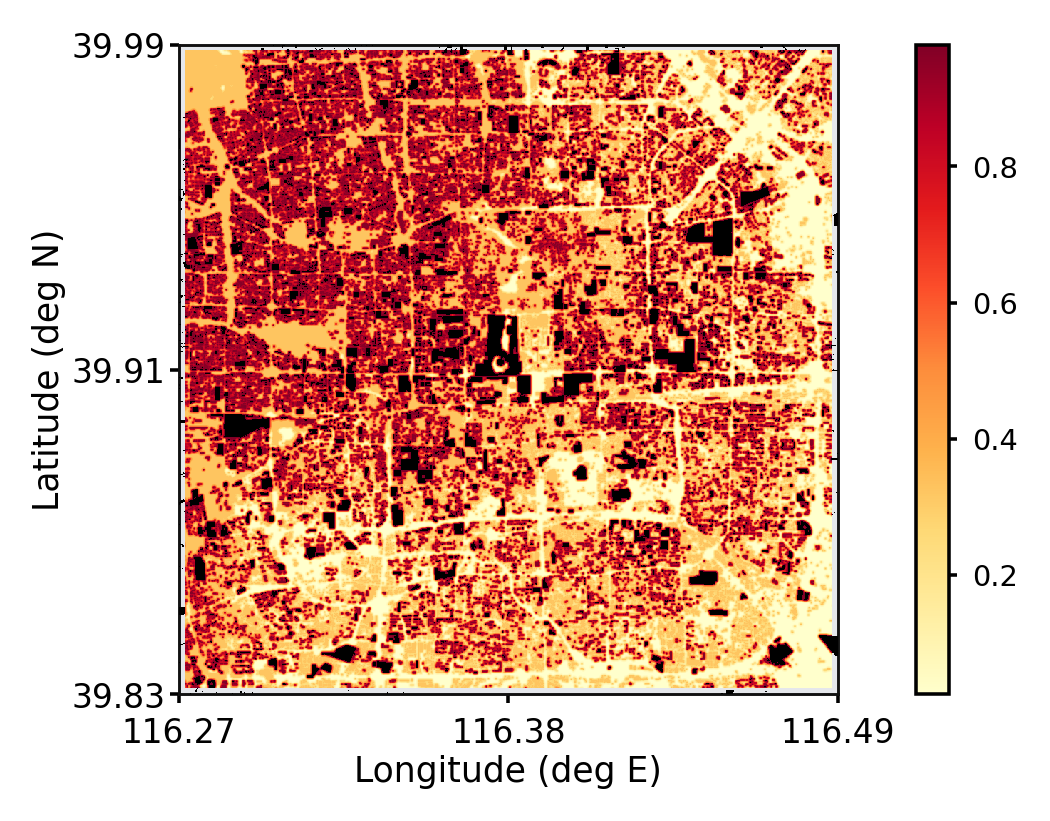} \WBCPanelGap
        \WBCPanel{0.47\textwidth}{\textbf{(c)} $B^0$ spatial-median profile}{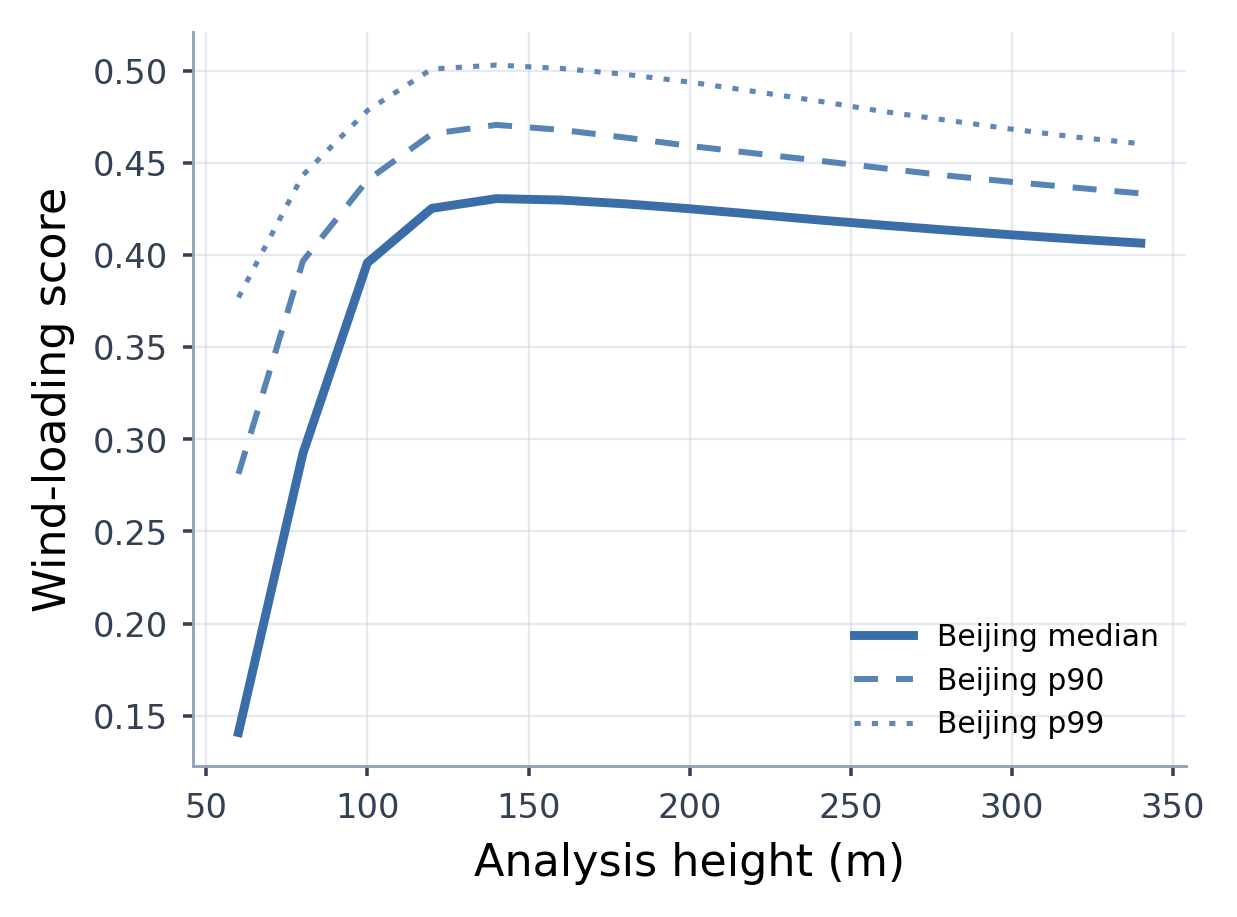} &
        \WBCPanel{0.47\textwidth}{\textbf{(d)} $C^0$ spatial-median profile}{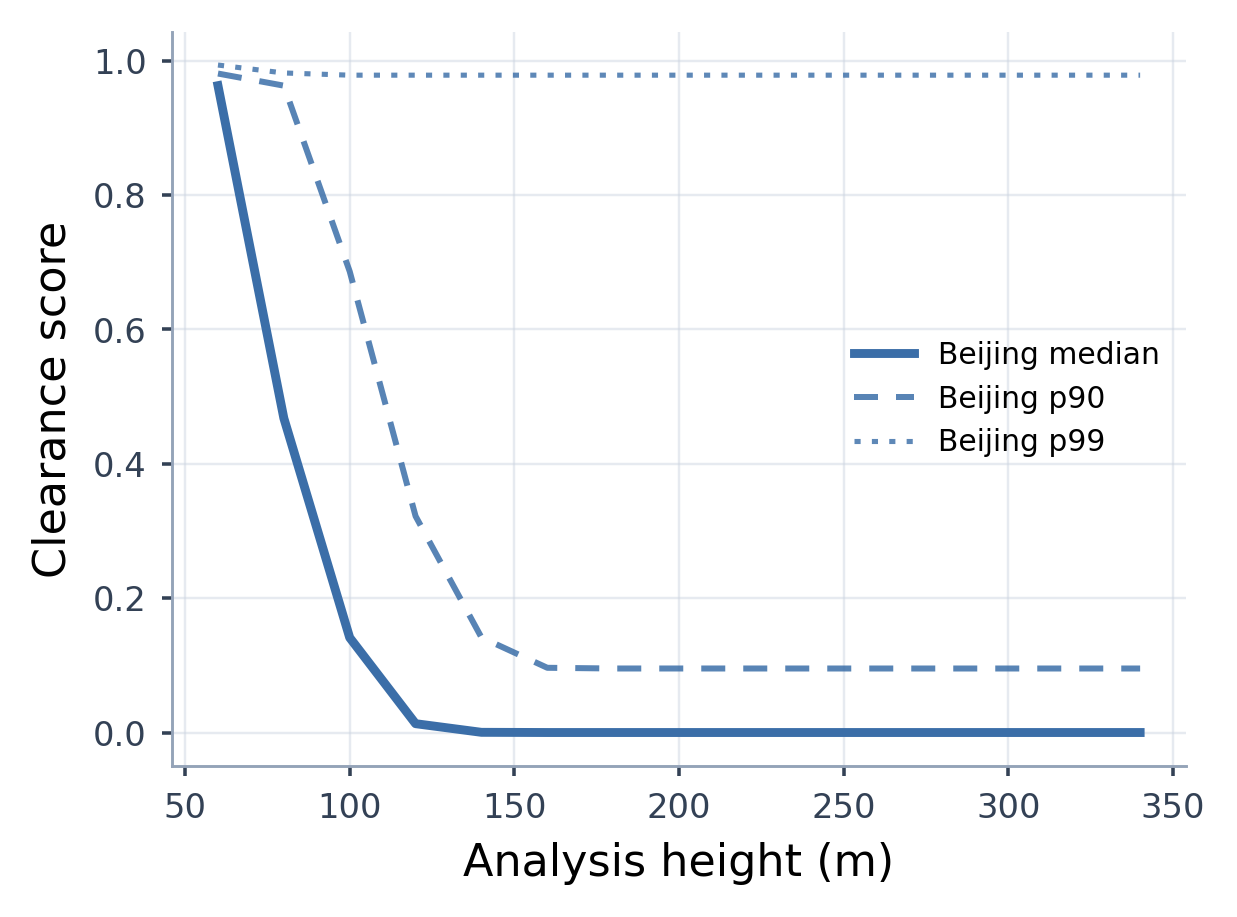}
    \end{tabular}
    \endgroup
    \caption{Beijing nominal-support baseline context from the $N=16384$ trilinear-sampling analysis. \textbf{(a-b)} Native-cell sample-mean fields at 80 m for fixed-support wind loading $B^0$ and nominal-support clearance exposure $C^0$; black cells are occupied or restricted cells outside the airspace-cell set. \textbf{(c-d)} Spatial-median height profiles across finite airspace cells, with upper spatial quantiles retained as distribution context.}
    \label{fig:app_beijing_baseline}
\end{figure}

The Beijing case is retained as a same-design check of the Shanghai main-text method. Figure~\ref{fig:app_beijing_baseline} reports the nominal-support wind-loading and clearance-exposure context. The maps report the cell-level sample-mean fields at the 80 m native scene height, while the profiles report spatial medians across airspace cells at each height.

The Beijing $B^0$ profile increases from a lower-altitude layer median of 0.141 at 60 m to a maximum median of 0.431 at 140 m, then decreases gradually to 0.406 at 340 m. The median $C^0$ is 0.963 at 60 m and 0.468 at 80 m, then drops to 0.141 at 100 m and approaches zero above 140-160 m. At the same time, the 90th and 99th percentile $C^0$ curves remain high through the upper reported layers, indicating that tall or locally constrained building clusters continue to produce high clearance exposure even when the domain-wide median support has moved away from terrain-building proximity.

\begin{figure}[!htbp]
    \centering
    \begingroup
    \setlength{\tabcolsep}{3pt}
    \renewcommand{\arraystretch}{0.86}
    \begin{tabular}{@{}cc@{}}
        \WBCPanel{0.47\textwidth}{\textbf{(a)} $\Delta B^{\mathrm{traj}}$ map at 80 m}{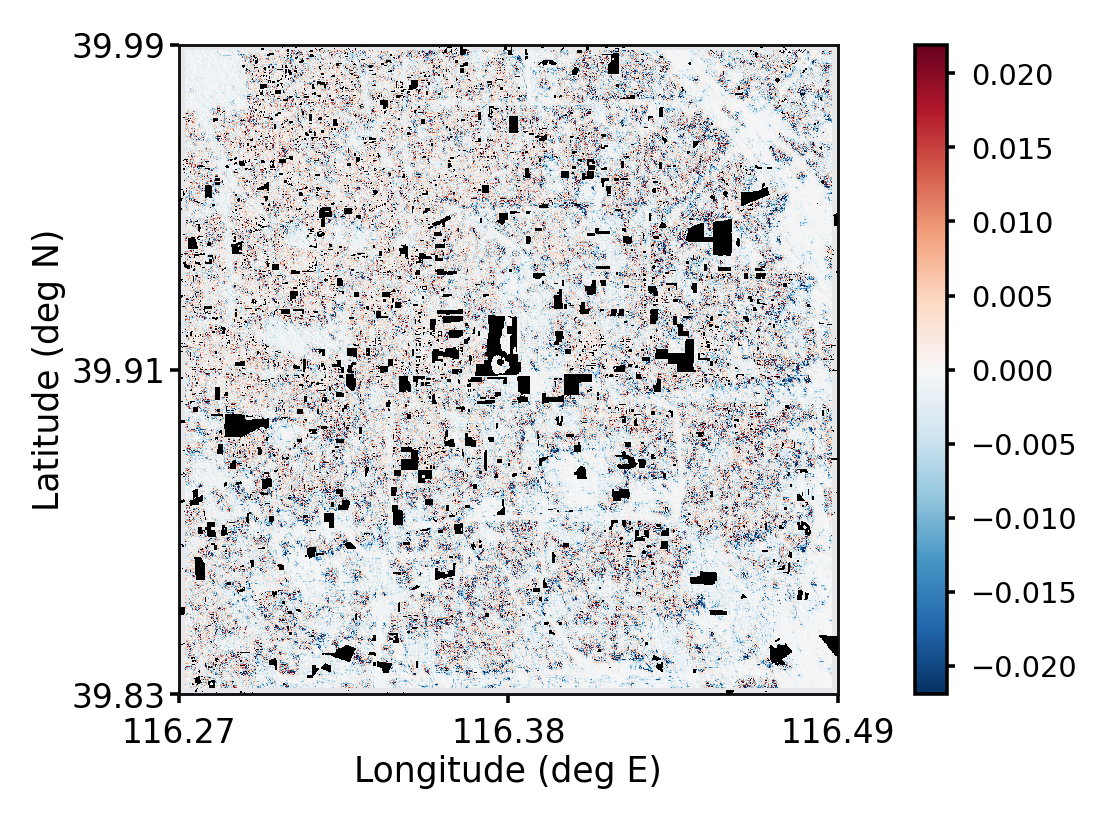} &
        \WBCPanel{0.47\textwidth}{\textbf{(b)} $\Delta B^{\mathrm{traj}}$ map at 160 m}{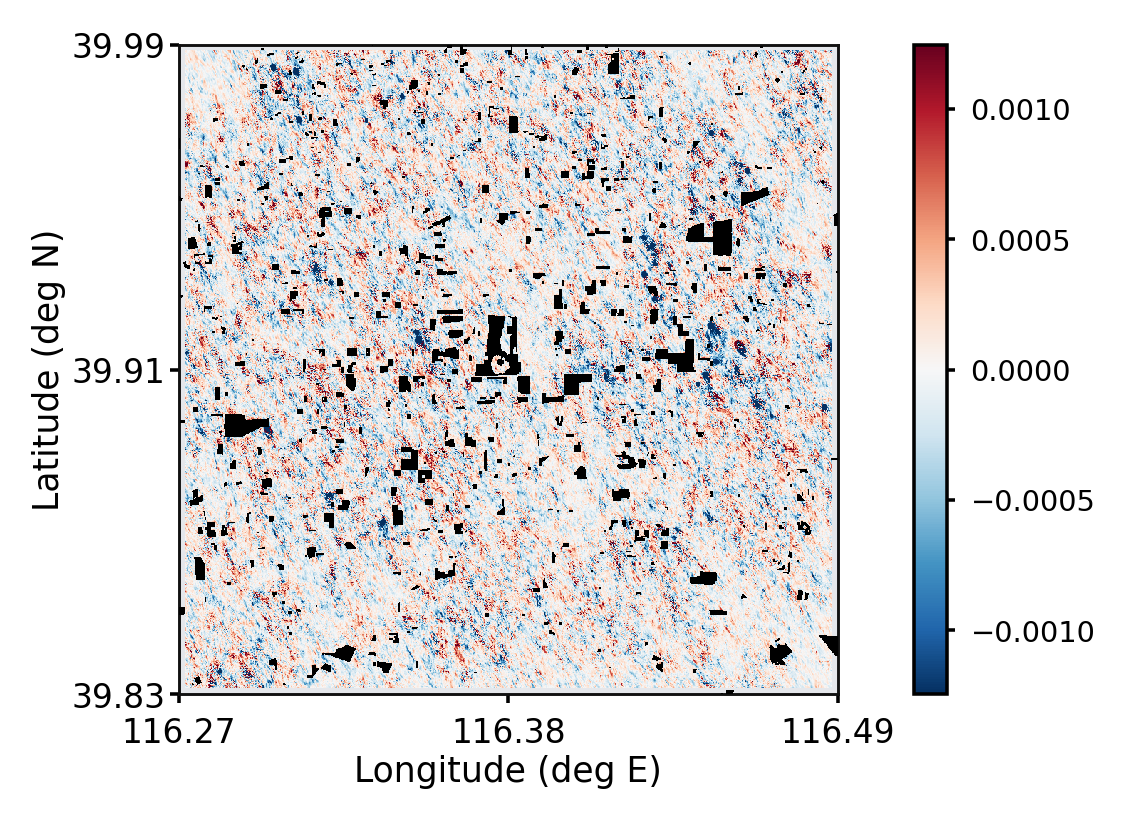} \WBCPanelGap
        \WBCPanel{0.47\textwidth}{\textbf{(c)} $\Delta C^{\mathrm{traj}}$ map at 80 m}{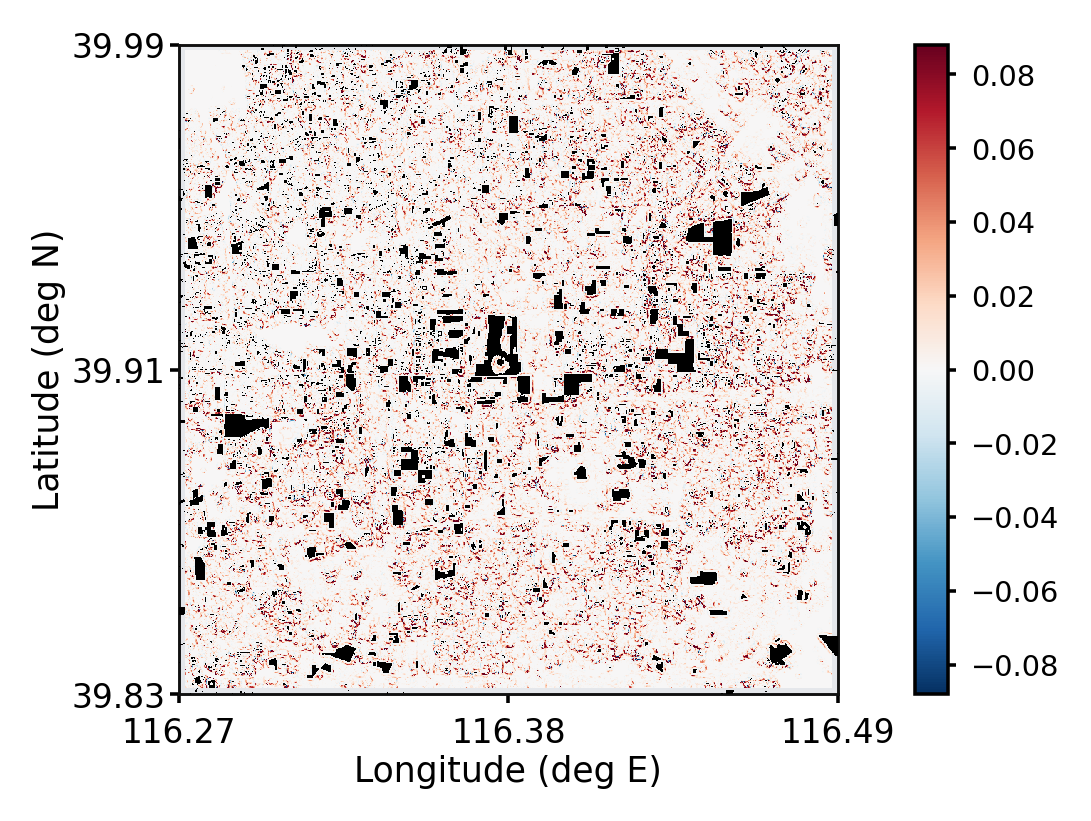} &
        \WBCPanel{0.47\textwidth}{\textbf{(d)} $\Delta C^{\mathrm{traj}}$ map at 160 m}{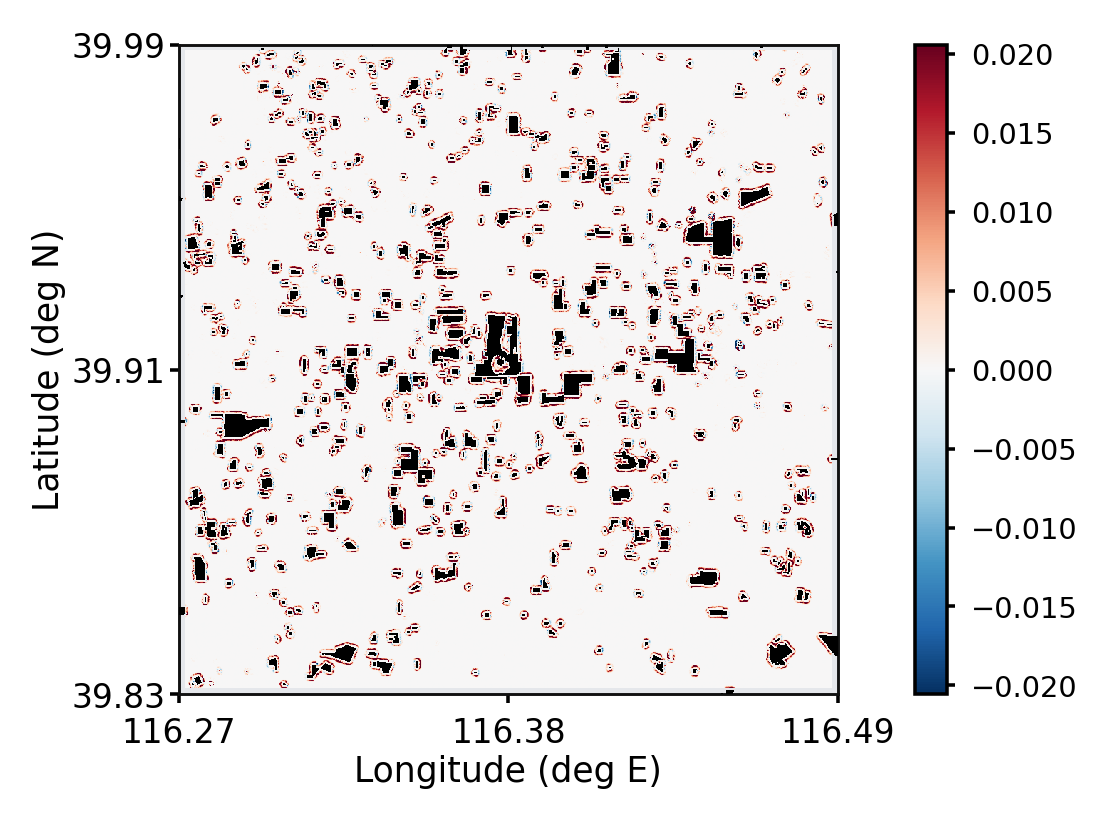}
    \end{tabular}
    \endgroup
    \caption{Beijing trajectory-induced wind-loading and clearance-exposure changes from the $N=16384$ trilinear-sampling analysis. \textbf{(a-b)} Native-cell sample-mean $\Delta B^{\mathrm{traj}}$ fields at 80 m and 160 m. \textbf{(c-d)} Native-cell sample-mean $\Delta C^{\mathrm{traj}}$ fields at the same heights. The maps use zero-centred diverging scales; black cells are occupied or restricted cells outside the airspace-cell set.}
    \label{fig:app_beijing_delta_maps}
\end{figure}
\FloatBarrier

\begin{figure}[!tbp]
    \centering
    \begingroup
    \setlength{\tabcolsep}{3pt}
    \renewcommand{\arraystretch}{0.86}
    \begin{tabular}{@{}cc@{}}
        \WBCPanel{0.47\textwidth}{\textbf{(a)} $\Delta C^{\mathrm{traj}}$ spatial-median profile}{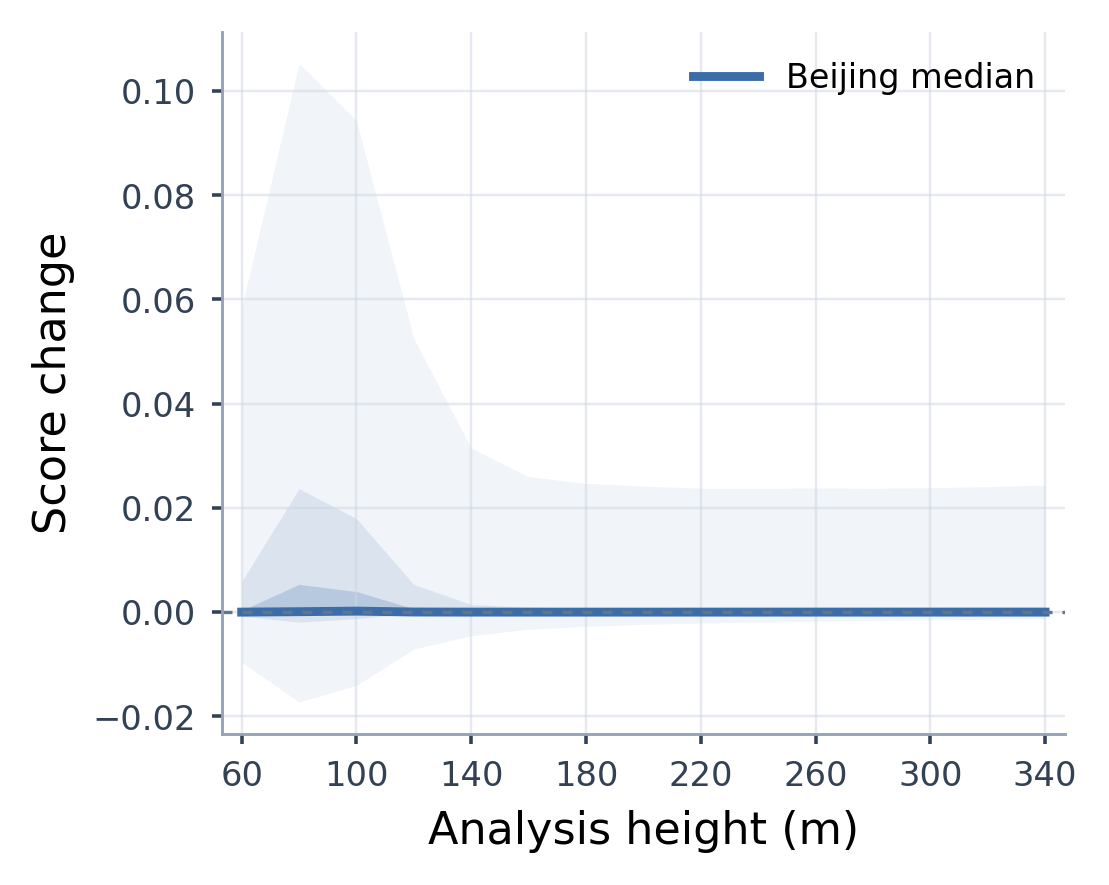} &
        \WBCPanel{0.47\textwidth}{\textbf{(b)} $\Delta B^{\mathrm{traj}}$ spatial-median profile}{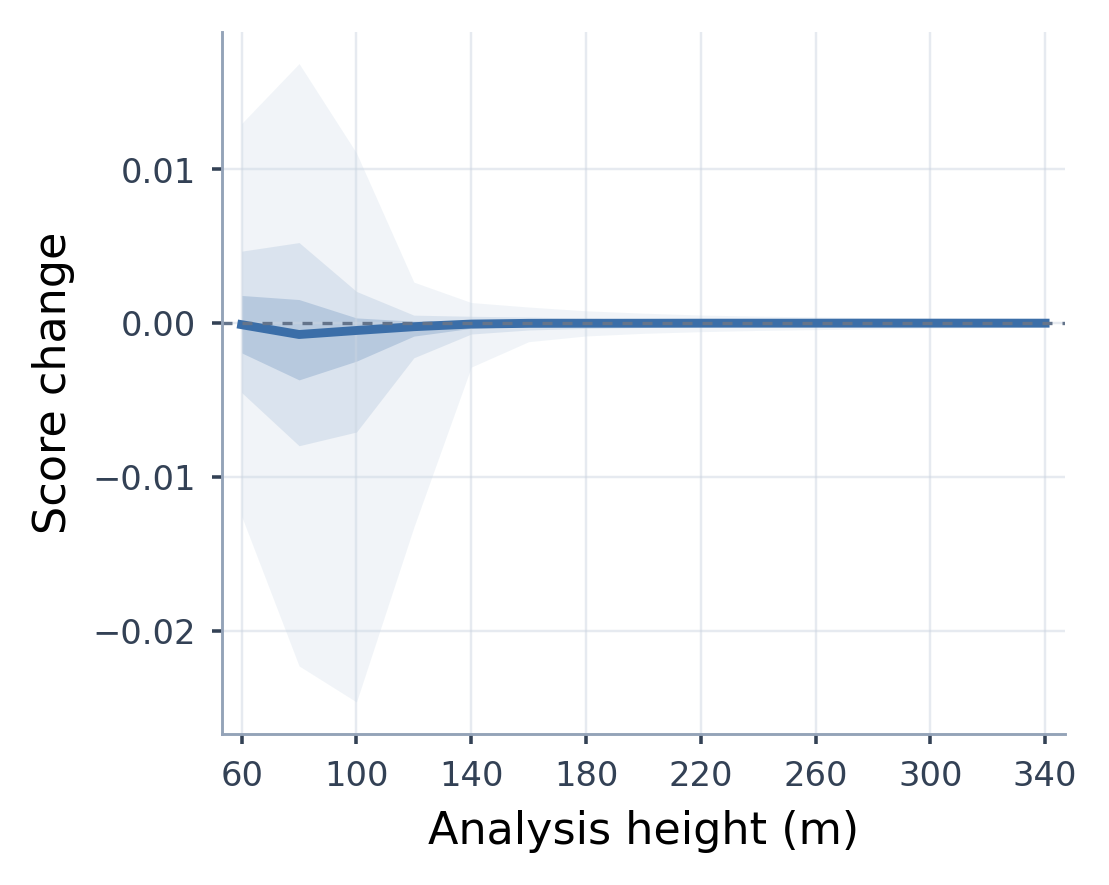} \WBCPanelGap
        \multicolumn{2}{c}{\WBCPanel{0.55\textwidth}{\textbf{(c)} Sign fractions for $\Delta C^{\mathrm{traj}}$}{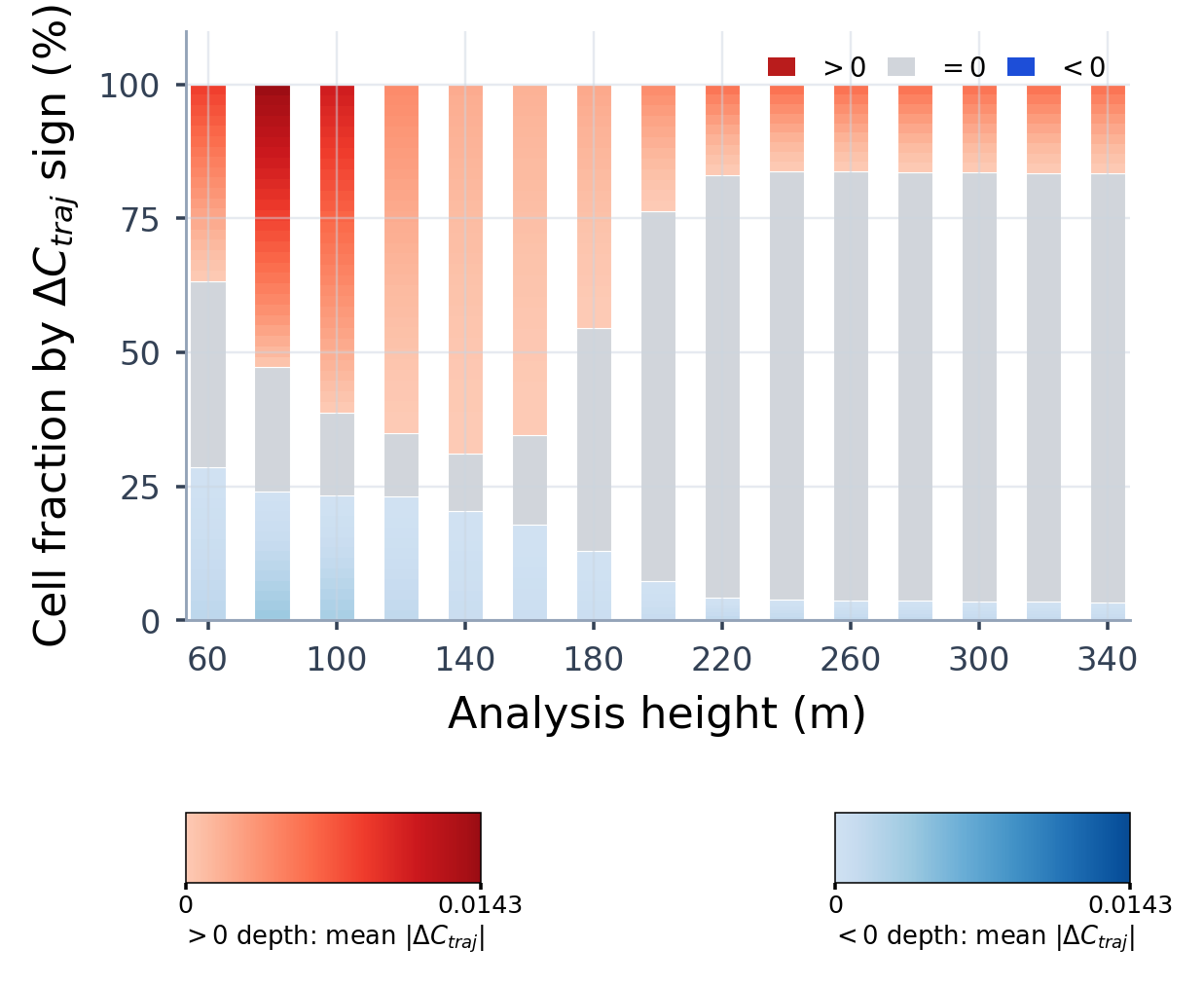}}
    \end{tabular}
    \endgroup
    \caption{Beijing vertical summaries for trajectory-induced contrasts. \textbf{(a-b)} Spatial-median profiles for $\Delta C^{\mathrm{traj}}$ and $\Delta B^{\mathrm{traj}}$ across reported heights, with interquartile, 10th-90th, and 1st-99th percentile bands. \textbf{(c)} Fraction of finite airspace cells with positive, zero, and negative clearance-exposure change by height.}
    \label{fig:app_beijing_delta_profiles}
\end{figure}

In Beijing, the $\Delta B^{\mathrm{traj}}$ panels have weak and sign-mixed wind-loading changes: at 80 m, the spatial median is $-7.4\times10^{-4}$, with 60.3\% of finite airspace cells negative and 39.7\% positive; by 160 m, the median is nearly zero ($-1.2\times10^{-5}$), with positive and negative cells almost balanced. The $\Delta C^{\mathrm{traj}}$ panels contain a small lower-altitude layer clearance response. At 80 m, the spatial median is $8.3\times10^{-5}$, the 90th percentile is 0.024, and 52.8\% of finite airspace cells are positive. At 160 m, the median is effectively zero; the 95th and 99th percentiles remain 0.0060 and 0.026, respectively. The vertical profile in Figure~\ref{fig:app_beijing_delta_profiles} follows the same pattern: the 90th percentile falls from 0.024 at 80 m to $9.1\times10^{-4}$ at 160 m, while zero-change cells increase to 68.9\% at 200 m and 80.0\% at 340 m.

\FloatBarrier


\clearpage
\begingroup
\hbadness=3000
\bibliographystyle{elsarticle-num}
\bibliography{references}
\endgroup

\end{document}